\documentclass[iop,numberedappendix]{emulateapj}
\usepackage{epsfig}
\usepackage{color}
\usepackage{natbib}  
\usepackage{graphicx}
\usepackage{subfigure}

\begin{document}

\font\tenbg=cmmib10 at 10pt
\def \rvecxi{{\hbox{\tenbg\char'030}}}
\def \rvecphi{{\hbox{\tenbg\char'036}}}
\def \rvecdelta {{\hbox {\tenbg\char'016}}}
\def \rvecepsilon {{\hbox {\tenbg\char'017}}}
\def \rvecmu{{\hbox{\tenbg\char'026}}}
\def \rvecOmega {{\hbox {\tenbg\char'012}}}


\title{Gas outflows in Seyfert galaxies: effects of star formation versus AGN feedbacks}

\author{C. Melioli\altaffilmark{1}, 
and E.M. de Gouveia Dal Pino\altaffilmark{1}}

\altaffiltext{1}{Department of Astronomy (IAG-USP),
                 University of Sao Paulo, Brazil;
                 claudio.melioli@iag.usp.br; dalpino@iag.usp.br}

\begin{abstract}
Large scale, weakly collimated outflows are very common in galaxies
with large infrared luminosities. 
In complex systems in particular,  where intense star formation (SF) coexists 
with an active galactic nucleus (AGN), it is not clear yet from observations 
whether the SF, the AGN, or both are driving these outflows.
Accreting supermassive black holes (SMBHs) are expected to influence their 
host galaxies through kinetic and  radiative feedback processes, but  in a
Seyfert galaxy where the energy emitted in the nuclear region is 
comparable to that of the body of the galaxy, it is possible that  stellar 
activity is also playing a key role in these processes. In order to achieve a 
better understanding of the mechanisms driving the gas evolution specially at 
the nuclear regions of these galaxies, we have performed high-resolution 
three-dimensional hydrodynamical simulations with radiative cooling 
considering  the feedback from both star formation regions including  
supernova (type I and II) explosions and an AGN jet emerging from the central 
region of the active spiral galaxy.
We  computed the gas mass lost by the system, separating the 
role of each of these injection energy sources on the galaxy evolution and 
found that at scales within one kiloparsec an outflow can be generally 
established considering intense nuclear star formation only. 
The jet alone is 
unable to drive a massive gas outflow, although it can 
sporadically drag and accelerate clumps of the underlying outflow to very 
high velocities.

\end{abstract}


\section{Introduction} 
 \label{sec:introduction}

Gas outflows extending for few kiloparsec scales (kpc) or larger are 
frequently observed in local galaxies with high
star formation rates (SFRs) and in active galactic nuclei (AGN). 
A large fraction of the outflows is detected in high-redshift, active star 
forming sources, and for this reason it is common to believe that outflows 
represent a stage of the galactic evolution \citep[see, e.g.,][]{sharpley}. 
Such outflows affect the multiphase distribution of the 
interstellar medium \citep[ISM; see, e.g.,][]{mckee95} and are responsible 
for the dynamical and chemical evolution of the galaxy, determining the 
enrichment of the intergalactic medium 
\citep[IGM; see e.g.][and references therein]{heckman90, melioli09, melioli13, 
melioli15} as well as the deficit of baryons seen in many galaxies.

Outflows may also quench star formation by heating up the cold gas component 
and ejecting it from the host galaxy \citep[see, e.g.,][]{binney04}. 
High energy outflows have been proposed as the main cause of the drop in AGN 
luminosity and the end of the galactic gas fuelling onto the central super 
massive black hole (SMBH), limiting the active phase of the nucleus to a time 
of about $10^8$ yr. Outflows are also particularly important for starburst 
(SB) galaxies whose intrinsically large SFR provides the perfect conditions 
for gas outflow to develop.

In complex systems where an intense star formation (SF) activity coexists 
with an AGN, it is unclear whether the 
SF, the AGN, or both are driving the outflows.
It is known that accreting SMBHs influence their host galaxies through 
kinetic/radiative feedback processes, but the details of these processes 
remain unclear \citep[e.g.][]{storchi09, storchi10, Morganti5063}. 
In particular, in Seyfert 
galaxies, where the nuclear source emits an amount of energy comparable to
that of the host galaxy, it would be possible that the outflows are driven 
predominantly by the stellar activity, which therefore would have a key role 
in the evolution of the system.

There are several studies that assume a thermal AGN feedback in the vicinity 
of the black hole \citep[e.g][]{diMatteo05, springel05, 
johansson.et.al.09}.
Observations indicate that there may be two distinct modes of AGN feedback 
\citep[see][for a review]{fabian12}.
The kinetic-mode (also denominated radio mode), thought to occur at low 
Eddington ratios ($L/L_{\rm Edd} <$ 0.1), is expected to operate through heat 
injection into the surrounding gas by a relativistic jet.
In the quasar-mode, likely linked to high Eddington ratios 
($L/L_{\rm Edd} >$ 0.1), the AGN power is believed to couple directly to the 
interstellar medium of the host galaxy via radiation pressure or accretion 
disc winds, resulting in fast outflows \citep[e.g.][]{aalto12, cicone14, 
genzel14, tombesi15}. The general conclusions of these 
studies is that AGN feedback can unbind a large fraction of the baryons 
provided that a few percent of the bolometric luminosity couples thermally to 
the ISM. 
This applies specially in the framework of quasar or 
radiative$-$mode AGN feedback 
\citep[see, e.g.,][]{king03, ostriker10, faucher12, zubovas12}.

\citet[][]{GaspMel11, GaspMel11b} and \citet[][]{GaspBrighTemi} performed 
three-dimensional simulations of the evolution of the gas in clusters, 
groups of galaxies and elliptical galaxies taking into account the AGN 
feedback and found that a mechanical feedback with these 
lower efficiencies would be enough to both constrain the growth of the black 
hole and balance the  radiative cooling  resulting in gas 
density and temperature distributions compatible  with the observation and 
with the results of similar studies \citep[see, e.g.,][]{DiegoGall, 
Falceta_Cap, ciotti10, ciotti11}.

However, these studies, in general, have focussed on the large (100 kpc $-$ 
1 Mpc) scale evolution of the gas, where most of the energy of the 
SMBH jet is directly transferred to the intra-group or intra-cluster medium 
carrying  out a large amount of gas and probably preventing cooling 
flows.
Almost no work has addressed the effects of the 
AGN on the host galaxy, where the highly collimated jet has possibly very 
weak interactions with the surrounding medium.
Of course, there is a theoretical interest in the ability of AGNs to affect 
the gas mass in the nuclear region and its star formation history 
\citep[e.g.,][]{croton06, schawinski07}, but the inclusion of effects like the 
interplay between star formation and the AGN activity, the merger of galaxies 
and secular processes, in general, is complex, and the parameter space is 
not yet fully mapped, mainly at small (1 kpc) scales. 

In a recent work, \citet[][]{wagner12}
investigated with help of HD simulations the detailed physics of the feedback 
mechanism between a SMBH jet and a two phase fractal ISM in the kpc-scale core 
of a galaxy, but their results refer only to the early stages of the jet 
evolution, when it propagates through the disk for the first time. 
In this case the dense embedded clouds of the ISM are accelerated by the ram 
pressure of the high velocity flow through the porous channels of the warm 
phase gas, but probably this regime cannot be maintained for much time after 
the jet digs a hole in the disk and beyond it and therefore, this cannot 
be taken as a stationary solution.

Generally, feedback of the ISM implies not only energy and momentum 
injection by supernovae and stellar winds \citep[e.g.][]{mckeeostriker77}, 
but also radiation from massive stars and from the SMBH 
\citep[][and references therein]{thompson05}. 
The coupling between photons and the interstellar gas is expected to be very 
weak if only Thompson cross-section is implied 
\citep[e.g.][]{ciotti01}, but this coupling is enhanced in 
the presence of dust. 
Dust grains are able to scatter and absorb UV radiation re-emitting it into 
infrared. Thus since the gas is coupled to the grains  hydrodynamically, 
radiation pressure on the dust may be able to push the gas against its 
self-gravity providing  an additional driving to  gas outflow.

When momentum from stellar radiation pressure and radiation pressure on
larger scales via the light that escapes from star-forming regions is 
considered \citep[see][]{hopkins11, hopkins12}, a more realistic 
multi-phase ISM develops, maintaining a reasonable fraction of the ISM at 
densities where the thermal heating from supernovae has a larger effect. 
The potential importance of radiative momentum feedback has been explored for 
some time now \citep[see, e.g.,][]{Haehnelt95, ciotti07}, and its
potential role in the context of AGN and SBs was first pointed out by 
\citet[][]{murray05} and \citet[][]{thompson05}. 
It might be considered in principle
at least in some situations to study the evolution of the gas of a galaxy 
hosting an AGN.

In this study we will consider kinetic (or radio) mode AGN feedback and
neglect the radiation and dust driven mechanisms
\citep[see also][]{martin05, sharma11}. The photo-ionizing radiation 
will be considered only to justify the minimum temperature of the diffuse gas, 
T$_{min,g} \sim 10^4$ K. In fact, we assume that all the mechanical energy and
momentum are effectively deposited into the ISM, so that the mechanical 
feedback will dominate over the radiative feedback. This is a reasonable 
assumption for Seyfert galaxies for which the column densities are about 
$\sim 10^{22}$ cm$^{−2}$ (or less), as pointed out by \citet[][]{ciotti11}. 
Moreover, although dusty structures, in the form of spiral and filaments at 
hundreds of parsecs scales are often observed in early-type active galaxies 
\citep[see, e.g.,][]{simoes07} indicating that a reservoir of dust is 
a necessary condition for the nuclear activity, in Seyfert galaxies the dust 
mass is lower than that observed in ultraluminous IR galaxies (ULIRGs), and a 
rapid destruction in supernova-generated shock waves may make the typical 
dust lifetime shorter than the time over which the galaxy is active 
\citep[][]{McKeeDust, JonesDust}.

In order to achieve a better understanding of the processes driving the 
nuclear gas evolution in a Seyfert galaxy, in this study we perform fully 
three-dimensional (3D) 
hydrodynamical simulations at high (1.9 pc) resolution taking into account 
the effects of radiative cooling and considering the feedback from  
both star formation regions with type I and II SNe and from a collimated 
SMBH jet propagating from the central region of an active 
spiral galaxy. We compute the gas mass lost by the system and separate the 
role of each energy injection source on the galaxy evolution.

The paper is organized as follows. In Section 2 we summarize some recent 
observational results about the gas dynamics in the inner kiloparsec scale 
of Seyfert galaxies. 
In Section 3 we describe the main physical processes in the nuclear
region of an active galaxy and present solutions for a steady state system.
In Section 4 we outline the main characteristics of our model, while in 
Section 5 we present the main results of the simulations and finally, in 
Section 6, we summarize our conclusions.

\section{Observations}

Seyfert galaxies are usually spiral galaxies whose nuclei are exceptionally 
bright. The original definition of the class \citep[][]{seyfert43} was 
primarily morphological, i.e., these are galaxies with high surface 
brightness nuclei, but today the definition has evolved so that Seyfert 
galaxies are now identified spectroscopically by the presence of strong, 
high-ionization emission lines.
\citet[][]{Khachikian74} were the first to realize that exist two distinct 
subclasses of Seyfert galaxies, Type I and Type II, which are distinguished 
by the presence or not of broad bases of  permitted emission lines. 
The spectra of Type I Seyfert galaxies show broad lines that include both 
allowed lines, like HI, HeI or HeII and narrower forbidden lines, like OIII
($broad$ $line$ $regions$, BLR), while the spectra of Type II Seyfert galaxies 
show only (permitted and forbidden) narrow lines ($narrow$ $line$ $regions$,
NLR). In some cases the spectra show both broad and narrow permitted lines, 
and these objects are then classified as an intermediate type between Type I 
and Type II, such as Type 1.5 Seyfert.
NLRs are characteristic of a low-density ionized gas with widths corresponding 
to velocities of several hundred kilometres per second, while BLR have widths 
of up to $10^4$ km s$^{-1}$, and the absence of broad forbidden-line emission 
indicates that the broad-line gas is of high density, so the 
non$-$electric$-$dipole transitions are collisionally suppressed. BLRs are 
associated to the nuclear region (r$_{BLR}$ $\le$ 1 pc) where the relativistic 
SMBH jet develops, while NLRs are observed between 10 pc and 1 kpc, and are 
associated to less dense clouds that may be optically thick and/or optically 
thin.

The physical conditions of the NLRs have been observed
accurately, and in the last years a large number of studies have shown that 
photoionization by the central continuum source plays a dominant role in
exciting the gas \citep[e.g.][]{kraemer00a, kraemerEtal00, bennert06a, 
bennert06b}, although shocks may play
an important role in localized regions \citep[][]{kraemer00b, dopita02, 
grovesdopita04}.
Generally, the NLR is the largest observable feature which 
may be affected by the AGN radiation and its dynamical forces. Therefore, its 
investigation is crucial
to understand the AGN structure and evolution, as well as the interaction 
between the nuclear engine and the circumnuclear ISM and the stellar activity 
of the host galaxies.
Despite a large number of ground-based long-slit spectroscopic studies,
up to now a consensus about the kinematics of the NLR
and its velocity distribution has not been reached yet. 
Several dynamical models have been proposed such as 
infall, rotation, outflow 
\citep[e.g][and references therein]{veilleux91,
fraquelli00, fraquelli03, fisher10, muller11}, and radial acceleration by 
radio jets or tangential expansion of the gas around the jets 
\citep[][]{winge97, axon98}.
However, observations with the Space Telescope Imaging Spectrograph indicated 
that the main component of the motion in the NLR seems to be radial and more 
recently, further observational constraints on the kinematic of the NLR
in Seyfert galaxies have been established
\citep[e.g.][]{muller11}.
Though biconical models can explain very well
the radial velocity profiles of the outflows, there are still
several local variations in the gas kinematics that remain unexplained.

Also recently, observations of Seyfert galaxies based on
integral-field spectroscopy have found evidences of radial outflows in the NLR 
\citep[][]{garcia01, storchi10, muller11}.
In these studies, it were measured the kinematics of 
the ionized and molecular gas surrounding the nucleus of the galaxy and it was
found that the ionized-gas kinematics may be the result of three velocity
components, namely, an extended emission at a systemic  velocity observed
in a circular region around the nucleus, an outﬂowing component along the 
bi-cone characterized by an angle of $\sim$ 60$^{\circ}$, and another component 
due to the interaction of the radio jet with the galactic disk.
The data indicate that the origin of the extended emission is gas from 
the galactic plane that could be ionized by the AGN and suggest that the NLR 
clouds may be accelerated very close to the nucleus ($\sim$ 10 pc), after 
which the flow moves at essentially constant velocity. 
The observed mass outflow rates  vary between 1 
and 10 M$_{\odot}$ yr$^{−1}$, being  $\sim$ 10$^2$ to 10$^3$ times larger than 
the expected accretion rates for an AGN, while the kinetic powers associated to 
the outflow are between 10$^2$ and 10$^4$ times smaller than the typical 
AGN bolometric luminosities \citep[][]{storchi10, muller11}.
Regarding the third kinematic component, that is, the one due to the 
interaction of the radio jet with the ambient gas moving at a systemic 
velocity, there are clear indications that the jet is launched close to the 
plane of the galaxy, pushing the gas it encounters on its way. 
The observed high-velocity gas in the nuclear
regions  of some galaxies as well as the large ratios between M$_{out}$ and 
M$_{acc}$ indicate that the outflow cannot  be attributed to a single 
AGN wind episode. It might be rather due to  a process in which the
interstellar matter around  the nuclear region is pushed away by a previous
AGN wind \citep[][]{muller11}.

Molecular hydrogen emission arises in extended regions
along the axis of the galaxy stellar bar, avoiding the region of
the bi-cone. The H$_2$ velocities are close to systemic with a small
rotation component, supporting an origin in the galactic plane. On the 
contrary, the maximum velocity of the outflowing component ranges
between 120 km s$^{-1}$ and 2000 km s$^{-1}$ and is reached at a typical 
distance of $\sim$ 180 pc from the AGN.

\section{Steady state solutions}

As described, the physical processes occurring in the nuclear region of a 
Seyfert galaxy are quite complex. 
SNI and SNII explode at a given rate injecting energy,
momentum and metals into the ISM. The supernova remnants (SNRs)
generated by the SN explosions expand through a turbulent multiphase medium 
in which clouds and filaments are embedded. At the same time, the gas heated 
by the SNe and by ionizing photons emitted by massive stars is also cooled 
via radiative losses that depend on the square of the gas density and on the
metal abundance affected by the stellar activity. At the center of the galaxy,
the presence of the SMBH also contributes to make more complex the evolution 
of the system. In fact, imaging studies have revealed that structures such as 
small-scale disks or nuclear bars and associated spiral arms are frequently 
observed in the inner kiloparsec of active galaxies 
\citep[][]{erwin99, pogge02, laine03}, transporting gas into the inner few
hundred parsecs. In the same central region, small-scale jets 
emerge from the base of the accreting disk of the black hole, transfering 
a large amount of energy into a very collimated region, producing strong 
shocks and bright emission knots typical of collisionally excited plasma. 
As stressed in \S 1, in this contest it is very hard to understand if the 
observed gas outflow is a consequence of the SMBH feedback or if SNe-driven 
hot superbubbles may fill out the core of the galaxy developing a wind rich 
in filamentary structures and turbulences, or both effects are prevailing.

\subsection{Energy balance}

A first analytical approach study of the evolution of this sort of system may 
be done by comparing the energy injected by the main sources (SNI, SNII, Jet) 
with the energy lost by radiative cooling and the energy required 
to drive a gas outflow.
The rate of SNI explosions is about 0.01(M$_{bulge,10}$) yr$^{-1}$ 
\citep[][]{pain96}, where 
M$_{bulge,10}$ is the stellar mass of the bulge in units of 10$^{10}$ M$_{\odot}$, 
while the rate of SNII explosions corresponds to about 0.01 (SFR$_{SB,1}$) 
yr$^{-1}$ \citep[see][]{melioli09}, where SFR$_{SB,1}$ is the star formation 
rate of a given SB region in units of 1 M$_{\odot}$ yr$^{-1}$. 
Therefore, assuming that each SN releases 10$^{51}$ erg, the injected luminosity
associated to the SN events is:

\begin{equation}
\label{eq:enin}
L_{in} = 3 \times 10^{41} \left[SFR_{SB,1} + M_{BH,7}\right] \ \ {\rm erg \ s^{-1}}
\end{equation}
\noindent
where M$_{BH,7}$ is the mass of the central SMBH in units of 10$^7$ M$_{\odot}$
and where we have considered the observed relation between the bulge and the 
SMBH mass.

\begin{figure}
\begin{center}   
\psfig{figure=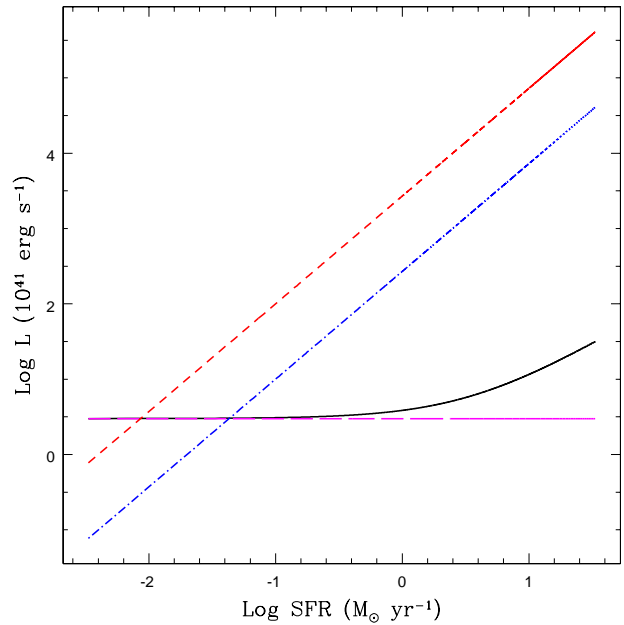,width=0.49\textwidth}    
\end{center}   
\caption{Rate of energy injected and lost in the nuclear region of a Seyfert 
galaxy considering an active region with cylindrical geometry 
characterized by a radius $R$=300 pc and a
thickness $h$= 200 pc as a function of the SFR. 
Solid (black) line: energy injected by SNI and SNII; 
long-dashed (magenta) line: energy injected by the SNI only; 
dashed (blue) line: energy lost by radiative cooling assuming a fixed value for
the cooling function $\Lambda(T) = 10^-{23}$ erg s$^{-1}$ cm$^{-3}$; 
dot-dashed (red) line:  energy lost by radiative cooling assuming a fixed 
value 
for the cooling function $\Lambda(T) = 10^-{22}$ erg s$^{-1}$ cm$^{-3}$.
The SFR is given in M$_{\odot}$ yr$^{-1}$ (log$_{10}$ scale) while the energy 
rates are given in units of 10$^{41}$ erg s$^{-1}$ (log$_{10}$ scale).}
\label{fig:sn_inject} 
\end{figure}

The total rate of radiative cooling in an optically thin plasma within a 
cylindrical region of radius R$_{SB,300}$ (in units of 300 pc) and thickness
h$_{SB,200}$ (in units of 200 pc) is:
\footnote{As remarked in the introduction we are neglecting dust 
driven mechanisms and radiation pressure because of the low dust column 
densities observed in the nuclear regions of the Seys and the expected high 
frequency of SN shocks waves that will help to destroy dust structures. 
At the same time, the shocks as well as photoionization and non-thermal 
heating processes by the AGN will ensure a high gas temperature (we will 
find below in the simulations values as large as $10^9$ K; 
see. e.g., Figures 10 and 17). 
Thus, the assumption of an optically thin gas in near 
ionization equilibrium at the pc-scales around the AGN is reasonable. In this 
case, the thermal radiative cooling function given in Eq. 2 should be a good 
approximation even in the presence of photoionization and non-thermal heating 
by the AGN. This will actually help to keep the surrounding interstellar gas 
at temperatures larger than or equal to $10^4$ K.}
\begin{equation}
\label{eq:enlost}
L_{cool} = 10^{42} \left(\Lambda(T)_{-23} \ R^2_{SB,300} \ h_{SB,200} \ 
n_{10}^2\right) \ \ {\rm erg \ s^{-1}}
\end{equation}
\noindent
where $n_{10}$ is the gas density in units of 10 cm$^{-3}$ at the height $z=0$, 
and $\Lambda(T)_{-23}$ is the cooling function in units of $10^{-23}$ erg 
s$^{-1}$ cm$^{3}$, corresponding to a temperature of about 10$^{4}$ K. 
Given the Kennicutt-Schmidt law \citep[][]{kennicuttLaw} that provides an 
excellent parametrization of the global SFR over a density range extending 
from the most gas-poor spiral disks to the cores of the most luminous SB 
galaxies, after some algebraic operations we obtain
\begin{equation}
\label{eq:sfr}
SFR_{kpc} \sim 10^{-1} \ \left(n_{10} \ h_{SB,200}\right)^{1.4} \ \ 
{\rm M_{\odot} yr^{-1} kpc^{-2}}
\end{equation}
where SFR$_{kpc}$ = 3.5 SFR$_{SB}$/R$_{300}^2$, and substituting Eq. \ref{eq:sfr} 
in Eq. \ref{eq:enlost} we may directly compare the energy injected with the 
energy lost as a function only of the SFR (per kpc$^2$) of a given galaxy.
Therefore, the rate of energy radiated away may be rewritten as:

\begin{equation}
\label{eq:enlostsfr}
L_{cool} = 2.7 \times 10^{43} \left(\Lambda(T)_{-23} \ R^2_{SB,300} \ h_{SB,200}^{-1}
 \ SFR_{kpc}^{1.43}\right) \ \ {\rm erg \ s^{-1}}
\end{equation}
\noindent
and the trend of the energy injected and lost (per second) by the system is 
shown in Fig. \ref{fig:sn_inject}.

Regarding the jet associated to the SMBH, an amount of energy
$E_{BH} \sim 0.1 M_{BH}c^2$ could in principle be injected in the 
surrounding gas. This means that about 5$\times 10^{44} M_{BH,7}$ erg s$^{-1}$ 
may be released over a time of $10^8$ yr, a value significantly larger 
than the binding energy of the interstellar medium in a typical spiral or
elliptical galaxy, or even in a group of galaxies. 
However, in a Seyfert galaxy the luminosity
associated to the central active jet is lower, between $10^{41}$ and $10^{42}$ 
erg s$^{-1}$, i.e., of the same order of the luminosity injected by the SNe 
(Eq. \ref{eq:enin}). It is also comparable to the energy carried out per 
second by the gas mass outflow, that is 
$\dot{E}_{out} \sim 0.5 \dot{M} v_{out}^2$. 
Moreover, unlike the events described above, near 
the central black hole, the jet is strongly collimated and its interaction
with the ISM of the host galaxy is limited to a small fraction of the total
gas mass of the system, of the order of $10^{-2} (R_{jet,20} / 
R_{300})^2$, where $R_{jet,20}$ is the width of the jet within the galaxy disk 
in units of 20 pc. Therefore, we can expect that, except for the very early 
stages of evolution, the SMBH jet will influence only weakly the increase  of 
the mass outflow, so that we can neglect its contribution in Eq. \ref{eq:enin}.

From Fig. \ref{fig:sn_inject}, we note that for a cooling function with a 
value of $\sim 10^{-23}$ erg s$^{-1}$ cm$^{3}$, which is a lower limit for a gas 
at T=10$^4$ K, the rate of energy injected is larger than the rate of energy 
radiated away only for a SFR smaller than $\sim$ 0.1 M$_{\odot}$ yr$^{-1}$ 
kpc$^{-2}$, and for a value of $\sim 10^{-22}$ erg s$^{-1}$ 
cm$^{-3}$, which is an upper limit for a gas at T=10$^4$ K, the rate of energy 
injected is larger than the lost energy rate only for SFR smaller than 
10$^{-2}$ M$_{\odot}$ yr$^{-1}$. 
In fact, according with the Kennicutt-Schmidt law, the SFR increases slower 
than the square of the density, and therefore there is a critical gas 
density value above which an outflow cannot develop because of the 
radiative cooling, as indicated in Fig. \ref{fig:sn_inject}.

\subsection{Hot gas and filaments}

The result above is mainly due to the idealized scenario adopted with a 
homogeneous distribution of the gas which misrepresents the expected and 
observed multi-phase ISM \citep[see, e.g.,][]{mckeeostriker77, dettmar90, 
hill14} in the disk of the galaxies.
Indeed, when the stellar winds and the SN energy are injected 
into the SB region, two competing processes occur: on one hand the winds and the
SNRs sweep the diffuse gas, thus lowering its mean density; on the other hand 
the interactions between the shells of the several SNRs determine the 
formation of dense and cold structures, generating 
turbulences and a network of filaments which may coexist with the 
high-temperature, low-density gas \citep[see, e.g.,][]{mel04, melioli05}.
Therefore, the importance of the radiative cooling depends on which of these 
two processes dominate. Besides, the radiative cooling is almost negligible 
in the hot bubbles generated inside the SNRs.

\begin{figure}    
\begin{center}   
\psfig{figure=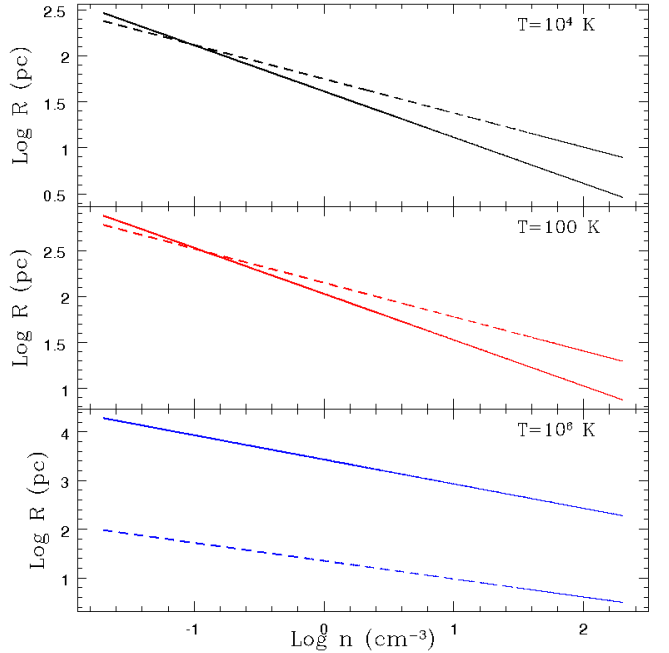,width=0.49\textwidth}    
\end{center}   
\caption{Average distance, $\lambda_{SN}$, between two SNe 
and SNR stall radius, $R_{stall}$, versus density $n$, for three different 
temperatures of the ISM, T=100 K (red lines), T=10$^4$ (black lines) and 
T=10$^6$ K (blue lines).
Solid line: $\lambda_{SN}$; dashed lines: $R_{stall}$. The distances are given 
in pc (log$_{10}$ scale).}
\label{fig:sn_distance} 
\end{figure}

The average distance crossed by a SNR before interacting with another one 
depends on its expansion velocity and thus on the density of the ISM, and 
also on the rate of SN explosions per volume. Assuming that a SNR expands for a 
maximum time t$_{exp} \sim 1.8 \times 10^5 n_{10}^{-0.35} T_4^{0.7}$ yr 
\citep[see equations 6 and 9 in][]{melioli06}, during this period the number
of SN explosions is:

\begin{equation}
\label{eq:Nsn} 
{\cal N}_{stall} \sim 5 \times 10^3 \ n_{10} \ R_{300}^2 h_{200}^{1.4} \ T_4^{0.7}
\end{equation}
\noindent
and the average distance between two SNe is:

\begin{equation}
\label{eq:distsn} 
\lambda_{SN} \sim 13 \ n_{10}^{-0.5} \ T_4^{0.35} \ h_{200}^{-0.7} \ \ {\rm pc}
\end{equation} 
\noindent
Therefore, when the average distance between two SNe is smaller
than the stall radius of the SNRs, these may interact developing a low 
density ambient at high temperature coexisting with filamentary structures. 
Figure \ref{fig:sn_distance} illustrates the average distance (solid lines) 
and the stall radius (dotted lines) as a function of the density and 
temperature of the ISM. 
We note that if the Kennicutt-Schmidt law is valid, 
when the temperature is 10$^4$ K or lower, SNRs always interact
for ambient densities higher than $\sim$ 0.1 cm$^{-3}$. When the temperature 
is about 10$^6$ K, each SN explosion may be considered
as a single event and there is no SNRs interaction mainly because of the 
high thermal pressure of the ISM.  

\begin{figure}    
\begin{center}   
\psfig{figure=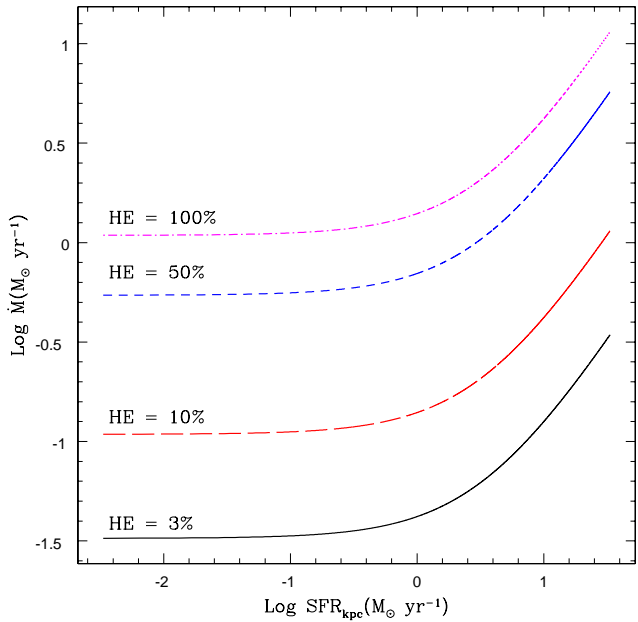,width=0.49\textwidth}    
\end{center}   
\caption{Mass loss rate for a nuclear region of 
a galaxy considering an active volume with a radius 
$R$=300 pc and a thickness $h$= 200 pc as a function of the SFR, for different
values of the heating efficiency, HE. 
Solid (black) line: HE = 3\%; long-dashed (red) line: HE = 10\%;
dashed (blue) line: HE = 50\%; dot-dashed (magenta) line: HE = 100\%.
The SFR and the mass loss rate are given in M$_{\odot}$ yr$^{-1}$ 
(log$_{10}$ scale).}
\label{fig:gas_out} 
\end{figure}

If there is interaction, the successive generations of SNRs expand 
in a medium more and more rarefied, thus becoming adiabatic and eventually
transferring most of their energy to the ISM; in this case
the heating efficiency (HE) of the SNe, that is, the fraction
of the SN explosion energy that remains effectively stored in the ISM 
gas and is not radiated away, will be close to unity. 
On the other hand, in a situation where filaments, clouds and clumps are
destroyed on short time-scales raising again very rapidly the ambient density, 
HE will be very small 
\citep[for a detailed study on the HE see, e.g.,][]{mel04}.   
Therefore, in a multiphase ambient medium characterized by diffuse low
density gas and denser filaments, if we assume an escape 
velocity of the order of $\sim \sqrt{2GM_{sys}/h_Z}$, the mass loss rate of 
the gas in the nuclear region will be approximately given by:

\begin{equation}
\label{eq:lost_mass}
\dot{M} \sim {{\dot{E} - \dot{\cal{L}}}\over
{\left({GM_{sys}}\over{h_z}\right)}} \sim {{{\rm HE} \times {\dot{E}}}\over
{\left({GM_{sys}}\over{h_z}\right)}}
\end{equation}
where $\dot{E}$ and  $\dot{\cal{L}}$ are the injected and radiated power, 
respectively.
Considering an average total mass of the system of $\sim 10^{11}$ M$_{\odot}$
and a height $h_z=1$ kpc, the outflowing bipolar gas mass rate, as a 
function of the global SFR for different values of  HE, is given 
in Fig. \ref{fig:gas_out}. 
The most striking result is that, for values of SFR smaller than or 
approximately equal to 1 M$_{\odot}$ yr$^{-1}$, 
the SB activity drives an outflow smaller than 1 
M$_{\odot}$ yr$^{-1}$, regardless of the SFR. For SFRs larger than 1 
M$_{\odot}$ yr$^{-1}$, different scenarios are possible. 
Depending on the efficiency of the radiative cooling,
a strong outflow may develop or not. If HE is high, between 0.5 and 1, 
a mass loss rate between 25\% and 50\% of the SFR is possible, but 
for smaller HE values either an outflow or an inflow can occur, or 
even both of them concomitantly. 
Of course, the higher the SFR, the density surface and/or the jet 
interaction with the ISM, more complex is the description of the evolution of 
the whole system using a simple approach. For this reason we 
need to perform hydrodynamical simulations in order to provide a better 
description of the main physical events in a nuclear region where a SB 
coexists with an AGN.

\section{The numerical model}
 \label{sec:model}

\subsection{Seyfert galaxy initial setup}

In our model we consider the central (kpc) region of a typical spiral 
(Seyfert) galaxy and set the initial conditions for the ISM following the 
procedure outlined in \citet[][]{melioli08, melioli09, melioli13}. 
We first assume a mass model for the galaxy, which includes the contribution 
of a bulge and a stellar disk, and then we set the ISM in equilibrium with the 
gravitational potential given by the summation of the dark matter halo, the 
bulge, and the disk contributions. 
The dark matter halo is assumed to follow the Navarro, Frenk \& White profile 
\citep{navarro96}:

\begin{equation}
\label{eq:phiblg}
\Phi_{\rm dm} (r)=-{{G M_{\rm vir}}\over{r_{\rm s}f(c)}}{\ln(1+x) 
\over x},
\end{equation}
\noindent
where $r$ is the spherical radius, M$_{\rm vir}$ is the mass at the virial 
radius, $r_{\rm vir}$ defined as the radius where the 
average density is $\sim 10^2$ times the cosmological critical density, 
$\rho_{\rm crit}$, $r_{\rm s}$ is a scale radius, $x$ is the
ratio between $r$ and $r_{\rm s}$, $c$ = $r_{\rm vir}/r_{\rm s}$ is the 
concentration and $f(c)$ = ln$(1+c)-c/(1+c)$.  
In this model we considered M$_{\rm vir}$ = 10$^{11}$ 
M$_{\odot}$, $r_{\rm vir}$ = 120 kpc, $r_{\rm s}$ = 10 kpc and $c = 12$, 
adopting a $\Lambda$ cold dark matter cosmological universe with 
$\Omega_{\rm M}$=0.27, $\Omega_{\Lambda}$=0.73 and $H_0$=71 km s$^{-1}$ Mpc$^{-1}$.
However, we note that due to the limited extension of the galaxy region here 
investigated, 
we verify that the gravitational potential of the dark matter halo has almost 
no influence in the determination of the initial distribution and the 
evolution of the gas.

The gravitational potential of the stellar bulge is given by a Plummer 
distribution \citep{plummer}:
 
\begin{equation}
\label{eq:phipl}
\Phi_{\rm b} (r)=-{{G M_{\rm b}}\over{r_{\rm c} + r}},
\end{equation}
\noindent
where M$_{\rm b}$ is the total mass of the bulge and $r_{\rm c}$ is the core 
radius defined as the radius containing 50\% of the stellar mass. 
The bulge mass is assumed to be always M$_{\rm b}$ = 10$^{10}$ M$_{\odot}$,
while the core radius varies according to the model.

Finally, the gravitational potential of the gas disk is given by a Miyamoto \& 
Nagai profile \citep{nagai}, that is:

\begin{equation}
\label{eq:rhodisk}
\Phi_{disk}(r,z) = - \frac{GM_{disk}}{\sqrt{r^2 + \left(a+
\sqrt{z^2+b^2}\right)^2}}
\end{equation}
\noindent
where M$_{\rm disk}$ is the mass of the disk (for the whole galaxy), while $a$ 
and $b$ are its radial and vertical scales, respectively. In this study we
assume M$_{\rm disk}$ = 5 $\times $10$^{10}$ M$_{\odot}$, $a$ = 8 kpc and $b$ = 
0.8 kpc.
Therefore the total gravitational potential is: 
$\Phi(r,z)$ = $\Phi_{\rm dm}$ + $\Phi_{b}(r,z)$ + $\Phi_{disk}(r,z)$.

The rotating gas in the disk is initially put in equilibrium with the 
galaxy gravitational potential. We obtain the rotation velocity resolving the 
equation $v(r,z) = \sqrt{(r \ {\rm d}\Phi(r,z))/{\rm d}r)_{z={\rm constant}}}$ 
\citep[see][]{melioli08, melioli09, melioli13, melioli15}. 
The values of the velocity obtained in this way mimic quite well the rotation 
curve of a spiral galaxy.

We assume three distinct galaxy models, $SyL$, $SyM$ and $SyH$ 
characterized by gas disk distributions with different total mass, density 
profile and  resulting column densities defined as 
$low$ ($N_H=10^{21}$ cm$^{-2}$), $medium$ ($N_H=10^{22}$ cm$^{-2}$) and $high$ 
($N_H=10^{23}$ cm$^{-2}$), respectively. 
All models have a multi-phase gas distribution, and each phase 
at $t=0$ is defined by a typical temperature $T_{i,0}$ and central density 
$\rho_{i,0}$, initially in equilibrium in the total gravitational potential 
described above. 
The isothermal density distribution of each phase will have the form:

\begin{equation}
\label{eq:disk}
\rho_{i}(r,z) = \rho_{0,i} \; exp \left[-\frac{\Phi(r,z) - 
e_i^2\Phi(r,0)-(1-e_i^2)\Phi(0,0)}{c_{disk,i}^2}\right]
\end{equation}
\noindent
where $c_{disk,i}^2$ is the isothermal sound speed of the $i$-phase of the gas 
and $e_i$ quantifies the fraction of rotational support of the ISM.
In our model the $i$-phase density is replaced by the $(i+1)$-phase density 
wherever the $(i+1)$-phase pressure is larger than the (coldest and densest) 
pressure of the $i$-phase.

\begin{figure}
     \begin{center}
        \subfigure{%
            \includegraphics[width=0.43\textwidth]{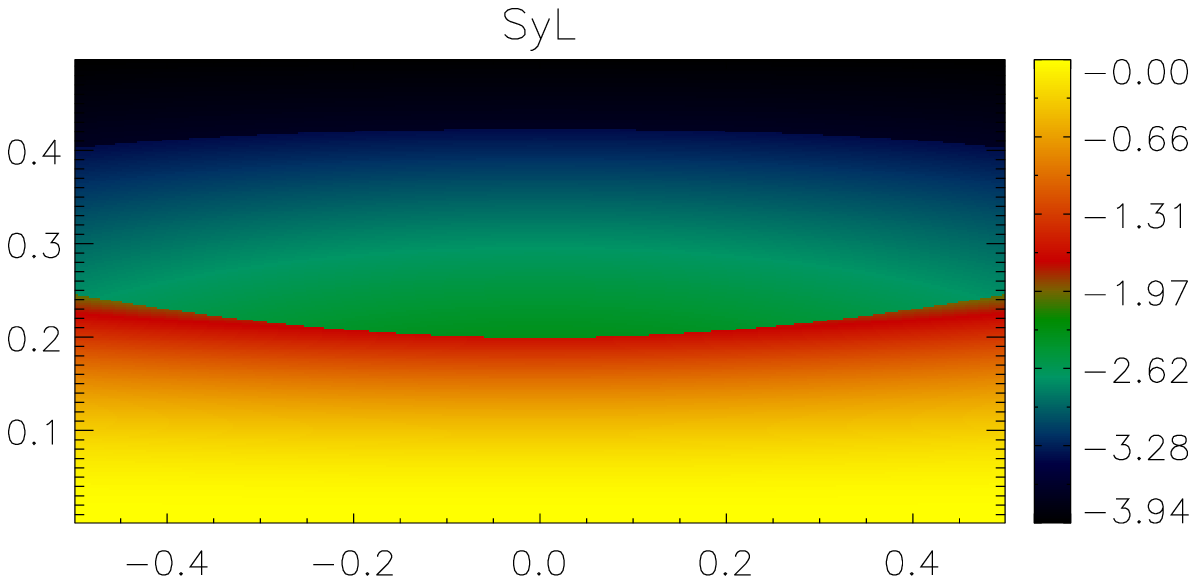}
        }\\%
        \subfigure{%
           \includegraphics[width=0.43\textwidth]{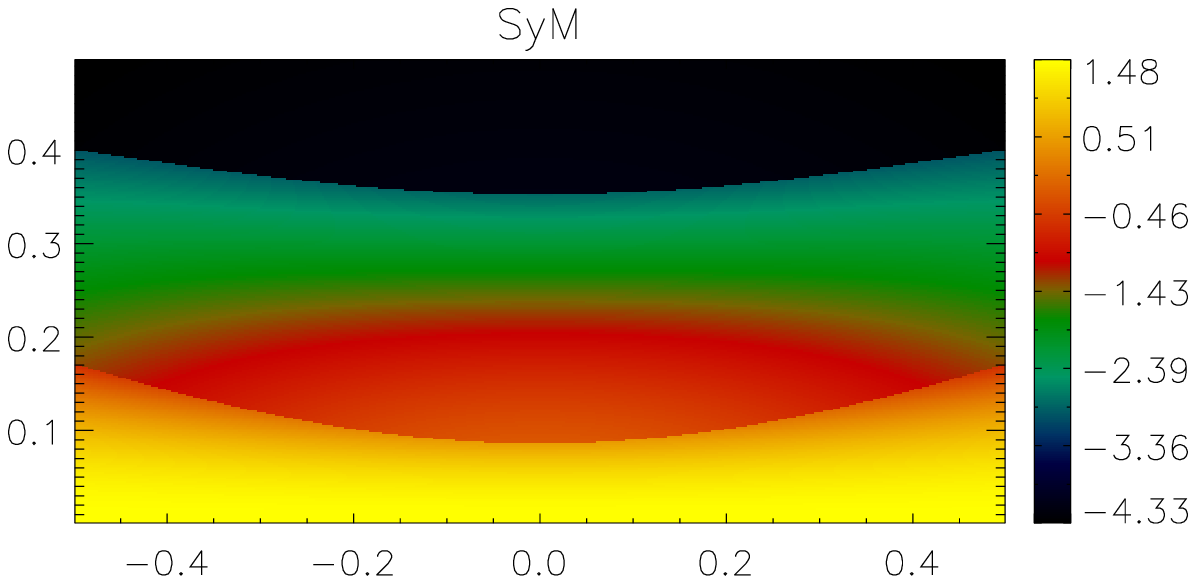}
        }\\%
        \subfigure{%
            \includegraphics[width=0.43\textwidth]{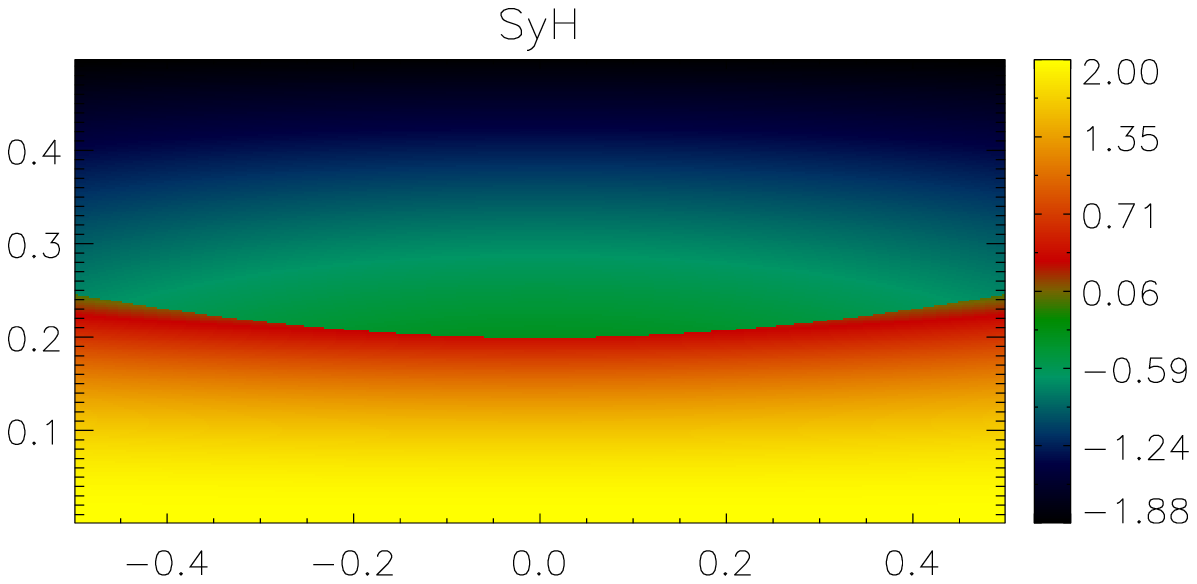}
        }%
    \end{center}
    \caption{Edge-on logarithmic density distribution 
(with density given in cm$^{-3}$) of the unperturbed ISM 
for the models $SyL$ (top panel), $SyM$ (middle panel) and $SyH$ 
(bottom panel). Distances are given in kpc.}
   \label{fig:SyASyB}
\end{figure}

The galaxy parameters adopted in this study are presented in Table 
\ref{tab:gal},
and the initial density 
and temperature distribution for each model is shown in Fig. \ref{fig:SyASyB}.

\begin{table*}
 \centering
 \begin{minipage}{140mm}
  \caption{Parameters for the galaxy models setup $SyL$ (thin disk), $SyM$ 
(intermediate disk) and $SyH$ (thick disk)}
  \label{tab:gal}
  \begin{tabular}{@{}ccccccccccccc@{}}
  \hline
Model & $r_{c}$ & $\rho_{0,1}$ & $\rho_{0,2}$ & $\rho_{0,3}$
& $T_{disk,1}$ & $T_{disk,2}$ & $T_{disk,3}$ & $e_1$ 
& $e_2$ & $e_3$ & $N_H$ & M$_{\rm 300 pc}$  \\
 & kpc & cm$^{-3}$ & cm$^{-3}$ & cm$^{-3}$ & K & K & K & & & 
& cm$^{-2}$ & $10^8$  M$_{\odot}$ \\
 \hline
SyL & 1.3 & 1 & 10$^{-2}$ & 10$^{-3}$ & $10^4$ & 
$5 \times 10^4$ & $10^5$ & 1 & 0.9 
& 0.8 & 2$\times 10^{21}$ & 0.03 \\
SyM & 0.7 & 30 & 1 & 3$\times 10^{-2}$ & $10^4$ & 
$5 \times 10^4$ & $10^6$ & 1 & 0.9 & 0.5 & 4$\times 10^{22}$ & 0.5 \\
SyH & 1.3 & 100 & 1 &  & $10^4$ & 
$5 \times 10^4$ & $10^5$ & 1 & 0.9 & 0.5 & 2$\times 10^{23}$ & 3.0\\
\hline
\end{tabular}
\end{minipage}
\end{table*}

\subsection{Energy injection}

As stressed before, the energy injected in the central region of a Seyfert 
galaxy comes from a large number of sources: stellar winds and protostellar 
jets, SN (type I and II) explosions, SMBHs, cosmic rays and spiral waves. 
However, the average energy injected by the protostellar jets is about 100 
times smaller than the energy injected by the SN explosions \citep{maclow04},
while only the Wolf-Rayet stars have winds that are energetically important 
\citep[][]{nugis00}, but last only 
$\sim 10^5$ years. Furthermore, since we are mainly interested in the gas 
outflow, cosmic rays and spiral waves may be also neglected 
because although they may play an important role in the star formation 
processes, they are expected to be less relevant to drive the gas out of the 
disk in active galaxies\footnote{See, however, the importance of cosmic rays 
to drive winds in normal spiral galaxies and dwarf galaxies 
\citep[see, e.g.,][]{Everett08, boothCR}}.

For these reasons, in this study we assume that the kpc-scale outflow 
observed in different wavelengths  is mainly driven by the 
Type I and Type II SN explosions and/or by the jet produced by the activity 
of the SMBH in the center of the galaxy.

The rate of type I SN explosions is proportional to the bulge mass ($M_b$),
which in turn is proportional to the mass of the central SMBH ($M_{BH}$) 
(see \S 3.1). 
Given the relation $M_{BH}/M_b$ $\sim$ $10^{-3}$ \citep[see, e.g.,][]{haring04},
and assuming a SNI rate in the bulge of $\sim$ 0.01 yr$^{-1}$ 
per $10^{10}$ M$_{\odot}$ \citep[][]{pain96}, we obtain:
\begin{equation}
SNI = 10^{-9} M_{BH} \ \ \ {\rm yr^{-1}}
\end{equation}
\noindent
where the mass of the SMBH, M$_{BH}$, is given in units of M$_{\odot}$.
Each $i$-th SN explosion is associated randomly to a given position $P_i$ in 
the stellar bulge of the galaxy at a random time $t_i$ in the interval 
$0<t_i<t_{\rm SF}$, where $t_{\rm SF}$=30 Myr is the time of stellar activity. 
This procedure generates points with a random spatial frequency proportional 
to the stellar density described by the Plummer profile \citep{plummer}:
\begin{equation}
\label{eq:starpl}
\rho_{\rm b,star} (r) = {{3 M_{\rm b}}\over{4 \pi r_{\rm c}}} \left(1 + {{r^2}\over{r_{\rm c}^2}}\right)^{-5/2},
\end{equation}

To introduce the SNII (characterized by a lifetime $\leq$ 30 Myr), we assume 
a SB in the central region of the host galaxy, in the disk
within a radius of $\sim$ 300 pc and with a typical half-height 
(above and below the disk) of 80 or 160 pc, depending on the model. 
We consider an event of star
formation in which a total mass of stars $\sim$ 3 $\times$ 10$^7$ or 
3 $\times$ 10$^8$M$_{\odot}$ (depending on the model) form. 
Considering a SN rate is $\sim 10^{-2}$ (or $10^{-1}$) yr$^{-1}$, which is in 
agreement with many observations of Seyfert galaxies 
\citep[see, e.g.,][]{Forbes98}, we expect to have $\sim$ 3 $\times 
10^5$ (or 3 $\times 10^6$) SNe injecting an average luminosity 
L$_{\rm SN}$ $\sim$ 3$\times 10^{41}$ (or 3$\times 10^{42}$) erg s$^{-1}$ in a 
time interval of 30 Myr. 
As in the case of the SNI, to set the SNII spatially and 
temporally within the SB region, we associate randomly to each $i$-th SNII a 
position $P^i=(r_i,h_i)$, where $0<r_i<300$ pc and $(-80,-160)$ pc 
$<h_i<(80,160)$ pc 
depending on the model, and a time $t^i$ in the range $0<t^i<30$ Myr.
Both for the SNe I and the SNe II, we inject the energy of each explosion 
over a time of 300 yr and with a SN heating efficiency HE = 100\% 
\citep{mel04} \footnote{We note that a SNR will develop only after 
the SN shock front enter the Sedov phase, that is, after a time $\sim$ 
400/n$^{1/3}$ yr. The SNR evolution may be considered adiabatic up 
to a time t$_{sedov} \sim 3 \times 10^4$/n$^{0.55}$ yr. 
This explains why we adopt HE = 100\% in the first 300 yr.}.
Therefore, each SN explosion injects mass and energy at rates $\dot M$ = 
M$_{\rm inj}$/(300 yr) and $L_{\rm SN}$ = $10^{51}$/(300 yr), respectively, 
where M$_{\rm inj}$ is the mean mass released by a single SN explosion, that
is, 16 M$_{\odot}$ for the SNII and 1 M$_{\odot}$ for the SNI.

Finally, we assume that the SMBH injects a total constant luminosity of 
$\sim 10^{42}$ erg s$^{-1}$ above and below the midplane of the 
galaxy in two collimated jets characterized in most of the 
models by a non-relativistic velocity perpendicular to the plane of the disk 
of 2.1 $\times 10^9$ cm s$^{-1}$ and by a total rate of injected 
matter of $\sim$ 6 $\times 10^{-3}$ M$_{\odot}$ yr$^{-1}$. 
The jet is injected in a central cylindric volume of about 14 pc$^3$.
In order to verify how the jet setup may affect the results, we
have also considered a model with a higher velocity and smaller density jet, 
with the same luminosity of the one described above, but with a relativistic 
velocity of $\sim$ 0.2 $c$ and a rate of injected matter of $\sim$ 5 
$\times 10^{-4}$ M$_{\odot}$  yr$^{-1}$.

Since in this study we are interested in understanding the role of each 
of the energy sources described above on the galaxy evolution 
we have performed different simulations where either  all or only part of the 
energy sources have been considered. 
In order to cover a parametric space as large as possible, we have run
16 different models and a summary of the characteristics and parameters 
adopted for each model are given in Table \ref{tab:mod}.

\subsection{Numerical methodology}

To simulate the evolution of the system descrived above, we employed a 
modified version of the Cartesian Godunov MHD code originally developed by G. 
Kowal \citep[see][]{kowalCode, diego08, reinaldo10, marciaCore}, 
using the Harten-Lax-van-Leer C 
(HLLC) solver and a second order Runge$–$Kutta for time integration. 
It also uses the message-passing interface 
(MPI) library to achieve portability and efficient scalability on a variety of 
parallel high$-$performance computing systems. 
The code is available upon request directly from the authors. 

In this version, the code includes a parametrized 
cooling function in the energy equation that allows the gas to cool down to 
10 K with errors smaller than 10$\%$ and which is calculated implicitly in 
each time step for each grid position. The cooling function considers an 
optically thin gas in ionization equilibrium and takes into 
account also the gas metallicity and the gas fraction of H$_2$, according to 
the methodology used by \citet[][]{raga00, raga02}.
We have run all the models with a maximum resolution of 1.9 pc per cell.
\footnote{We have also performed several tests with a larger 
resolution of 0.95 pc per cell for about 2 Myr and found that the results are 
very similar to those with a 1.9 pc resolution per cell. 
For this reason, we adopted the latter scheme for the study presented here as 
it allowed us to save computation time without loosing information and 
quality in the results.}
The adopted box has physical dimensions of 1$\times$1$\times$1 kpc in the 
x, y and z directions, respectively, and is covered by 512$^3$ cells.

The energy associated to each SN explosion is injected as thermal energy in a 
single cell, while the SMBH jet is injected in a single cell above and below 
the midplane disk, at the center of the system.

In all simulations the initial metallicity at the midplane is assumed 
to be equal to the solar one ($z/z_{\odot}$=1) and decreases inversely 
proportional with the height of the disk, up to a minimum value 
$z_{min}/z_{\odot}$=0.1. We note that the metallicity can influence the value of 
the cooling function up to a factor 10 specially at gas temperatures  
$\sim 10^5$ K, but it is not so important for the gas cooling at temperatures 
between $10^4$ and $5 \times 10^4$, or above $10^7$ K. For this reason, we have 
adopted for the injected SNe a metallicity that increases according to their 
evolution as described, e.g., 
in \citet[][see also references therein]{melioli13}.
We further notice that we have also run a few models (not presented here) 
with different initial values of the metallicity in the disk 
($z/z_{\odot}$=0.3; $z/z_{\odot}$=5) and found no significant modifications in 
the results with regard to the reference model ($z/z_{\odot}$=1).

Each simulation was run on 512 processors of the LAI supercomputing centre at
IAG-USP, using about 10$^5$ CPU hours per simulation.

\begin{table*}
 \centering
 \begin{minipage}{140mm}
  \caption{Initial conditions adopted in the hydrodynamic simulations for
16 different models. For each Model: $a)$ Name; $b)$ Host galaxy setup;
$c)$ SNI rate (yr$^{-1}$); $d)$ SF rate (M$_{\odot}$ yr$^{-1}$); 
$e)$ SB region half-hight (pc); $f)$ SMBH Jet luminosity (erg s$^{-1}$);
$g)$ SMBH Jet velocity (v$_{jet}$/$c$)}
  \label{tab:mod}\centering
  \begin{tabular}{@{}llrrrrr@{}}
\hline
  Model$^a$ & Gal. Type$^b$ & $\dot{\rm SNI}^c$ & SFR$^d$ & h$_{\rm SB}^e$ & 
  Jet Lum.$^f$ & Jet Vel.$^g$ \\
\hline
  SyL-SNI  & SyL  & 0.01 & $ - $ & $ - $ & $ - $ & $ - $\\
  SyL-SNI-SB  & SyL & 0.01 & 1 & 160 & $ - $ & $ - $ \\
  SyL-SNI-SB-JET  & SyL & 0.01 & 1 & 160 & $10^{42}$ & 5$\times 10^{-2}$ \\
  SyL-SNI-JET  & SyL & 0.01 & $ - $ & $ - $ & $10^{42}$ & 5$\times 10^{-2}$ \\
  SyM-SNI  & SyM  & 0.01 & $ - $ & $ - $ & $ - $ & $ - $ \\
  SyM-SNI-SB  & SyM & 0.01 & 1 & 80 & $ - $ & $ - $ \\
  SyM-SNI-SB-JET  & SyM & 0.01 & 1 & 80 & $10^{42}$ & 5$\times 10^{-2}$ \\
  SyM-SNI-JET  & SyM & 0.01 & $ - $ & $ - $ & $10^{42}$ & 5$\times 10^{-2}$ \\
  SyM-SNI-10SB  & SyM & 0.01 & 10 & 80 & $ - $ & $ - $ \\
  SyH-SNI & SyH & 0.01  & $ - $  & $ - $ & $ - $ & $ - $ \\
  SyH-SNI-SB & SyH & 0.01 & 1 & 80 & $ - $ & $ - $ \\
  SyH-SNI-SB-large & SyH & 0.01 & 1 & 160 & $ - $ & $ - $ \\
  SyH-SNI-SB-JET & SyH & 0.01 & 1 & 80 & $10^{42}$ & 5$\times 10^{-2}$ \\
  SyH-SNI-JET & SyH & 0.01 & $ - $ & $ - $ & $10^{42}$ & 5$\times 10^{-2}$ \\
  SyH-SNI-SB-JET-light & SyH & 0.01 & 1 & 160 & $ - $ & 2$\times 10^{-1}$ \\
  SyH-SNI-10SB & SyH & 0.01 & 10 & 80 & $ - $ & $ - $ \\
\hline
\end{tabular}
\end{minipage}
\end{table*}

\section{Results}
 \label{sec:results}

In this section, we present the main results for the different models listed
in Table 2. 
All the models were followed for a time interval between 2 and 12 Myr 
(depending on the model), which is an appropriate time to investigate the 
global properties of the wind build-up, from the first SN explosions up to a 
nearly steady-state phase, when no gas inflow (from the external regions to 
the center of the galaxy) is detected.

Although we analysed all models of Table 2, we will focus our attention on 
four of them, namely, the models SyM-SNI-SB, 
SyM-SNI-SB-JET, SyH-SNI-SB and SyH-SNI-SB-JET which combine one or more of the 
potential outflow-driving mechanisms.
Observations \citep[see, e.g.,][]{GaspRod} indicate that a large 
fraction of Seyfert galaxies has an amount of gas in the 
inner region of the order of 10$^8$ M$_{\odot}$, so that models  
with the $SyH$ setup (Table 1) can be taken as main references in this study.
On the other hand, models with the $SyL$ setup, which have
lower values of gas column density in the center, are more rare to observe
and for this reason their results will be only briefly discussed in
the Appendix.

\subsection{Seyfert galaxy models with intermediate column density (SyM)}
\subsubsection{SyM-SNI}

In this model, we consider a disk galaxy with the $SyM$ setup (see Table 1), 
characterized by a total mass in the central half kpc$^3-$ volume of about 5 
$\times 10^7$ M$_{\odot}$, a maximum density of 30 cm$^{-3}$ and a column
density, $n_{\rm H}$ = 10$^{22}$ cm$^{-2}$.
The energy injected comes only from the SNI explosions, which are expected to 
occur in any galaxy regardless of the presence or not of a SB region or a 
SMBH jet. 
In this case we find that the SNe I, alone, are unable to 
remove the gas from the center of the galaxy. 
Figure \ref{fig:M_evol_V1} depicts the mass evolution for this model 
(dashed-green line) and we clearly see in the top panel 
that the mass of the whole system remains nearly constant over the time 
simulated, that is, 6.5 Myr for this model\footnote{We
note that in this and all diagrams representing the global evolution  
of the system, there is a small decrease in the gas mass mainly due to the 
rotation of the galaxy which causes an artificial mass loss  through the 
lateral boundaries of the computational domain. For this reason, in our 
simulations, in  order to compute the final mass loss due to the stellar (or 
jet) feedback we calculate the difference between the total mass in a model 
with gas outflow and its counterpart without gas outflow.} 
In the core of the galaxy, within a radius of 40 pc, the mass of the gas 
varies by about $\pm$ 15\% compared to its initial value, while in the disk 
up to a radius of 400 pc the total amount of gas is reduced by only a few per 
cent (see Fig. \ref{fig:M_evol_V1}, dashed-green lines in the middle and 
bottom panels).

\begin{figure}
     \begin{center}
        \subfigure{%
            \label{fig:first}
            \includegraphics[width=0.43\textwidth]{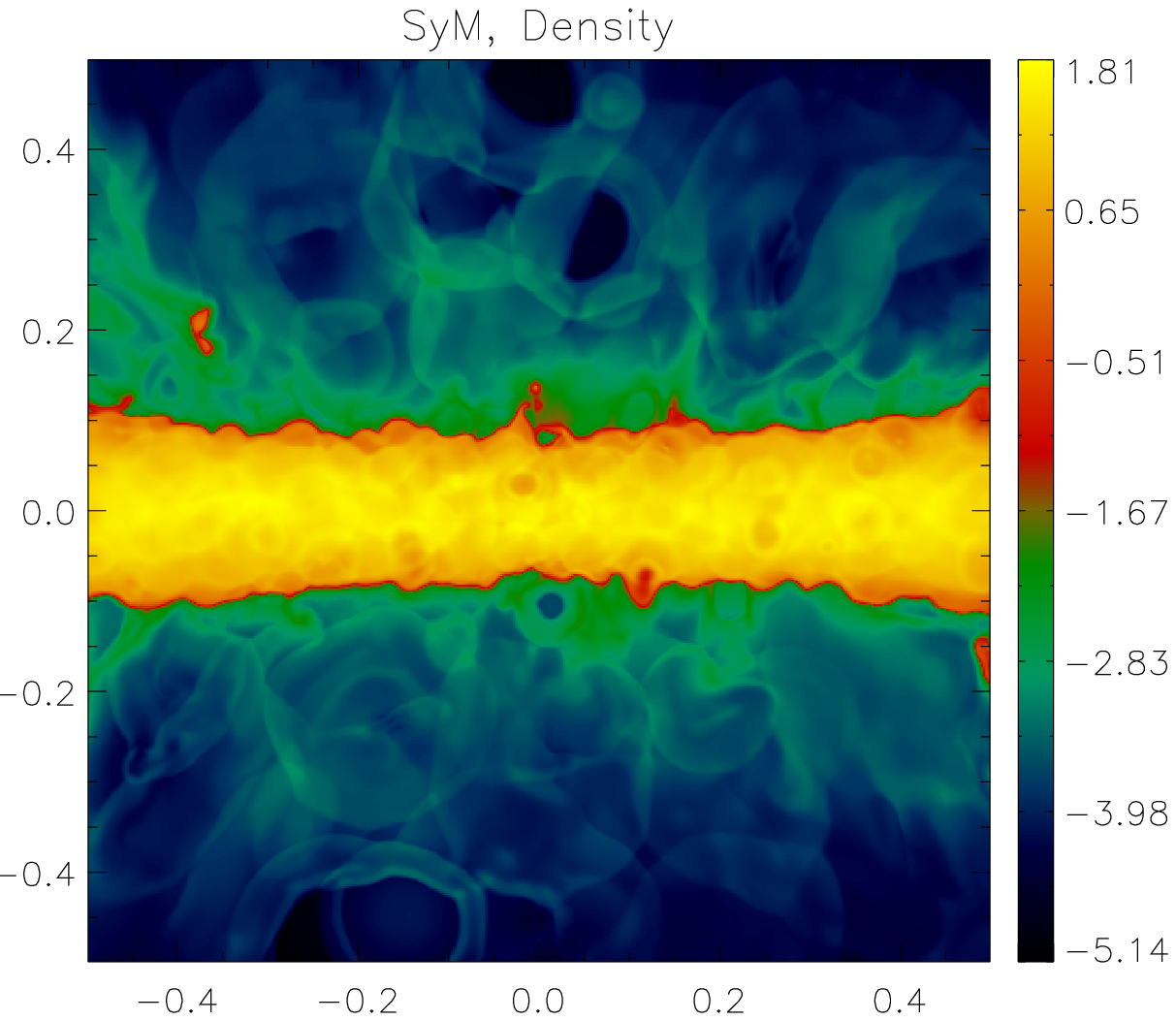}
        }\\%
        \subfigure{%
           \label{fig:second}
           \includegraphics[width=0.43\textwidth]{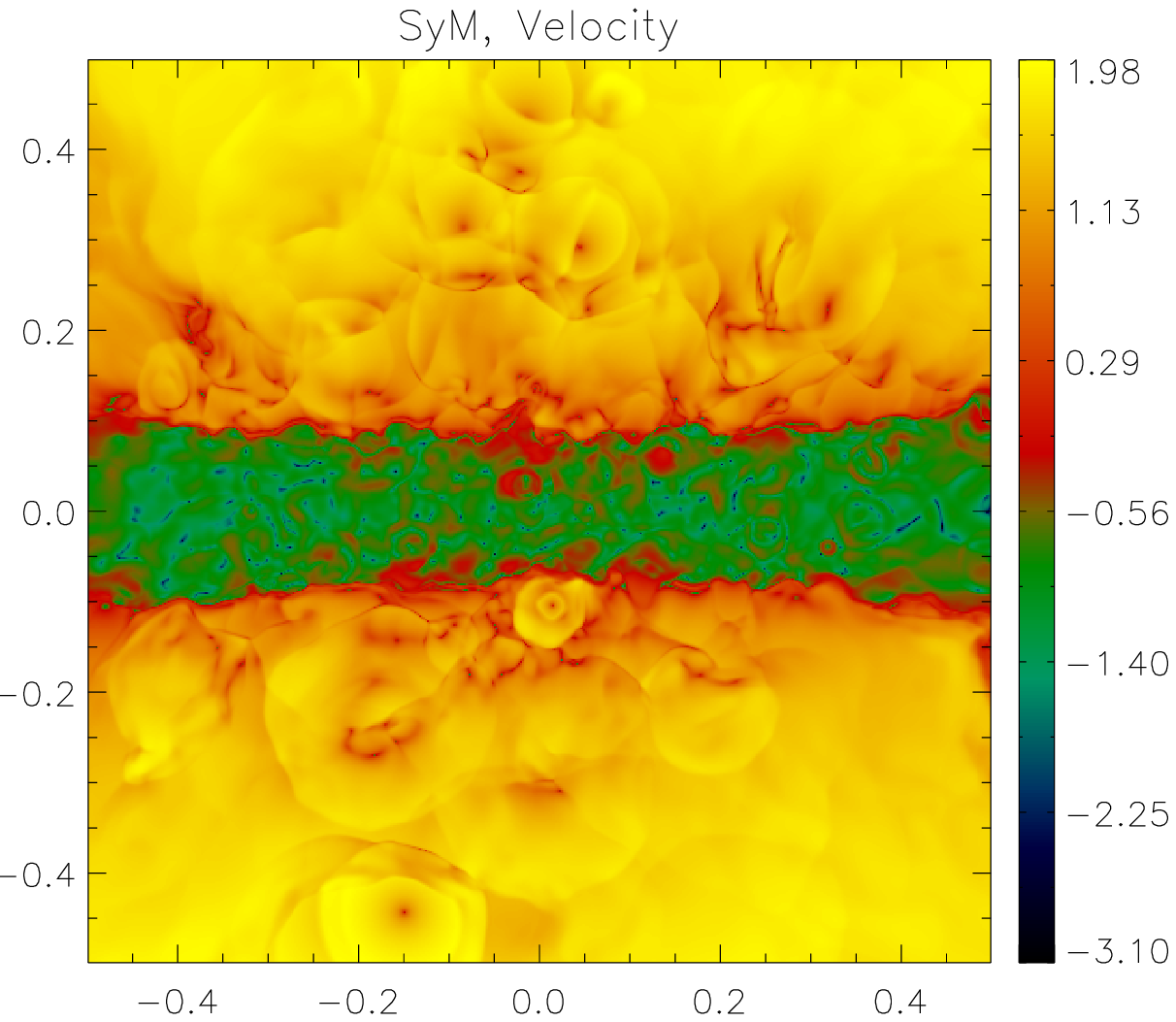}
        }%
    \end{center}
    \caption{Edge-on logarithmic gas density (top panel) and velocity 
(absolute value, bottom panel) distribution at 
$t$ = 4.2 Myr, for the model $SyM-SNI$. 
Distances are given in kpc, density is in cm$^{-3}$ and 
velocity is in units of the reference sound speed computed at 
T=5$\times 10^4$ K, $c_{s,5\times 10^4}$ = 33 km s$^{-1}$.}
   \label{fig:SyA_SNI}
\end{figure}

\begin{figure}    
\begin{center}   
\psfig{figure=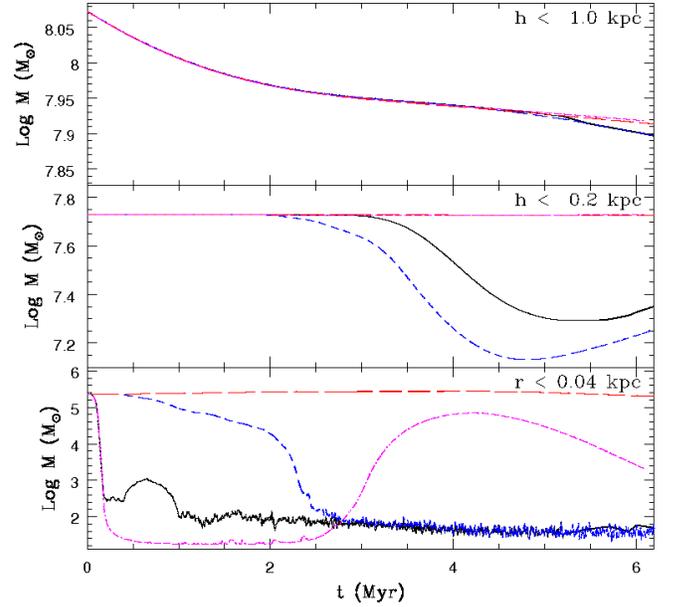,width=0.49\textwidth}    
\end{center}   
\caption{Time evolution of the mass of the gas within the whole system (z $\le$ 
500 pc, upper panel), the thick disk (z $\le$ 200 pc, middle panel) and the
central core of the galaxy ($r \le$ 40 pc, bottom panel) for the models 
SyM-SNI (long dashed-red lines), SyM-SNI-SB (dashed-blue lines), 
SyM-SNI-JET (dot-dashed-magenta lines) and SyM-SNI-SB-JET (solid-black lines).
Time is in Myr and mass is in units of M$_{\odot}$, logarithmic scale.}
\label{fig:M_evol_V1}
\end{figure}

\begin{figure}    
\begin{center}   
\psfig{figure=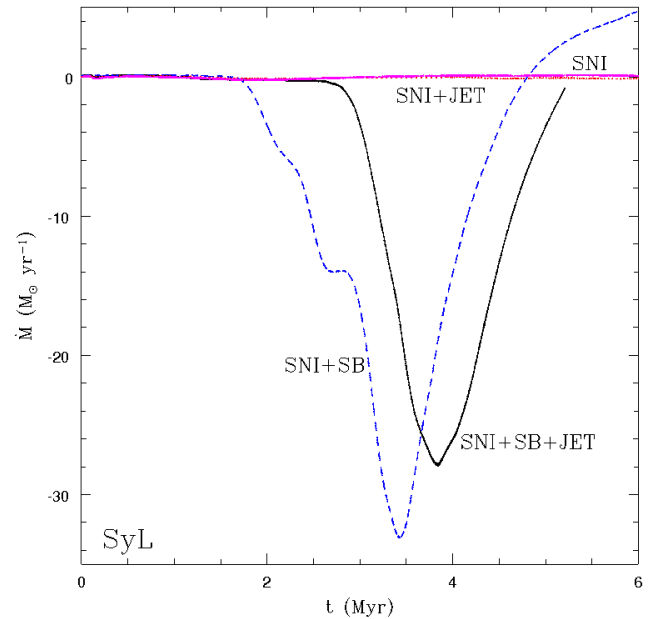,width=0.49\textwidth}    
\end{center}   
\caption{Time evolution of the gas mass transfer and loss rate of the thick 
disk (z$\le$ 200 pc) for the models SyM-SNI (long dashed-magenta lines), 
SyM-SNI-SB (dashed-blue lines), SyM-SNI-JET (dot-dashed-red lines) and 
SyM-SNI-SB-JET (solid-black lines).
Time is in Myr and mass loss rate is in units of M$_{\odot}$ yr$^{-1}$.}
\label{fig:M_lost_V1} 
\end{figure}

\begin{figure}    
\begin{center}   
\psfig{figure=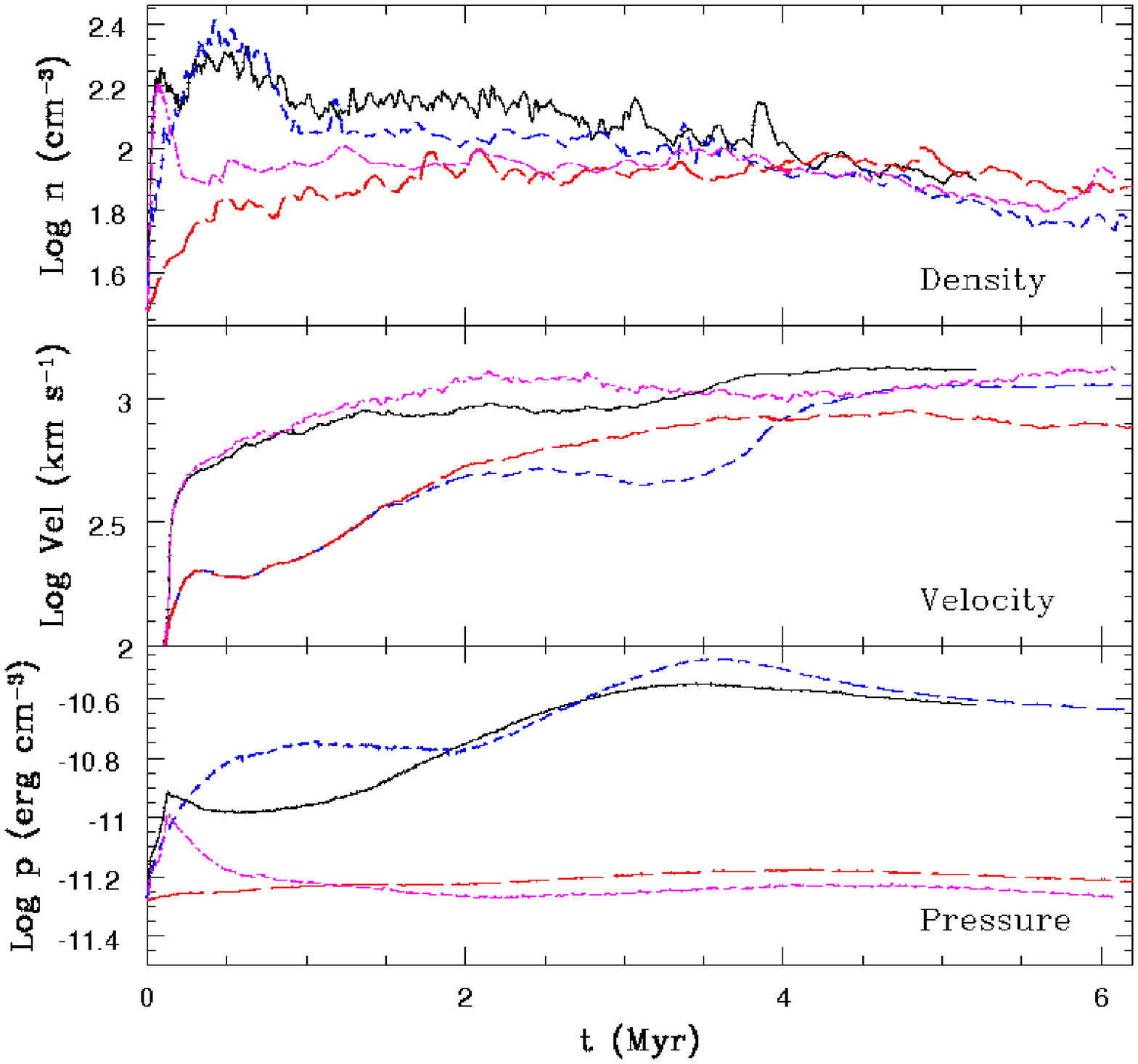,width=0.49\textwidth}    
\end{center}   
\caption{Time evolution of the main physical variables within the system: 
maximum number density (upper panel), average velocity (middle panel) and 
average pressure (bottom panel). As in Figs. \ref{fig:M_evol_V1} and 
\ref{fig:M_lost_V1}, we have the  models SyM-SNI (long dashed-red 
lines), SyM-SNI-SB (dashed-blue lines), SyM-SNI-JET (dot-dashed-magenta 
lines), and SyM-SNI-SB-JET (solid-black lines). Density and pressure are in
cgs, velocities are in km s$^{-1}$ (all variables are in log-scale).}
\label{fig:stat_V1} 
\end{figure}

Figures \ref{fig:SyA_SNI} and \ref{fig:stat_V1} (dashed-green line) 
indicate that the SNI explosions only induce the growth of turbulence in 
the disk and the halo. However, there is no impact in 
the evolution of the thermal pressure or the maximum density (which are 
almost constant along the simulation), so that no star formation process, 
gas outflow or inflow occurs in this case (see also Fig. \ref{fig:M_lost_V1}).
We note that in Figure \ref{fig:stat_V1}, the average velocity refers to 
the whole box and this average (performed over the volume and not over the 
mass) is strongly affected by the high velocities in the halo, above and below 
the disk. However, the halo (due to its very low density) does not contribute 
to the evolution of the disk gas and for this reason, despite the high 
average velocity, the system does not generate any significant outflow.

\subsubsection{SyM-SNI-SB}

\begin{figure}
     \begin{center}
        \subfigure{%
            \label{fig:first}
            \includegraphics[width=0.43\textwidth]{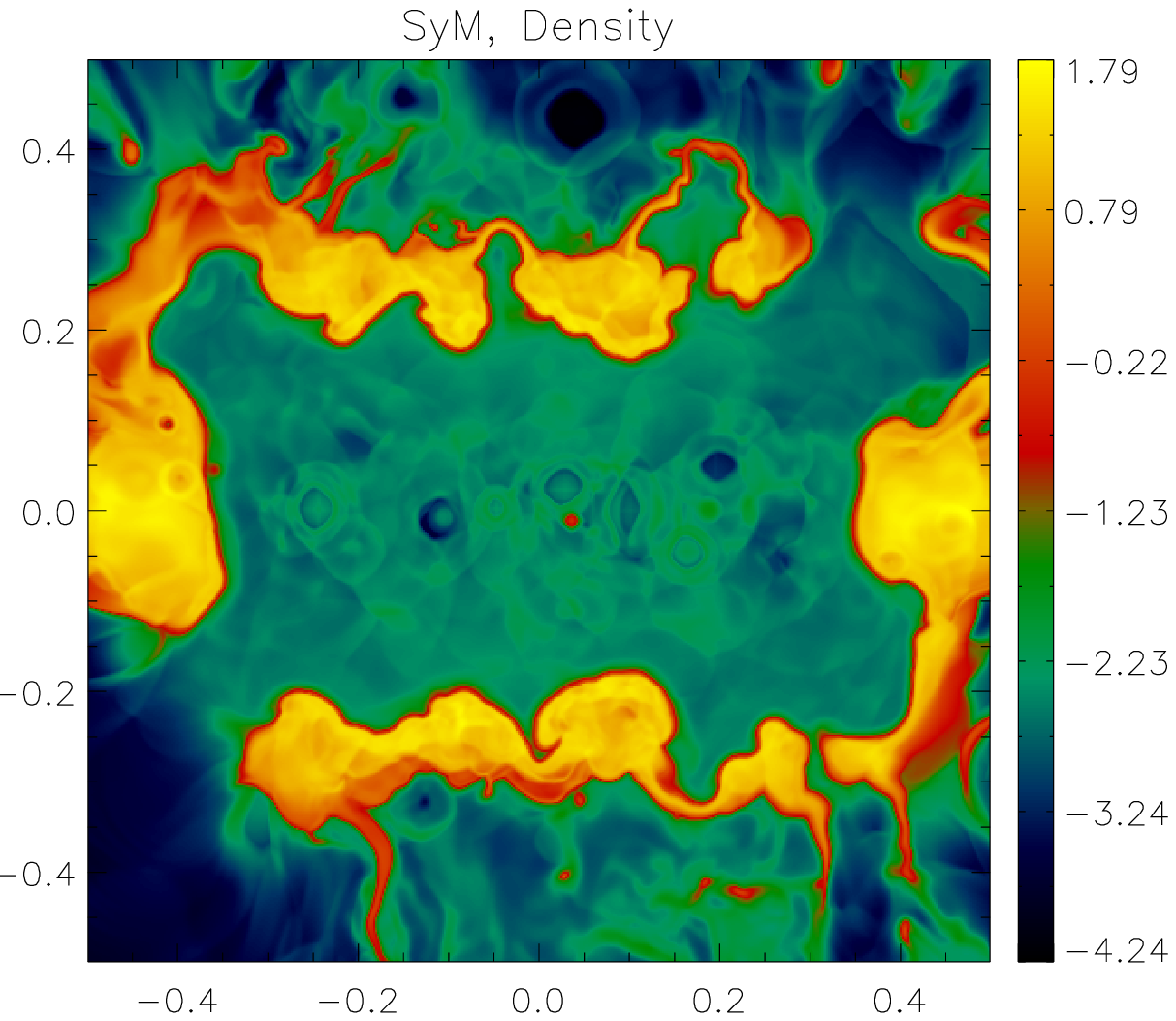}
        }\\%
        \subfigure{%
           \label{fig:second}
           \includegraphics[width=0.43\textwidth]{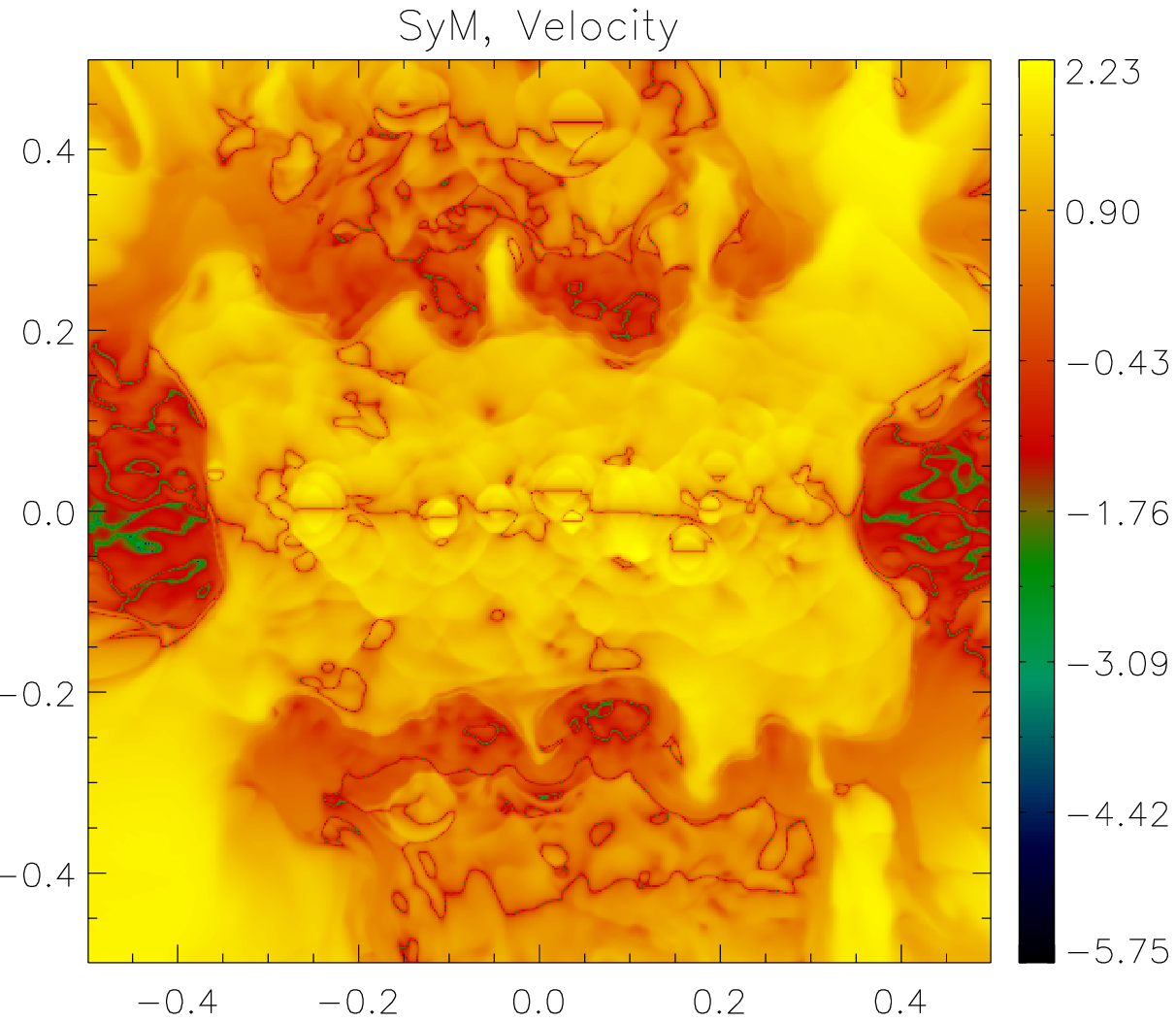}
        } 
%
    \end{center}
    \caption{Edge-on logarithmic gas density (upper panel) and velocity 
(absolute value, vertical component, bottom panel) distribution for the 
model SyM-SNI-SB at $t$ = 5 Myr.
Density in cm$^{-3}$ and
velocity in units of the reference sound speed ($c_{s,5\times 10^4}$ = 33 km 
s$^{-1}$).}
   \label{fig:SyA_SNI_SB}
\end{figure}

Also in this model the disk galaxy is built with the $SyM$ setup (see Table 1),
but the energy sources now include both regular SNI explosions and those of a 
SB region characterized by a SFR of 1 M$_{\odot}$ 
yr$^{-1}$. 
We calculate the disk evolution over a time of 12 Myr 
(Fig. \ref{fig:SyA_SNI_SB}) and find that between 2 and 6 Myr an 
average gas outflow of about 15 M$_{\odot}$ yr$^{-1}$ is driven when the SB 
region is active (see Figures \ref{fig:M_evol_V1} and \ref{fig:M_lost_V1}, 
dashed blue lines). 
The mass of the disk decreases by about 4.5 $\times$ $10^7$ M$_{\odot}$ in the 
first 5 Myr and then, since there is no gas infall coming from the external 
region of the galaxy (not considered in this study) the gas density drops to 
values of about $10^{-2}$ cm$^{-3}$ (see Fig. \ref{fig:stat_V1}, dashed blue 
lines), and a fraction of the gas raised by the
SN explosions falls back to the disk, leading to a final average density, after
12 Myr, of about 1 cm$^{-3}$ (Fig. \ref{fig:SyA_SNI_SB}). After 2 Myr the gas 
in the core of the galaxy (within a radius of 40 pc) is completely removed, 
and globally the kpc$^3$ volume here considered looses about 3.5 $\times$ 10$^6$
M$_{\odot}$ of gas along 2 Myr, corresponding to a total mass transfer rate of 
1.75 M$_{\odot}$ yr$^{-1}$.

Observing Fig. \ref{fig:SyA_SNI_SB} we also see that 
between the hot low density gas outflow and the highly energized gas in the 
disk, there is a third warm phase characterized by a temperature of 10$^4$ K, 
the minimum allowed in our model. This gas, which enters in the hot wind and 
flows out of the system, far away from the ionizing sources of the galaxy may
be able to cool to low temperatures and eventually, in presence of dust, may 
form molecular gas at few kpc above the disk, as it has been observed in some 
systems \citep[e.g.,][; see also the discussion section 6.2]{Morganti5063}.

\subsubsection{SyM-SNI-SB-JET}

\begin{figure}
     \begin{center}
        \subfigure{%
            \label{fig:first}
            \includegraphics[width=0.43\textwidth]{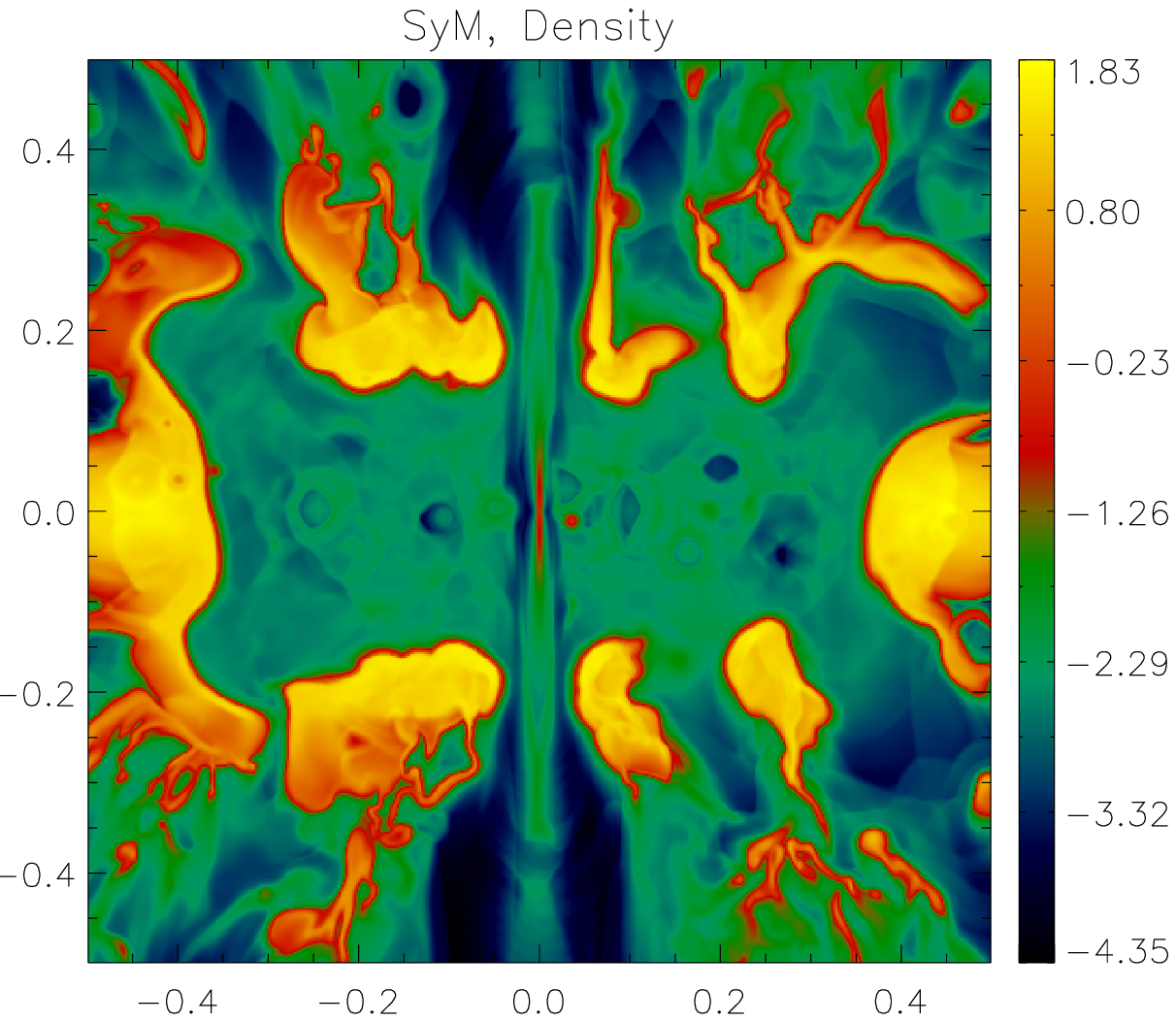}
        }\\%
        \subfigure{%
           \label{fig:second}
           \includegraphics[width=0.43\textwidth]{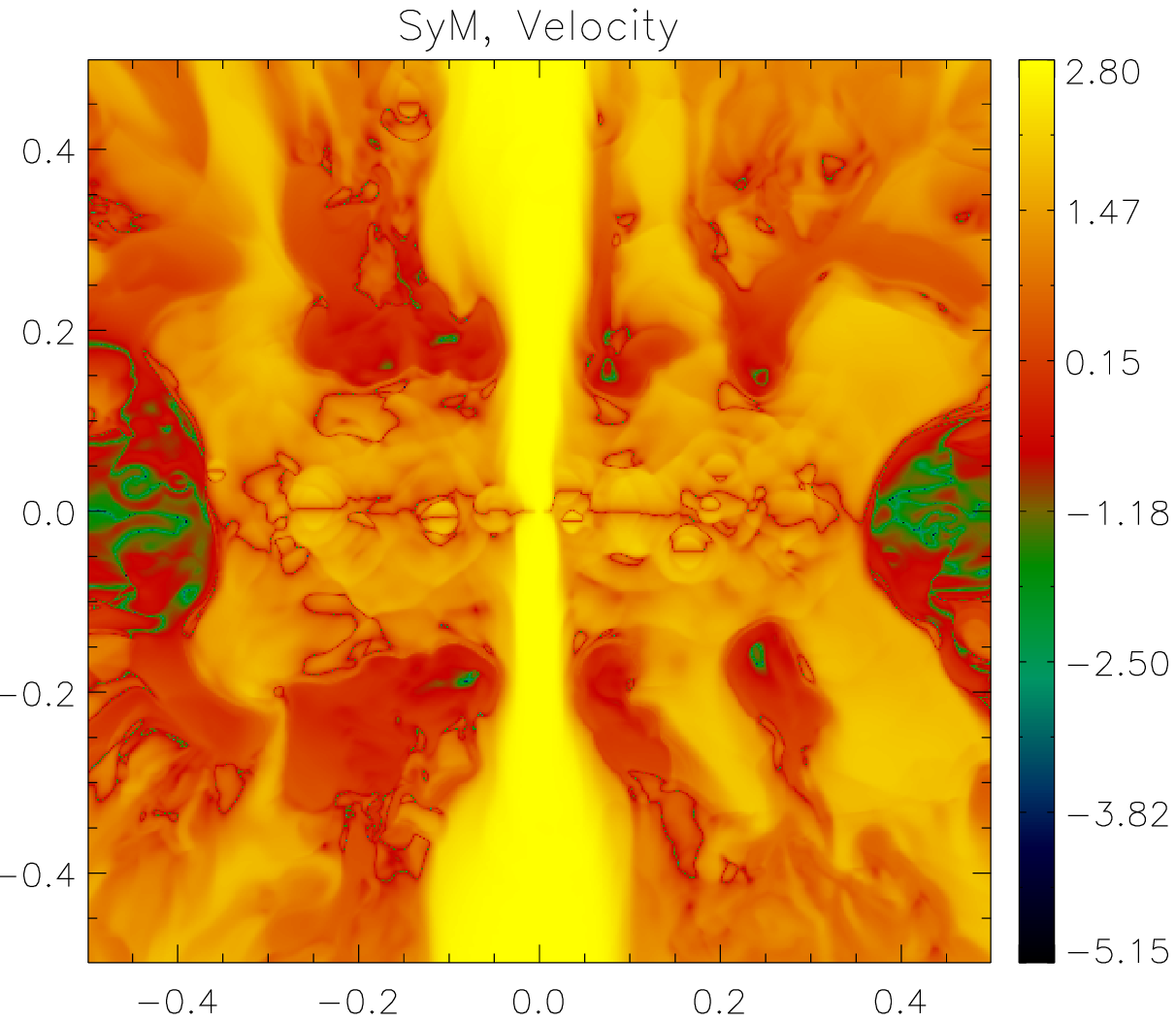}
        }\\ 
%
    \end{center}
    \caption{Edge-on logarithmic gas density (top panel) and velocity 
(absolute value, vertical component, bottom panel) distribution for the 
model SyM-SNI-SB-JET at $t$ = 5 Myr.
Density is in cm$^{-3}$ and
velocity is in units of the reference sound speed ($c_{s,5\times 10^4}$ = 33 km 
s$^{-1}$)}
   \label{fig:SyL-SNI-SB-JET}
\end{figure}

This model is equal to the previous one (SyM-SNI-SB), but in this case we 
considered also the presence of a highly collimated jet (Table 1) injected in a 
cylindrical volume characterized by a height of 1.9 pc above and below the 
midplane of the disk, a mass at a rate of about $6 \times 10^{-3}$ M$_{\odot}$
yr$^{-1}$, a (non-relativistic) velocity of about 2$\times$10$^9$ cm
s$^{-1}$ 
and a total kinetic luminosity of 10$^{42}$ erg s$^{-1}$. 
We calculated the evolution of the system over a time of 6 Myr.
The results, shown in Figs. \ref{fig:M_evol_V1}, \ref{fig:M_lost_V1} and 
\ref{fig:stat_V1} (solid-black lines) and  Fig. \ref{fig:SyL-SNI-SB-JET}, 
demonstrate that the overall gas evolution is basically the same obtained
without the jet. 
The only significant difference is in the central core of 
the disk, within 40 pc, where the gas mass is removed faster, in about 10$^5$ 
yr. The jet at the beginning of its evolution generates a transverse 
shock wave that propagates radially until reaching an equilibrium position 
at about 30 pc from the galaxy center. This makes the gas in this region
denser and less vulnerable to the action of the SN  explosions, which reduces
and delays the total gas outflow.
The presence of the jet, as expected, also causes a higher mean velocity 
in the whole system and a very high velocity signature in the low density gas 
above the disk. However, the mass,  density,  pressure and  temperature 
evolution of the disk are comparable with those of the model SyM-SNI-SB, and 
therefore we may conclude that in this case the jet does not affect 
significantly the global evolution of the system.   

\subsubsection{SyM-SNI-JET}

In this model we consider a disk with a $SyM$ setup, the presence of SNI 
explosions in the bulge and a central highly collimated jet with a total 
kinetic luminosity of 10$^{42}$ erg s$^{-1}$. Differently from model 
SyM-SNI-SB-JET, in this case we do not consider a SB region.
The results of the evolution of this model over 6 Myr (Figs. 
\ref{fig:M_evol_V1}, \ref{fig:M_lost_V1} and \ref{fig:stat_V1}, short-dashed 
purple lines) show, as in the previous case, that the jet alone is unable to 
remove from the system a significant amount of gas. 
Indeed, the fraction of gas shocked by the
jet along the axis of propagation corresponds to about 2\% of the gas in the 
nuclear region of the galaxy (within R $\le$ 300 pc), but there is a higher 
increase in density and pressure in the galactic disk induced by the jet in 
its normal direction.
As in the previous model, during the first 2 Myr only, the gas in the core of 
the galaxy (within R $\le$ 40 pc) is swept away by the shock wave generated 
at the beginning of the jet propagation. However, this phenomenon is local 
and transient, and does not affect much the global evolution of the system. 
After 2 Myr, about 30\% of the gas falls back into the nuclear region, and 
the evolution of the galaxy continues to be the same as in  Model SyM-SNI.

\subsection{Seyfert galaxy models with high column density (SyH)}
\subsubsection{SyH-SNI}

In this set of models we consider a disk galaxy with the $SyH$ setup (see 
Table 1), characterized by a total mass in the central half kpc$^3$ of about 3 
$\times 10^8$ M$_{\odot}$, a maximum density of 100 cm$^{-3}$ and a column
density $n_{\rm H}$ = 2 $\times 10^{23}$ cm$^{-2}$. In this 
configuration the disk is thicker and more massive than in the previous setup,
resembling the gas distribution observed in the nuclear region of Seyfert 
galaxies \citep[][]{GaspRod} and also of our own Galaxy 
(see, e.g., Vergani et al. 2004).

\begin{figure}
     \begin{center}
        \subfigure{%
            \label{fig:first}
            \includegraphics[width=0.43\textwidth]{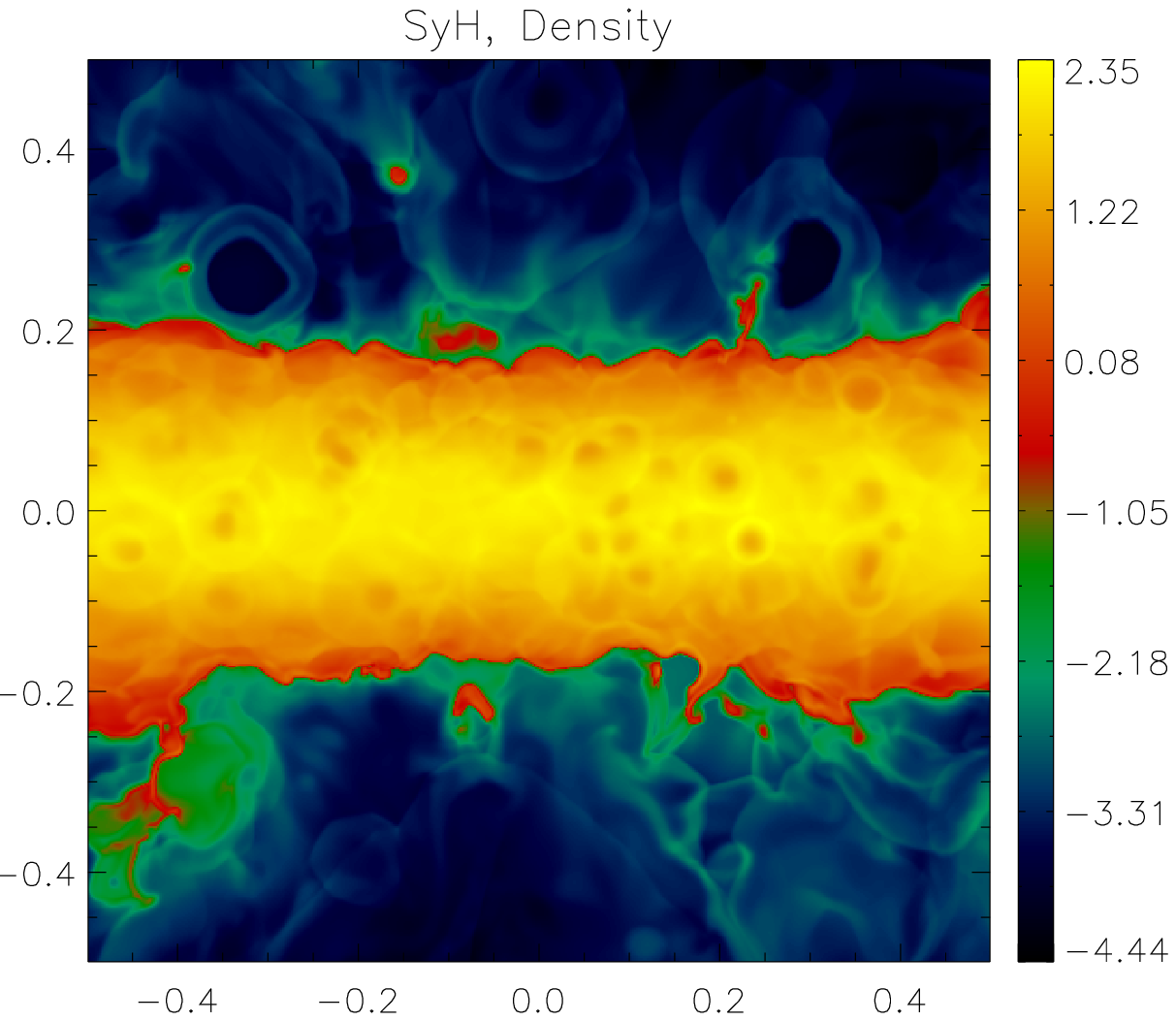}
        }\\%
        \subfigure{%
           \label{fig:second}
           \includegraphics[width=0.43\textwidth]{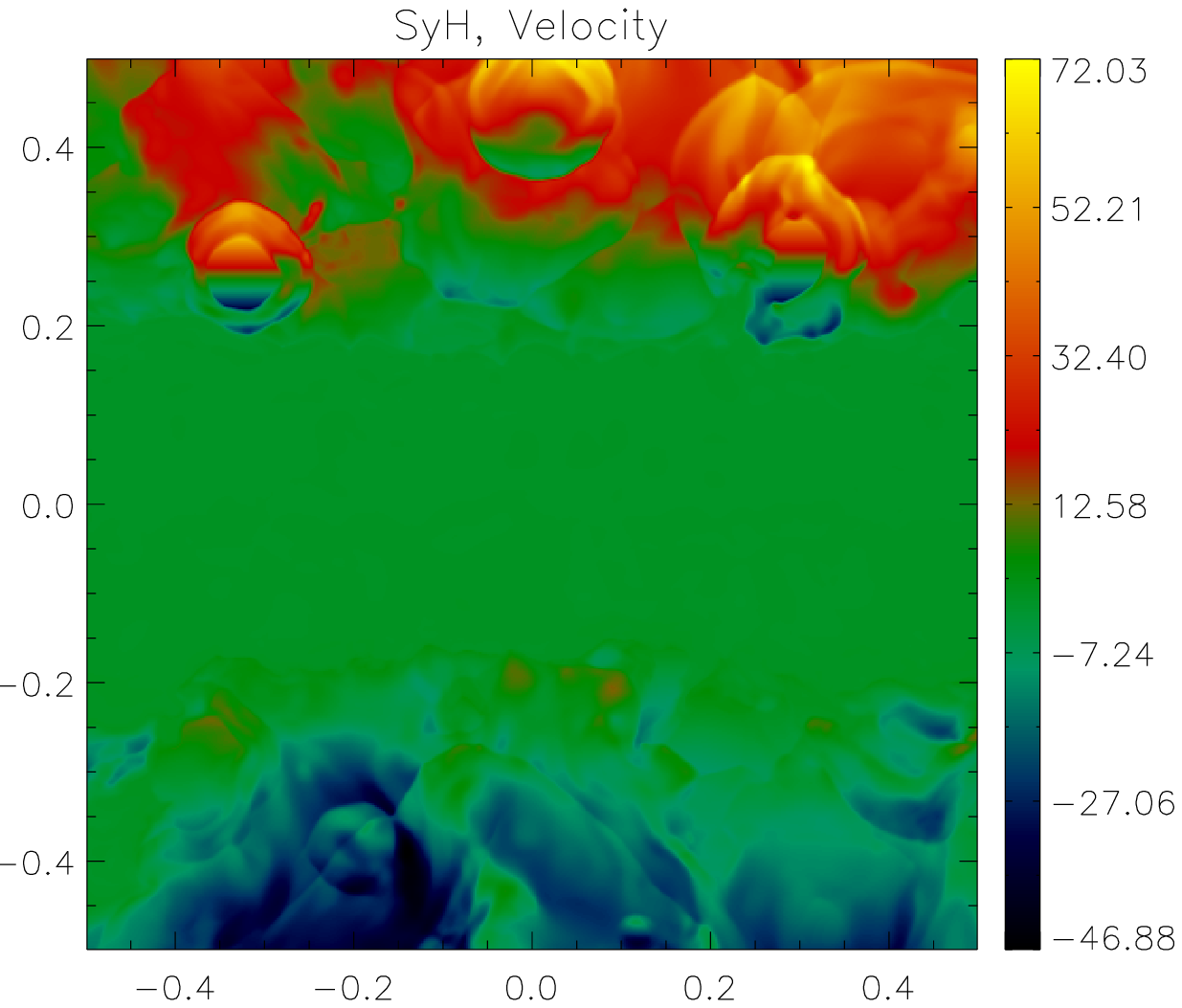}
        }%
    \end{center}
    \caption{Edge-on logarithmic gas density (top panel) and 
gas velocity (bottom panel) distribution at $t$ = 4.2 Myr, for the model 
SyH-SNI. Distances are given in kpc, density is in cm$^{-3}$ and 
velocity is in units of the reference sound speed computed at 
T=5$\times 10^4$ K, $c_{s,5\times 10^4}$ = 33 km s$^{-1}$. We note that here the 
velocity signal characterizes the real direction of the velocity 
(rather than its absolute value). It is generally positive above the disk and 
negative below it, characterizing an outflow motion. 
We also identify a few patches where the velocity has an opposite signal with 
regard to the surrounding gas, characterizing regions of fallback flow.}
   \label{fig:SyW-SNI}
\end{figure}

\begin{figure}    
\begin{center}   
\psfig{figure=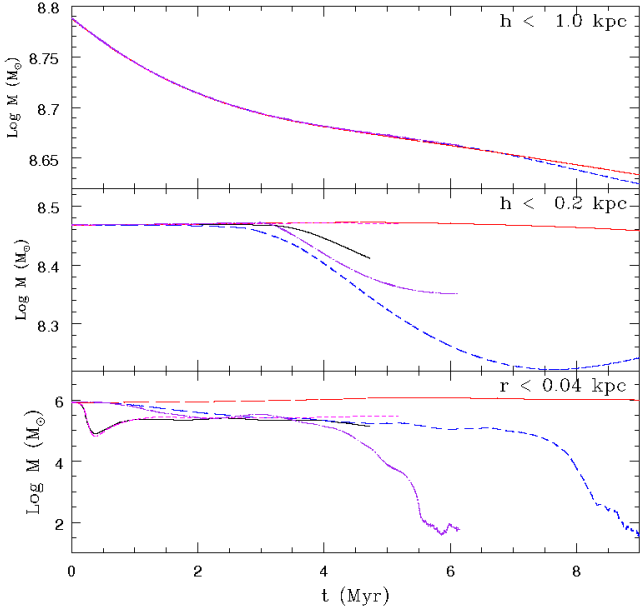,width=0.49\textwidth}
\end{center}   
\caption{Time evolution of the mass of the gas in the whole system 
(z $\le$ 500 pc, upper panel), in the thick disk (z $\le$ 200 pc, middle panel) 
and in the central core of the galaxy ($r \le$ 40 pc, bottom panel) for the 
models SyH-SNI (long dashed-red lines), 
SyH-SNI-JET (short-dashed-magenta lines), SyH-SNI-SB-JET (solid-black lines), 
SyH-SNI-SB $large$ (dashed-blue lines), and SyH-SNI-SB (dot-dashed-purple 
lines). Time is in Myr and the mass is in units of M$_{\odot}$, logarithmic 
scale.
We note that the earlier truncation in some 
tests was due to computational time constraints. From earlier tests with 
lower resolution we have verified  that the results did not change 
substantially, then we truncated the longer lasting runs earlier.}
\label{fig:M_evol_V2} 
\end{figure}

\begin{figure}    
\begin{center}   
\psfig{figure=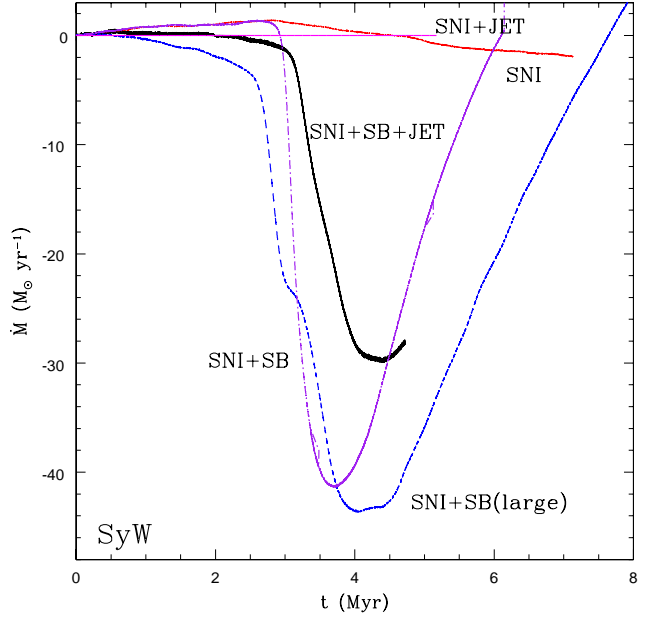,width=0.49\textwidth}       
\end{center}   
\caption{Time evolution of the gas mass transfer and loss rate of the 
thick disk (z$\le$ 200 pc) for the models SyH-SNI (long-dashed-red lines), 
SyH-SNI-JET (short-dashed-magenta lines), SyH-SNI-SB-JET (solid-black lines), 
SyH-SNI-SB $large$ (dashed-blue lines), and SyH-SNI-SB (dot-dashed-purple 
lines). Time is in Myr and mass loss rate is in units of M$_{\odot}$ yr$^{-1}$.}
\label{fig:M_lost_V2} 
\end{figure}

\begin{figure}    
\begin{center}
\psfig{figure=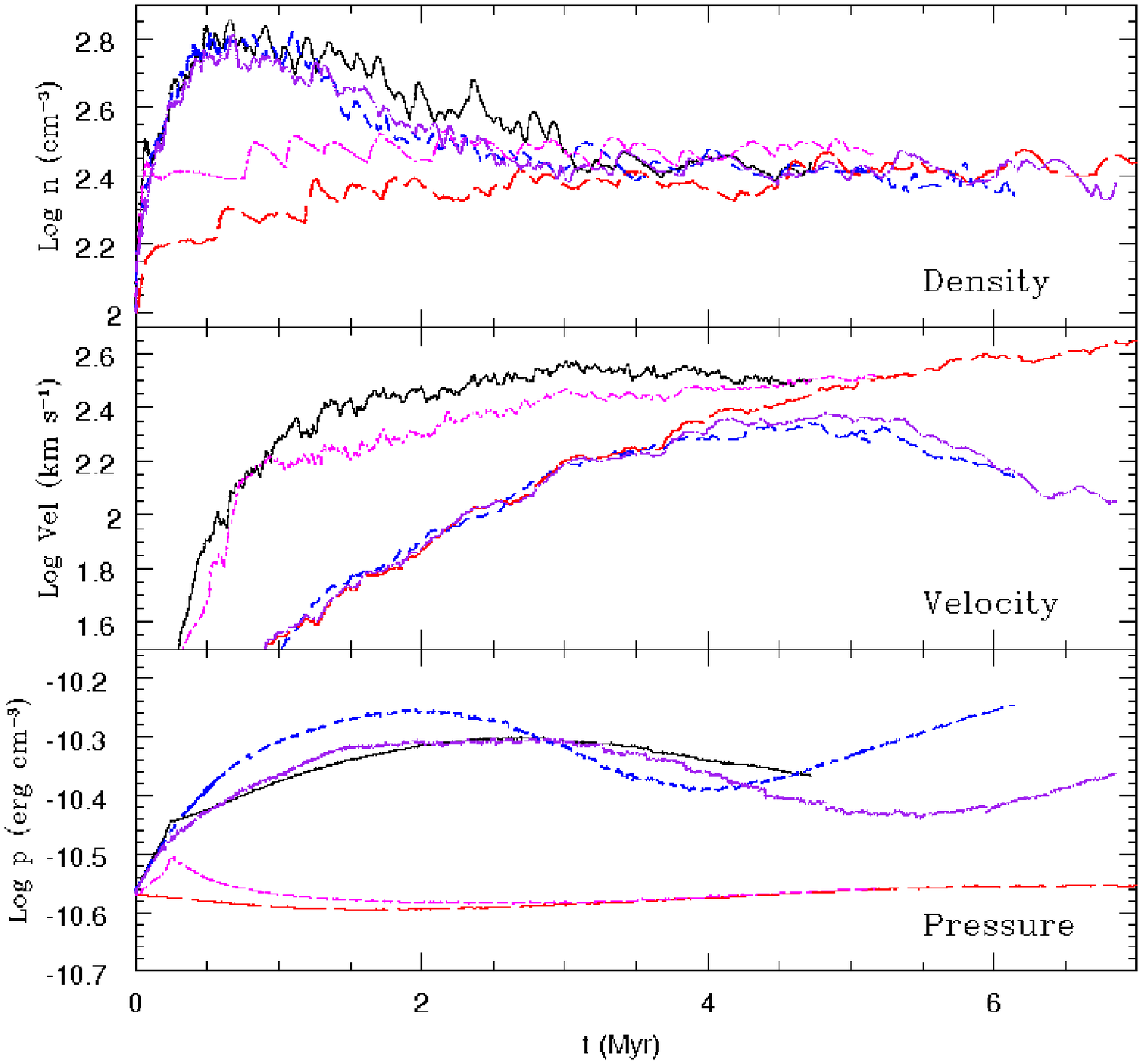,width=0.49\textwidth}    
\end{center}   
\caption{Time evolution of the main physical variables within the system: 
maximum number density (upper panel), average velocity (middle panel) and 
average pressure (bottom panel) for the models SyH-SNI (long-dashed-red lines), 
SyH-SNI-JET (short-dashed-magenta lines), SyH-SNI-SB-JET (solid-black lines), 
SyH-SNI-SB $large$ (dashed-blue lines), and SyH-SNI-SB (dot-dashed-purple 
lines). Density and pressure are in cgs, velocities are in km s$^{-1}$ 
(all variables are in log-scale).}
\label{fig:stat_V2}
\end{figure}

As in  model SyM-SNI, in the model SyH-SNI the only energy source comes 
from the SNI explosions that alone are unable to remove the gas from the 
center of the galaxy (Fig. \ref{fig:SyW-SNI}). The mass of the system is 
nearly constant over the maximum simulated time (9 Myr) and no significant 
mass transfer from the core of the galaxy to above the disk ($h \ge$ 200 pc) 
is detected, as depicted in Figures \ref{fig:M_evol_V2} and Fig. 
\ref{fig:M_lost_V2}. 
Therefore, as expected, the thick disk in a nuclear region of a galaxy, even if
completely photoionized by the central source (a SMBH in the case) and thus 
characterized by a minimum temperature of 10$^4$ K, is not destroyed or 
even partially removed by  the SNI explosions, which however contribute for 
maintaining the turbulence of the ISM, mainly above the disk 
(Fig. \ref{fig:stat_V2}).

\subsubsection{SyH-SNI-SB}

\begin{figure*}
     \begin{center}
        \subfigure{%
            \label{fig:first}
            \includegraphics[width=0.43\textwidth]{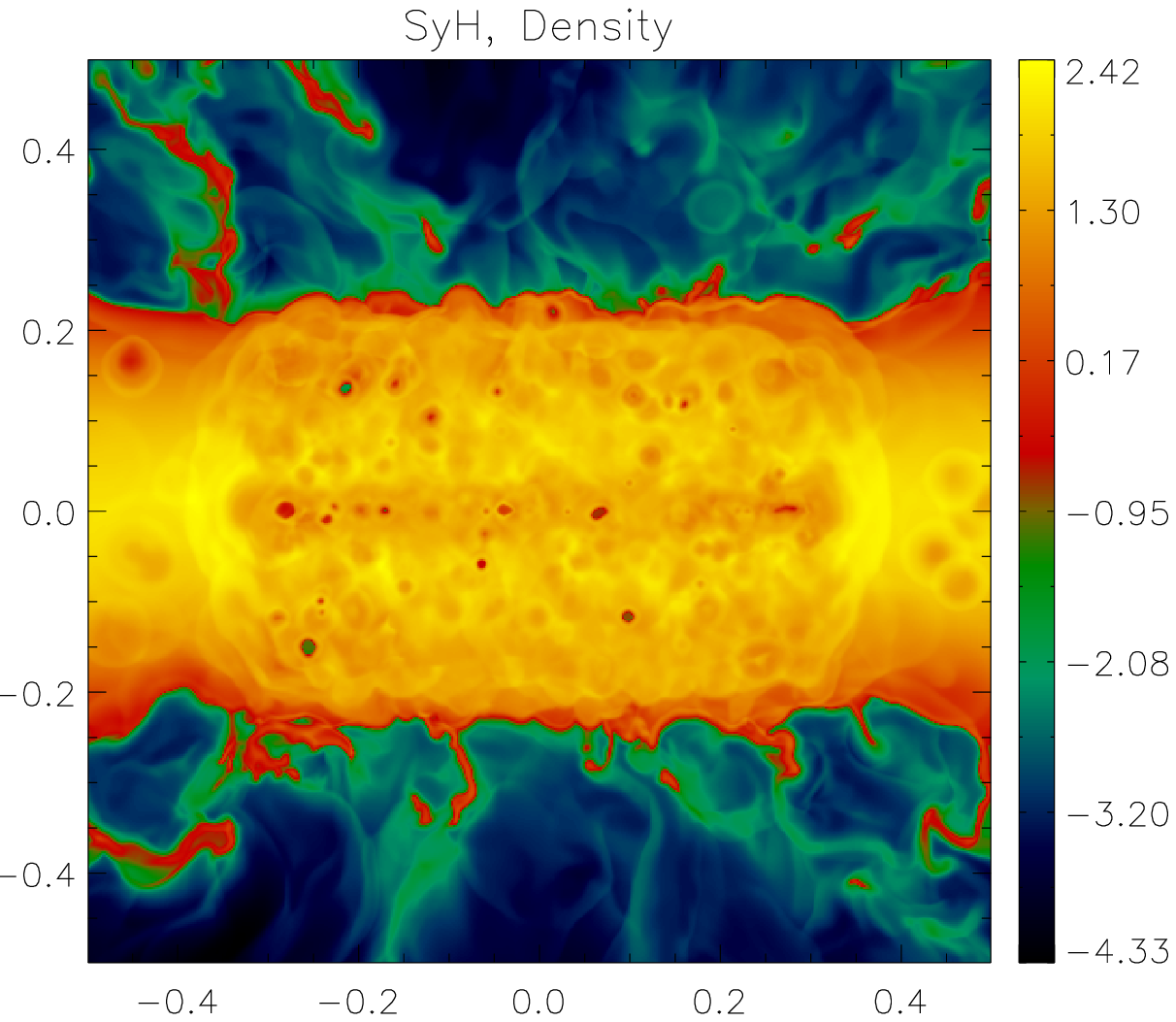}
        }%
        \hspace{0.7cm}
        \subfigure{%
           \label{fig:second}
           \includegraphics[width=0.43\textwidth]{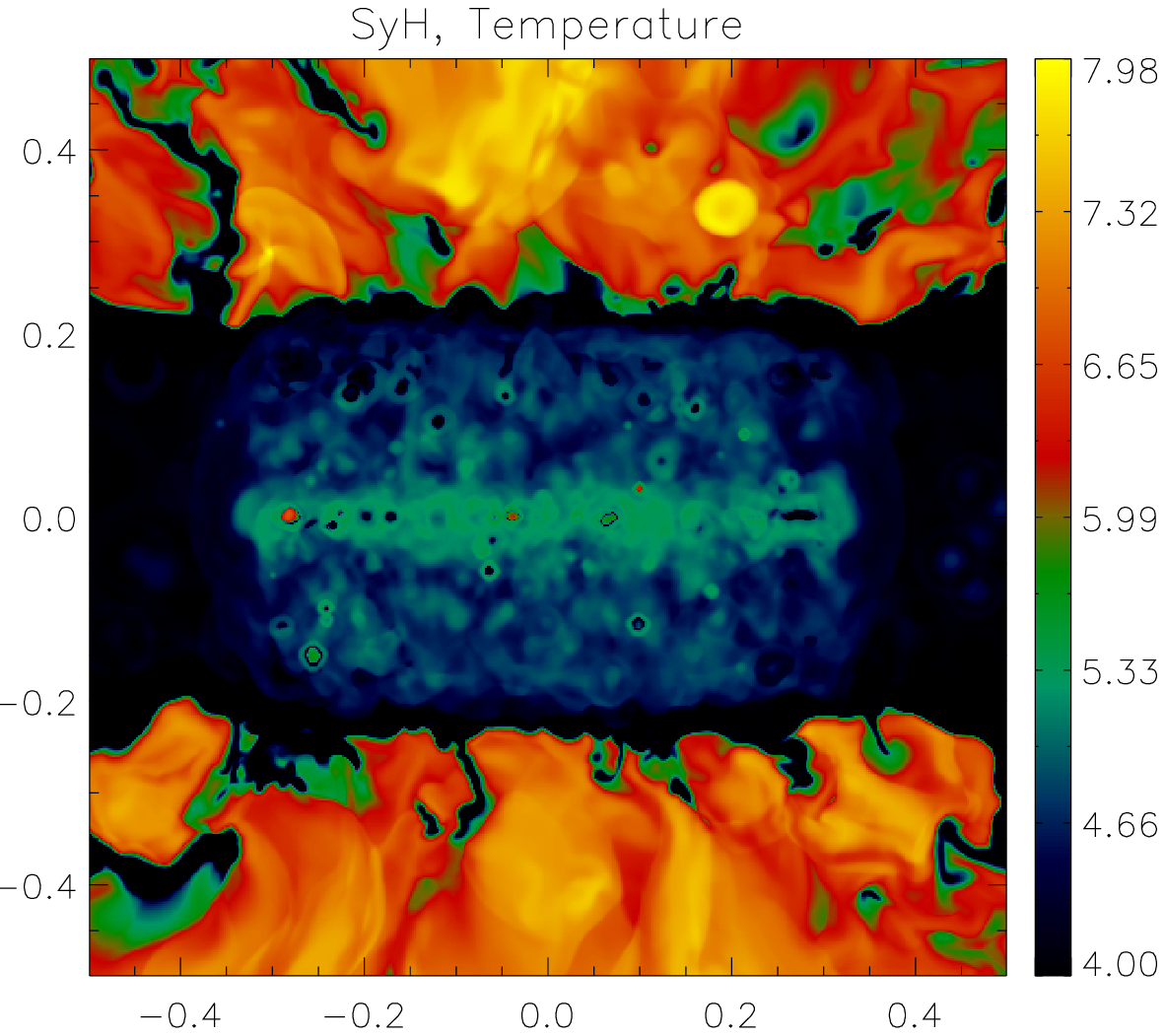}
        }\\ 
        \subfigure{%
            \label{fig:third}
            \includegraphics[width=0.43\textwidth]{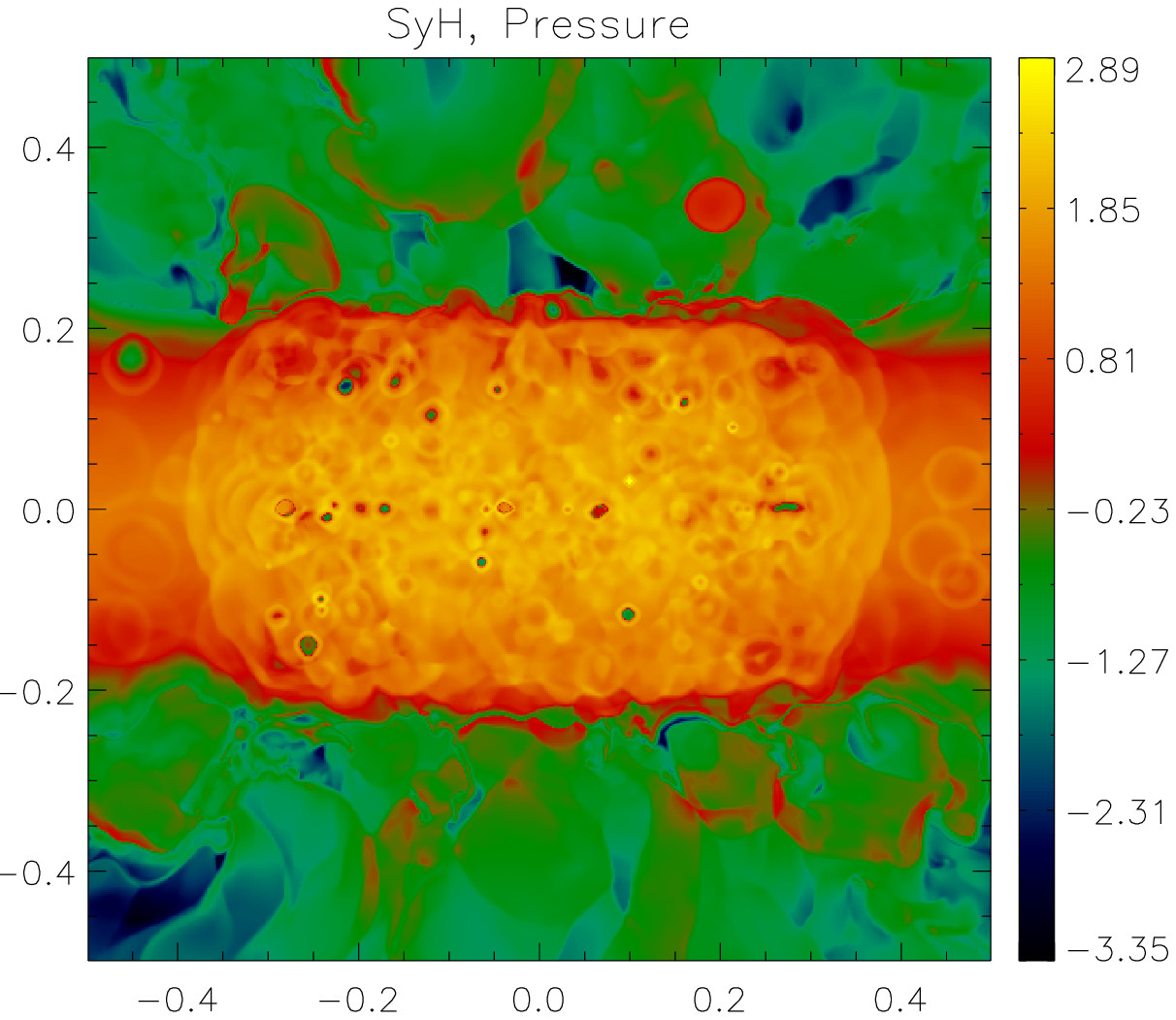}
        }%
        \hspace{0.7cm}
        \subfigure{%
            \label{fig:fourth}
            \includegraphics[width=0.43\textwidth]{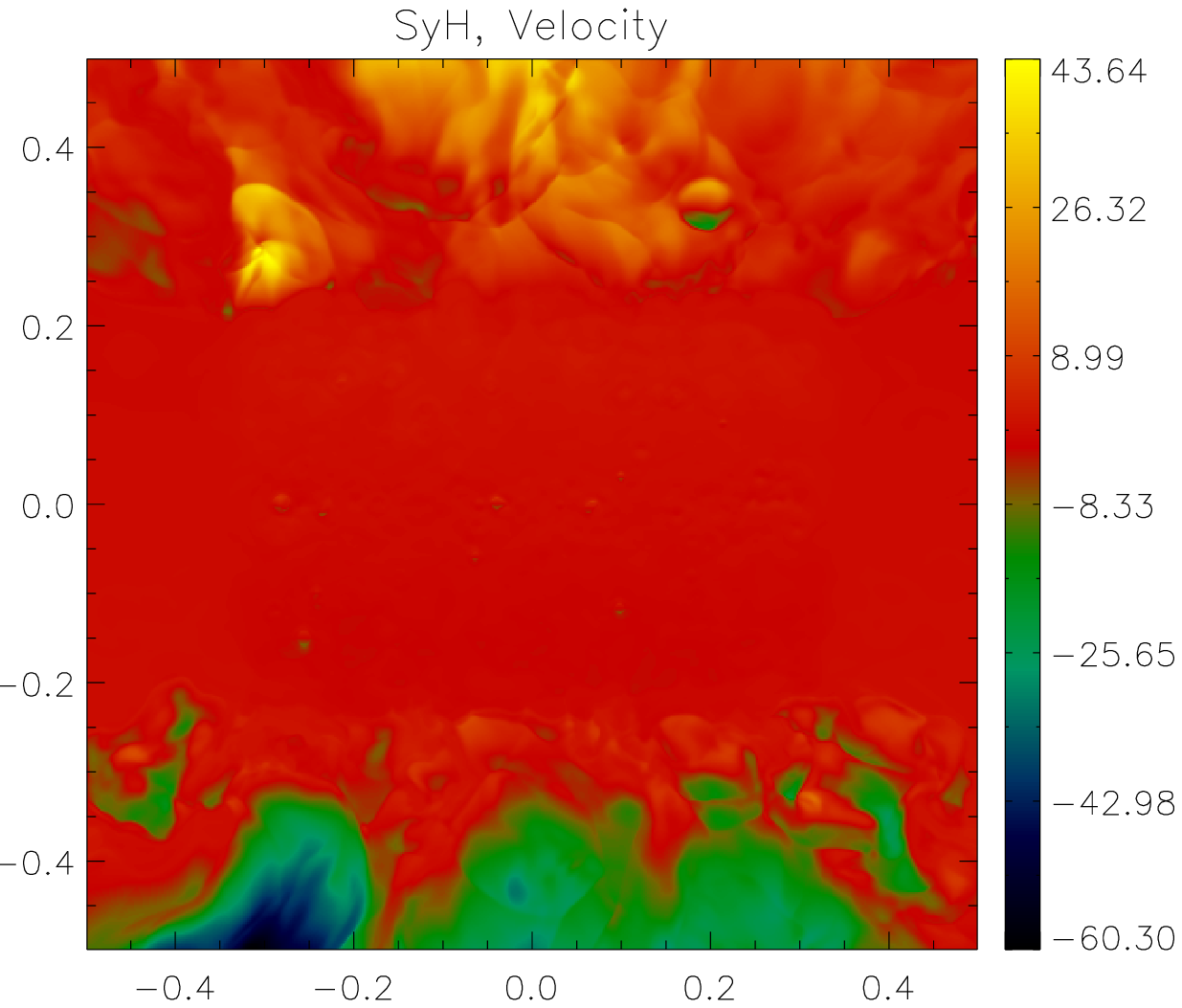}
        }%
    \end{center}
    \caption{Edge-on logarithmic gas density (upper left panel), 
temperature (upper right panel), pressure (bottom left panel ) and 
velocity (vertical component, bottom right panel) distribution for the 
model SyH-SNI-SB $large$ at $t$ = 3.75 Myr.
Density is in cm$^{-3}$, temperature in K, pressure in units of 
7$\times 10^{-12}$ erg cm$^{-3}$ (code units) and
velocity in units of the reference sound speed ($c_{s,5\times 10^4}$ = 33 km 
s$^{-1}$). Here the velocity signal characterizes the real direction of the 
flow velocity.}
   \label{fig:SyW_SNI_SB_LARGE}
\end{figure*}

\begin{figure}
     \begin{center}
        \subfigure{%
            \label{fig:first}
            \includegraphics[width=0.43\textwidth]{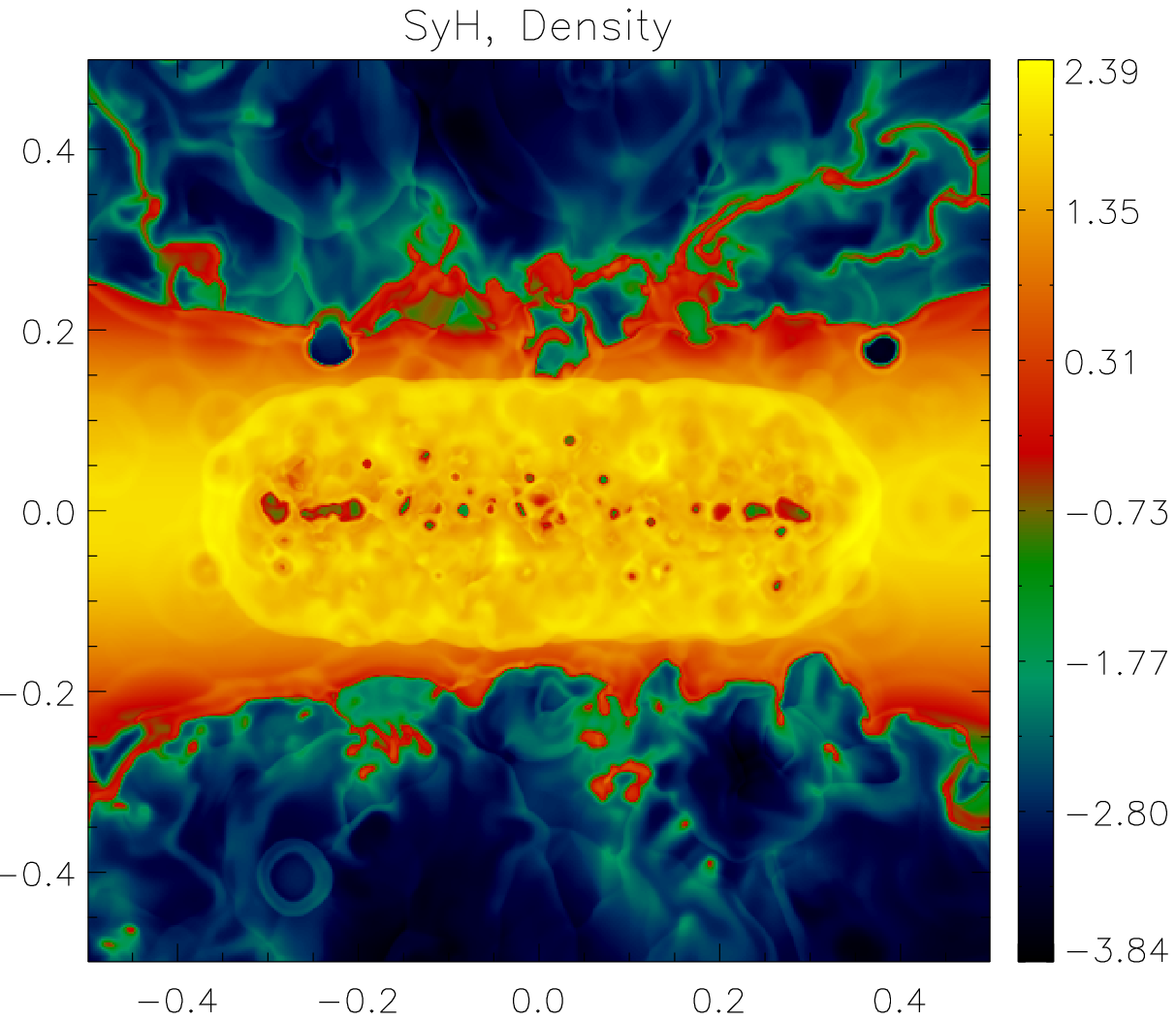}
        }\\%
        \subfigure{%
           \label{fig:second}
           \includegraphics[width=0.43\textwidth]{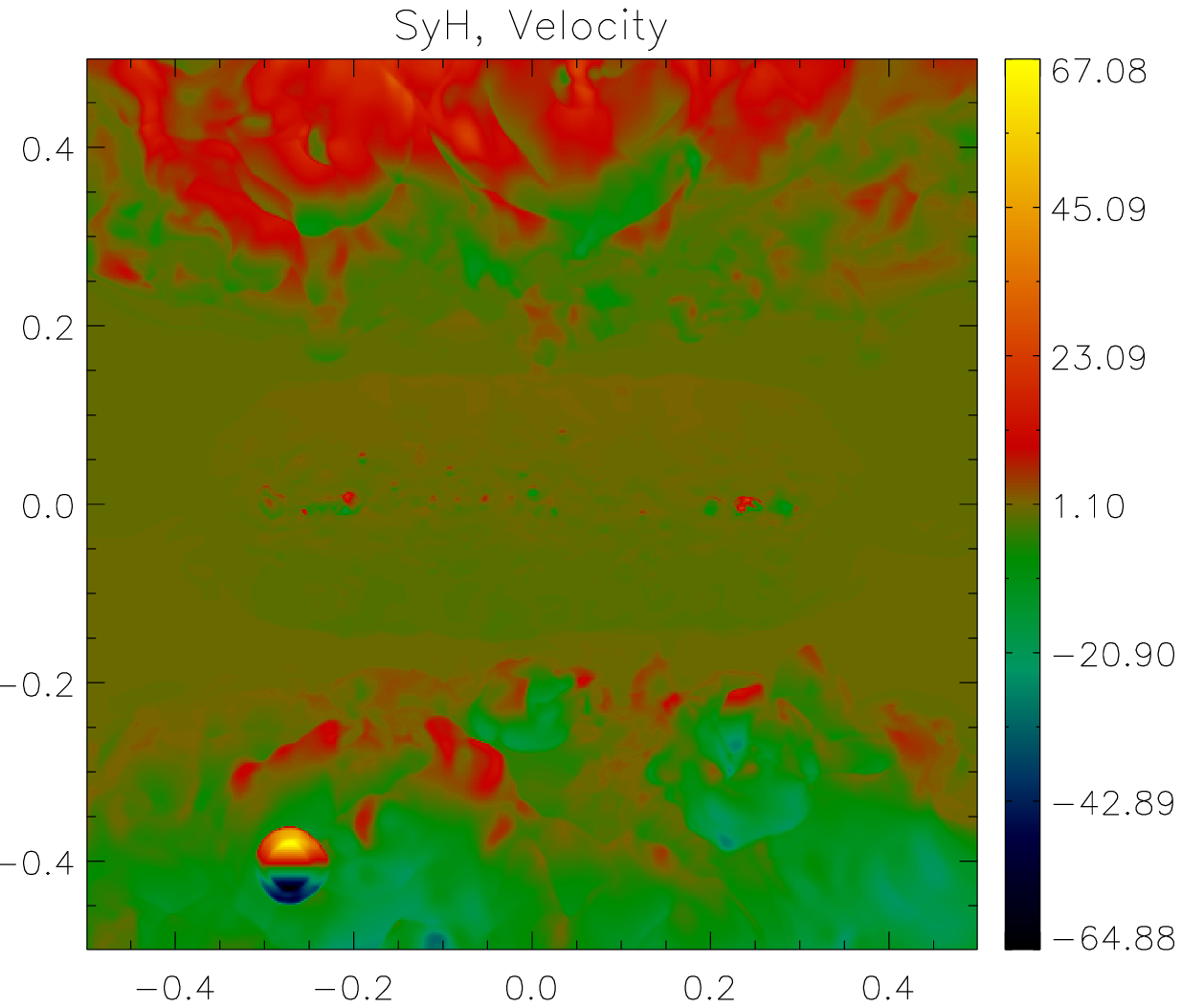}
        } 
%
    \end{center}
    \caption{Edge-on logarithmic gas density (top panel), 
and velocity (vertical component, bottom panel) distribution for the 
model SyH-SNI-SB at $t$ = 3.75 Myr.
Density is in cm$^{-3}$ and velocity is in units of the reference sound speed 
($c_{s,5\times 10^4}$ = 33 km s$^{-1}$). Here, the velocity signal charaterizes 
the real velocity direction (rather than its absolute value).}
   \label{fig:SyW_SNI_SB}
\end{figure}

In this case the energy injected in the thick disk ($SyH$) comes from
the SNI and SNII explosions associated to the bulge and to a SB 
region with a SFR of 1 M$_{\odot}$ yr$^{-1}$, respectively. 
Since the thick disk extends up to a height $h_z \sim$ 180 pc, the stars within
the SB region may be distributed in different ways. 
In this study we consider two different stellar distributions in a nuclear 
region with a radius $R_{SB}$ = 300 pc. In one case they are uniformly
distributed between a height -80 $\le h_{SB} \le$ 80 pc, while in the second 
case (here defined $large$) the stars are distributed with a power law 
$\propto (|h_z|/h_0)^{3}$ between a height -160 $\ge h_{SB} \ge$ 160 pc 
and  with $h_0$ = 160 pc, that is, the SB region is active throughout the 
thickness of the disk.
We have calculated the disk evolution over a time of 9 Myr and we see that the 
global evolution is quite similar in both cases (Figs. 
\ref{fig:SyW_SNI_SB_LARGE} and \ref{fig:SyW_SNI_SB}). 
This means that the SNII vertical distribution, within 
the disk, is not so important as the total SNII number and/or the SNII 
density surface, and for this reason in the next models we will consider 
only the $large$ stellar distribution within the SB which is possibly a more 
realistic case.

In both models above, after about 2.5$-$3 Myr, the gas energized by the SN 
explosions begins to rise from the disk to the halo at an average rate of 
$\sim$ 20 M$_{\odot}$ yr$^{-1}$, and  a maximum rate  of 45 M$_{\odot}$ yr$^{-1}$ 
at $t \sim$ 4 Myr.
Along 7 Myr the mass of the disk in the $large$ case decreases by about 
$10^8$ M$_{\odot}$, that is, $\sim$ 30\% of its initial mass is lifted above the 
denser disk to heights $\ge$ 200 pc. Similarly, when the SNII are uniformly 
distributed within a thinner region, the mass of the disk is reduced by $\sim$ 
23\%, with an active gas outflow starting at $t$ = 3 Myr and ending after
3 Myr (i.e., at $t$ = 6 Myr). Also the density, the pressure and the velocity 
evolutions are very similar in both cases (Fig. \ref{fig:stat_V2}. The
only difference is observed in the removal of gas from the nuclear region, 
clearly more efficient in  model SyH-SNI-SB. In this case, all the gas
is dispersed to larger radius, while in the $large$ case about 10\% of the
gas remains in the central 40 pc. However, these differences are not enough to 
affect the evolution of the system and we can state that the choice of 
the vertical distribution of the SNs does not seem to be so important in this 
study.
Close to the midplane of the disk, the SB activity generates a 2-phase ISM
characterized by a diffuse gas with density of $\sim$ $10^{-1}$ $-$ $10^{-2}$ 
cm$^{-3}$ and temperature of about $10^7$ K, where dense clumps and filaments 
are embedded with a density of $\sim$ 10 cm$^{-3}$ and temperature of 10$^4$ K.

Globally, the mass of the system here considered (with a kpc$^3$ volume) 
decreases by about 10$^7$ M$_{\odot}$ at an average rate of few M$_{\odot}$ per 
year (top panel of Fig. \ref{fig:M_evol_V2}), but since the region we have 
simulated extends only up to $\pm$ 500 pc (on each side of the disk), we are 
unable to predict the final fate of this gas, i.e., whether it will fall back
to the disk or definitively leave the galaxy as a wind. An estimate of the 
inner galactic disk escape velocity ($(2 GM/r)^{1/2}$) indicates that it is of 
the same order of the outflow velocity of  part of the energized gas, 
so that both fates are possible.

\subsubsection{SyH-SNI-SB-JET}

\begin{figure*}
     \begin{center}
        \subfigure{%
            \label{fig:first}
            \includegraphics[width=0.43\textwidth]{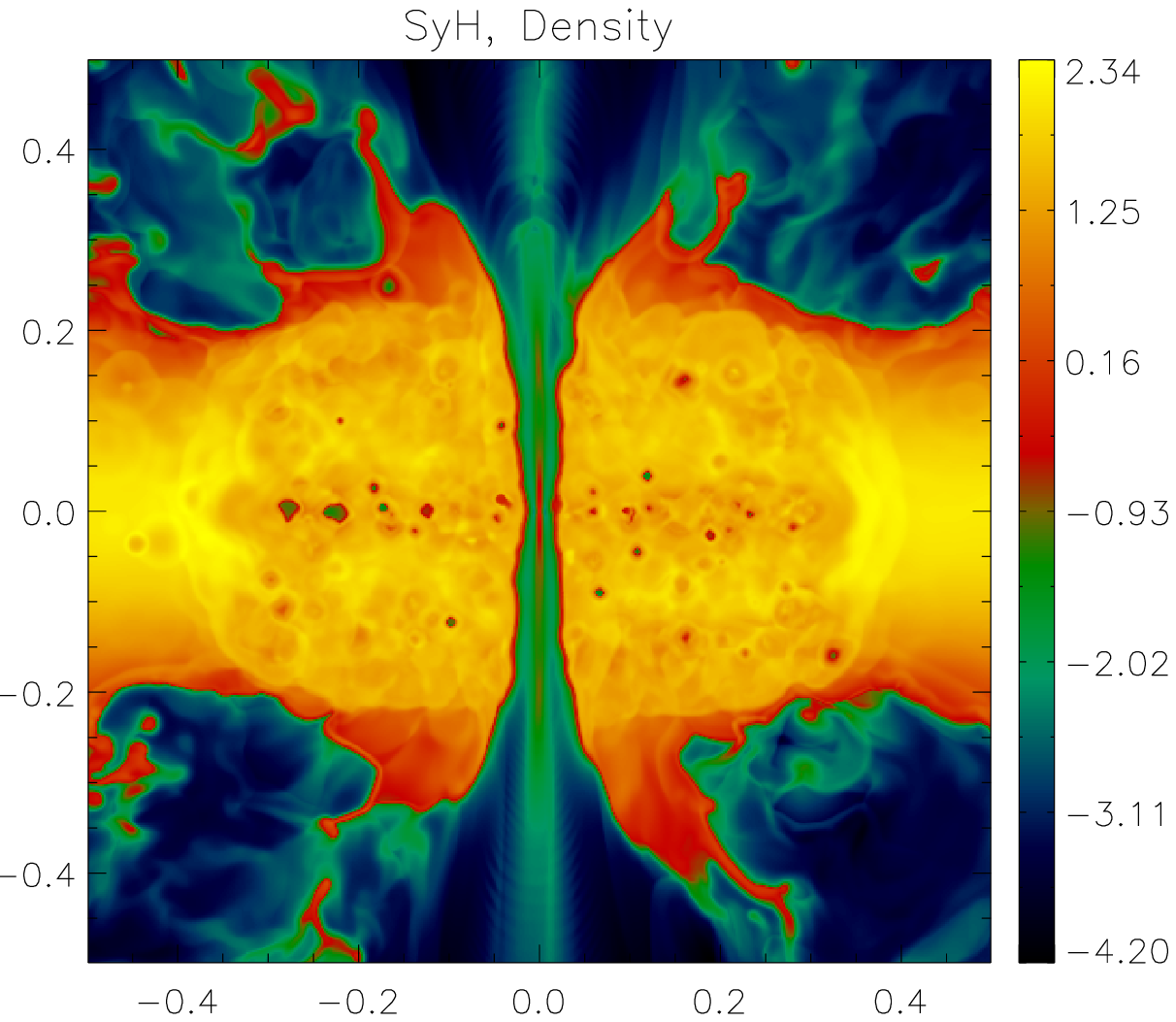}
        }%
        \hspace{0.7cm}
        \subfigure{%
           \label{fig:second}
           \includegraphics[width=0.43\textwidth]{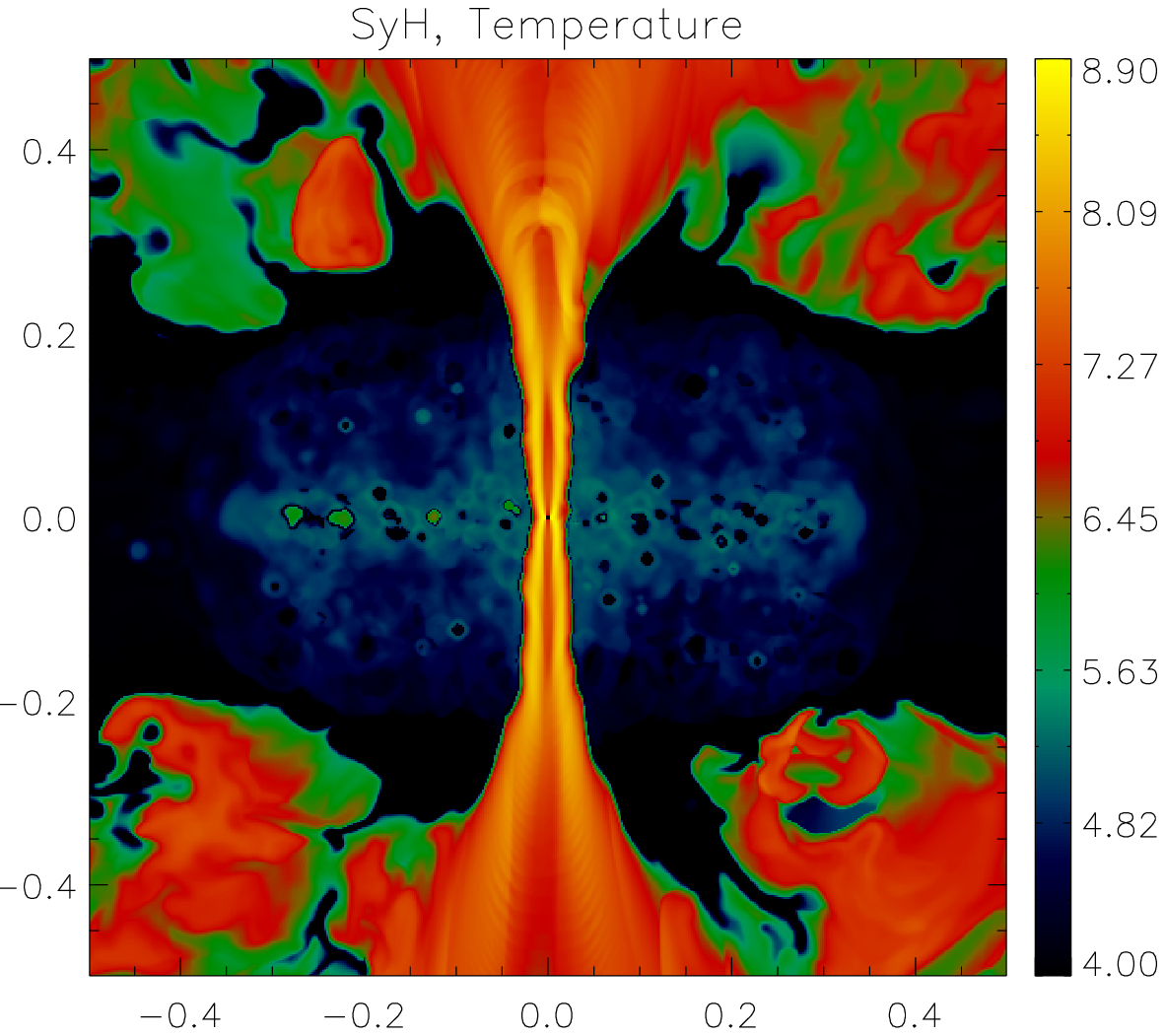}
        }\\ 
        \subfigure{%
            \label{fig:third}
            \includegraphics[width=0.43\textwidth]{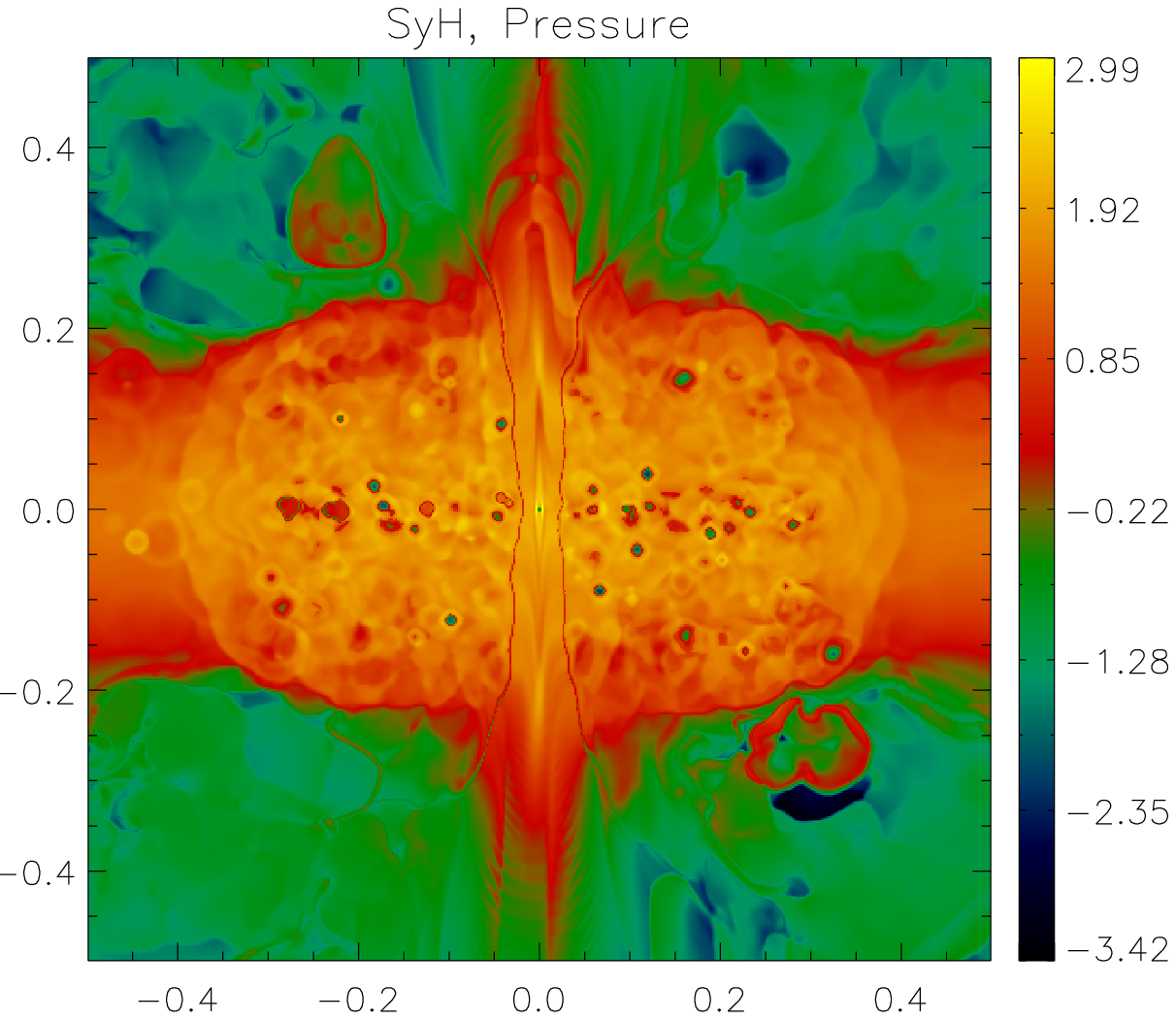}
        }%
        \hspace{0.7cm}
        \subfigure{%
            \label{fig:fourth}
            \includegraphics[width=0.43\textwidth]{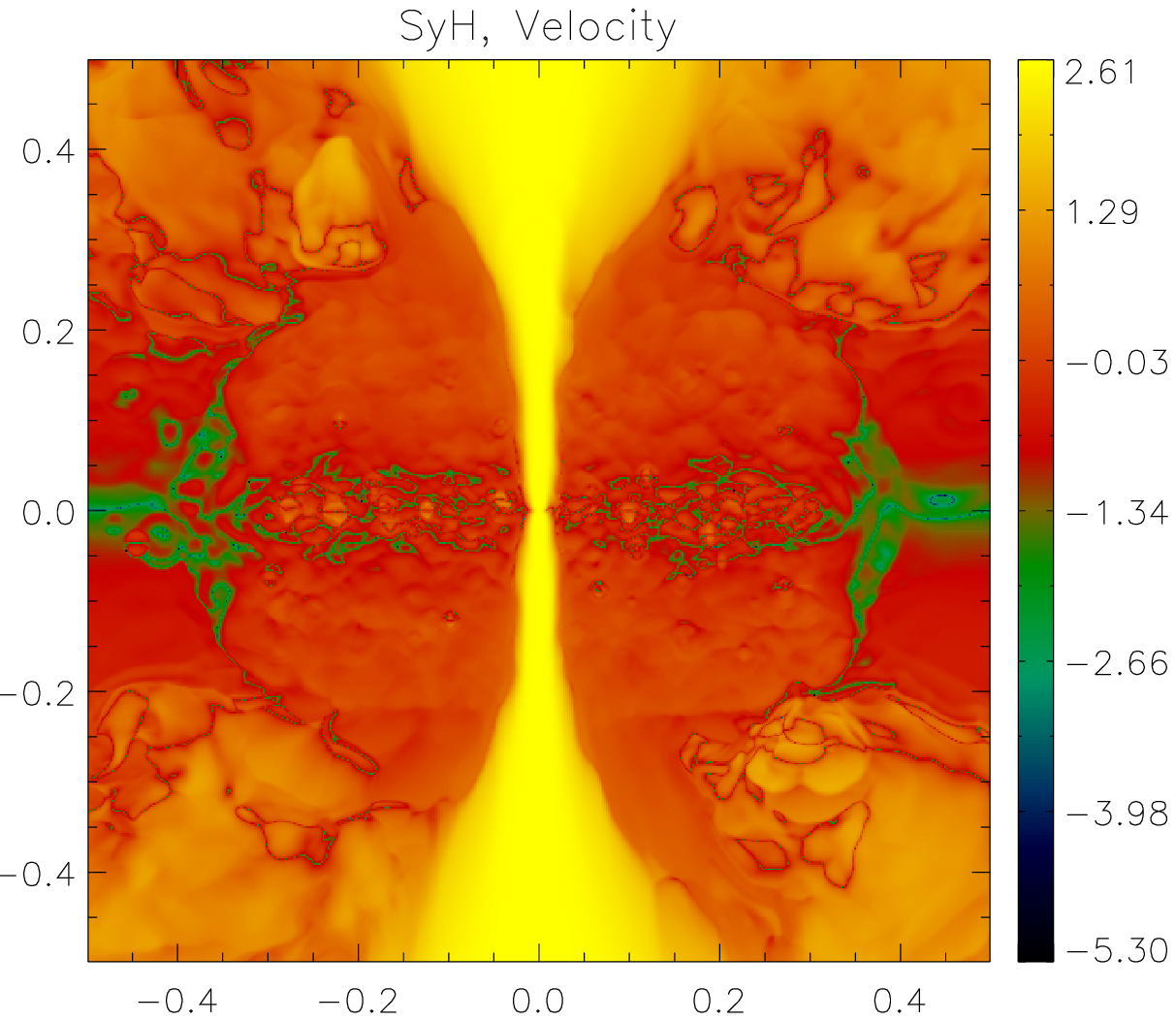}
        }%
    \end{center}
    \caption{Edge-on logarithmic gas density (upper left panel), 
temperature (upper right panel), pressure (bottom left panel) and velocity 
(absolute value, vertical component, bottom right panel) distribution for the 
model SyH-SNI-SB-JET $large$ at $t$ = 3.75 Myr.
Density is in cm$^{-3}$, temperature in K, pressure in units of 
7$\times 10^{-12}$ erg cm$^{-3}$ (code units) and
velocity in units of the reference sound speed ($c_{s,5\times 10^4}$ = 33 km 
s$^{-1}$).}
   \label{fig:SNI-SB-JET}
\end{figure*}

As in the study of the $SyM$ setup, this model has similar initial conditions 
to the previous one (SyH-SNI-SB), but includes also the presence of a highly 
collimated jet with the same characteristics described above (Table 1 and 
\S 4.2 4.2). We calculated the evolution of the system over a time of 6 Myr and
the results are shown in Figures \ref{fig:M_evol_V2}, \ref{fig:M_lost_V2}, 
\ref{fig:stat_V2} (solid-black lines), and  \ref{fig:SNI-SB-JET}. 
Also in this case, the gas evolution does not differ much from the model above
without the jet. The gas transport above the disk is somewhat less efficient 
than in the SyH-SNI-SB $large$ model, with a mean mass transfer rate of about 
15 M$_{\odot}$ yr$^{-1}$ (and a maximum of $\sim$ 30 M$_{\odot}$ yr$^{-1}$), 
while in the central core of the disk, within 40 pc, about 90\% of the gas is 
removed during the first 4$\times 10^5$ yr. due to the presence of the jet.  
However, this trend does not represent a steady state  and we note that after
2 Myr the gas evolution is very similar to that of the previous model.
Also the density, pressure and  temperature evolution of the disk are 
comparable with those of the model SyH-SNI-SB $large$, so that we may 
conclude that also in a system with properties that resemble those of a 
Seyfert galaxy with a thick disk, the SMBH jet is not expected to affect the 
global evolution of the gas in the inner regions, up to $\sim$ 500 pc. 

\begin{figure}    
\begin{center}   
\psfig{figure=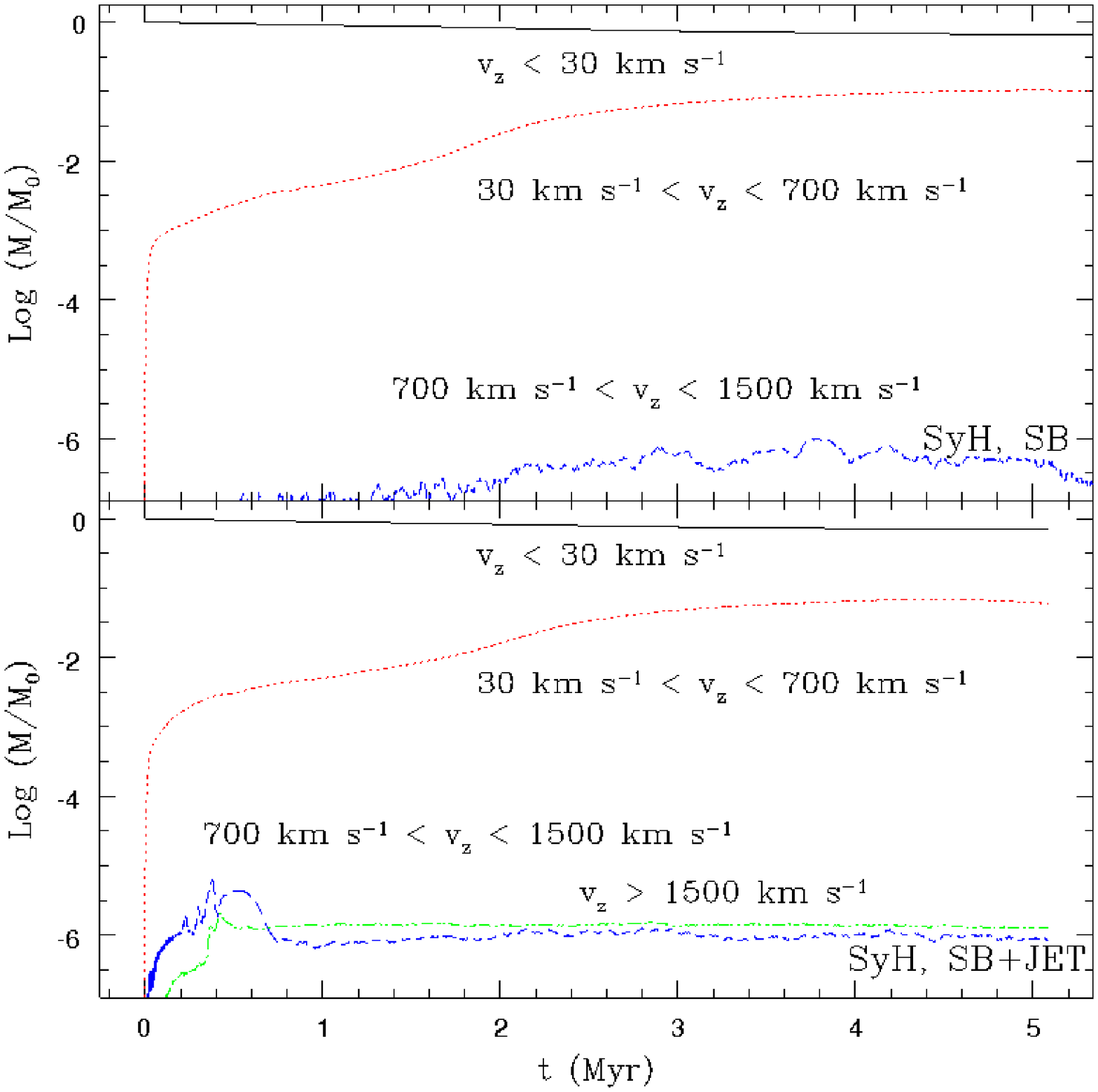,width=0.49\textwidth}    
\end{center}   
\caption{Time evolution of the mass of gas within the whole
system for the models SyH-SNI-SB (top panel) and SyH-SNI-SB-JET
(bottom panel) for different vertical velocities. Time is in Myr
and masses are in units of M$_{\odot}$, logarithmic scale.}
\label{vel_V2} 
\end{figure}

\begin{figure}    
\begin{center}
\psfig{figure=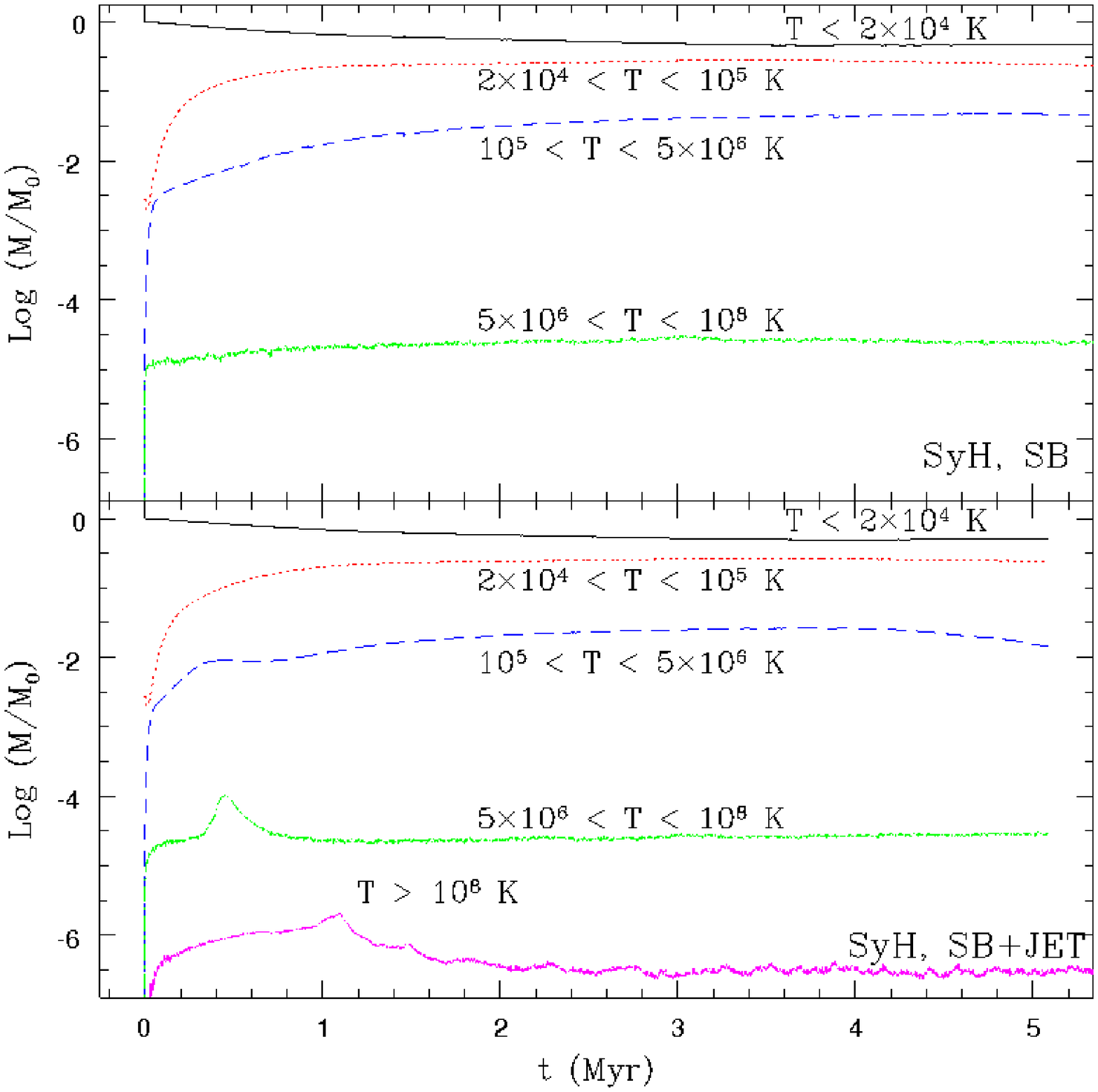,width=0.49\textwidth}    
\end{center}   
\caption{Time evolution of the mass of gas within the whole
system for the models SyH-SNI-SB (top panel) and SyH-SNI-SB-JET
(bottom panel) for different gas temperatures. Time is in Myr
and masses are in units of M$_{\odot}$, logarithmic scale.}
\label{tem_V2} 
\end{figure}

To understand better the evolution of this system SyH-SNI-SB-JET compared 
to the previous one without the jet (SyH-SNI-SB $large$), we have plotted in 
Figs. \ref{vel_V2} and \ref{tem_V2} the time evolution of the gas mass for 
different values of the vertical velocity and the temperature, respectively.
We note that the gas mass related to a given velocity and temperature in the 
model including the jet represents only a few per cent of the total mass of 
the nuclear region. This explains why we see no significant differences in 
the gas evolution between the two models.
In other words, though the presence of the jet may change a little the 
velocity and temperature profiles and thus the emission structure, it does 
not seem to be efficient enough to remove much gas at the galactic scales.

\subsubsection{SyH-SNI-JET}

To further stress the results and conclusions above and verify whether or not 
in the absence of a SB the remaining injecting mechanisms are still able to 
sweep large amounts of gas from the thick disk to the halo, we have also run a 
model including only the jet and the SNI. The magenta short-dashed lines of 
Figs. \ref{fig:M_evol_V2}, \ref{fig:M_lost_V2} and \ref{fig:stat_V2} indicate
that the only effect of the jet on the evolution of the system occurs in the 
nuclear region, within 40 pc, as in the previous model. No gas mass is
transported to outside and no gas outflow from the disk to the halo
is observed. The jet crosses the disk affecting it just in the very beginning
of its evolution. After a few hundreds of thousands years the jet interaction 
with the disk ceases and it continues to flow freely through the low-density 
channel previously opened by it.

\subsubsection{SyH-SNI-SB-JET-light}

\begin{figure*}
     \begin{center}
        \subfigure{%
            \label{fig:first}
            \includegraphics[width=0.43\textwidth]{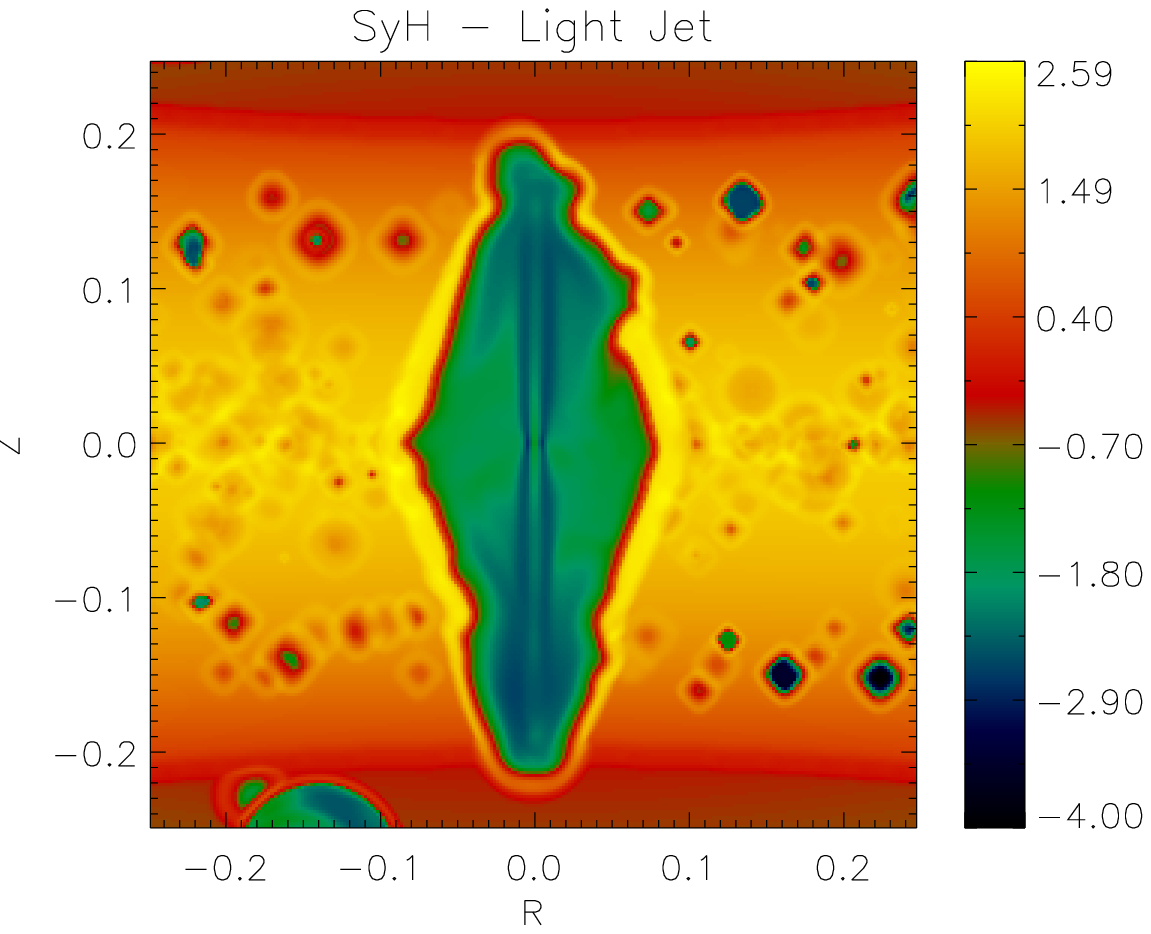}
        }%
        \hspace{0.7cm}
        \subfigure{%
           \label{fig:second}
           \includegraphics[width=0.43\textwidth]{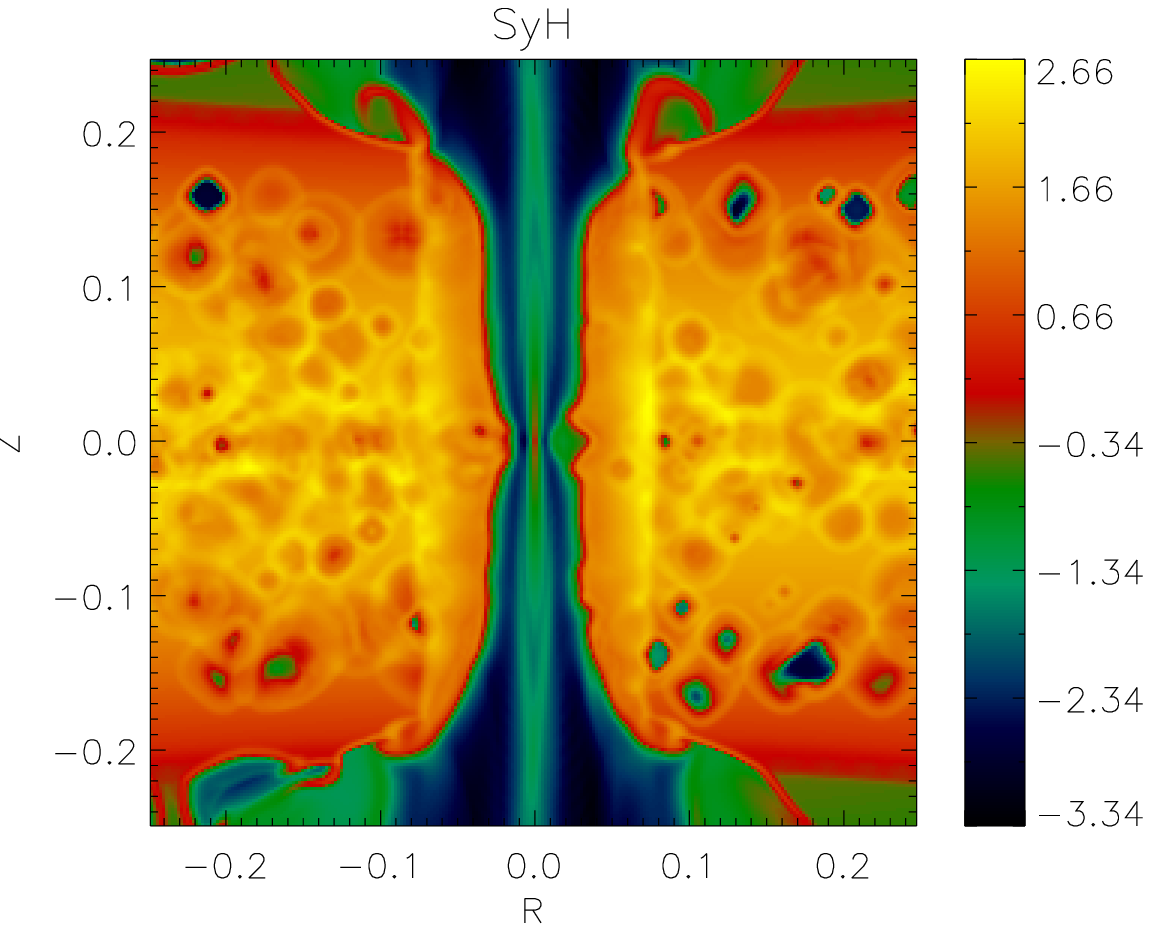}
        }\\ 
        \subfigure{%
            \label{fig:third}
            \includegraphics[width=0.43\textwidth]{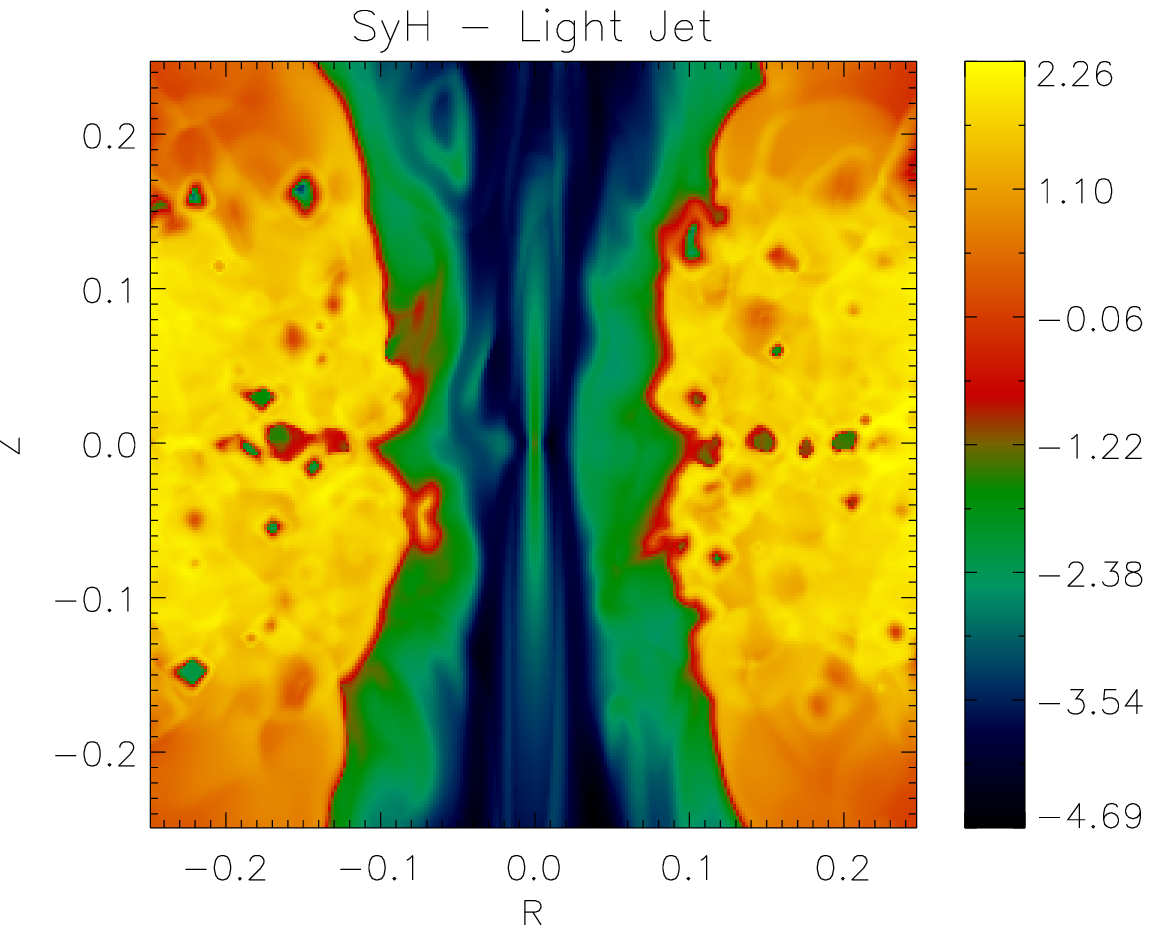}
        }%
        \hspace{0.7cm}
        \subfigure{%
            \label{fig:fourth}
            \includegraphics[width=0.43\textwidth]{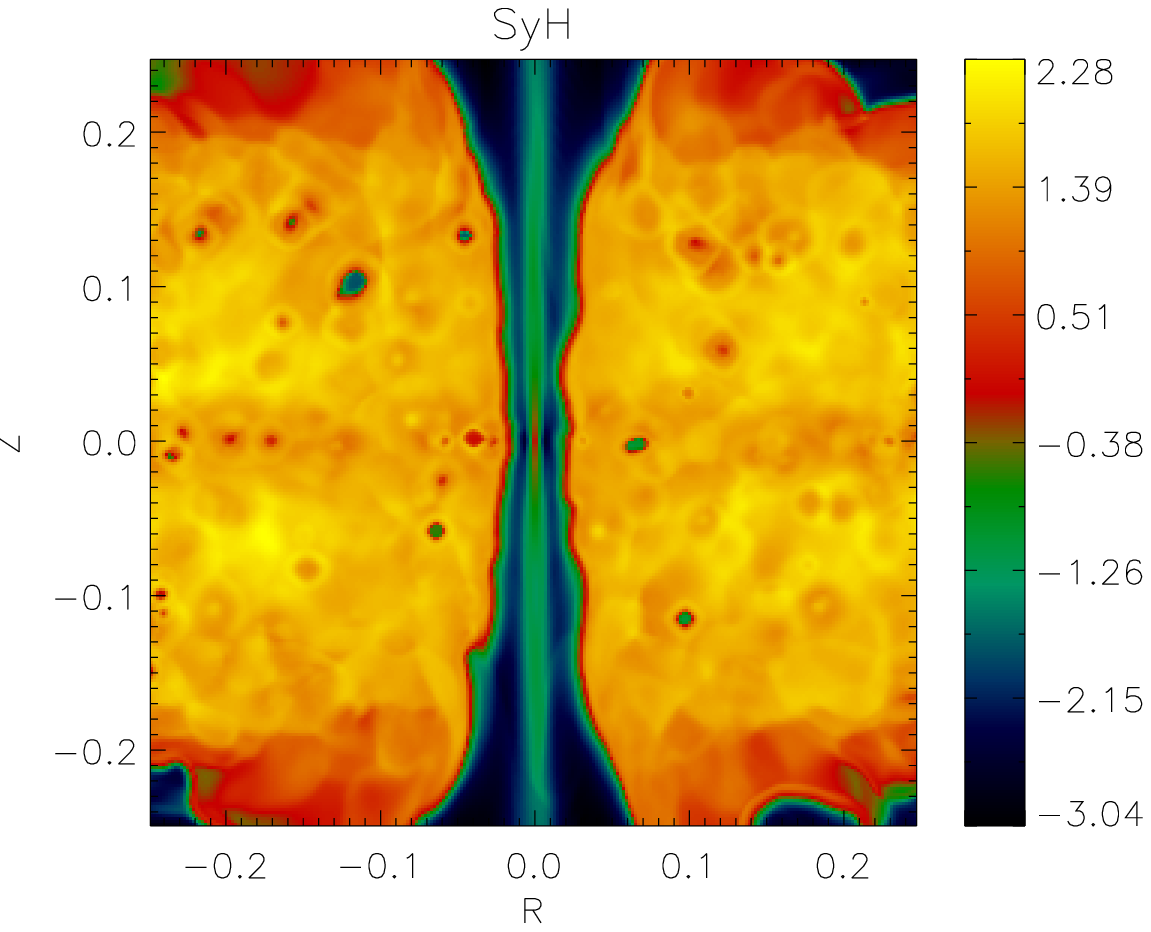}
        }%
    \end{center}
    \caption{Edge-on logarithmic gas density distribution for the models 
SyH-SNI-SB-JET $light$ (left panels) and SyH-SNI-SB-JET (right panels) at 
the nuclear region between -0.25 and 0.25 kpc (along R and z directions), at 
$t$ = 0.4 Myr (upper panels) and $t$ = 3 Myr (bottom panels). 
Distances are given in kpc and densities are in cm$^{-3}$.}
   \label{fig:SNI-SB-JETlight}
\end{figure*}

As remarked in \S 4.2, aiming at verifying how the characteristics of the jet 
can affect the evolution, we have also considered a $lighter$ jet with a 
relativistic velocity of $\sim$ 0.2 $c$ and a rate of injected matter of 
$\sim$ 5 $\times 10^{-4}$ M$_{\odot}$ yr$^{-1}$. In this case, though the jet 
speed, $v_j$, is larger, the density contrast between the jet and the disk 
near the base is much smaller ($\eta= \rho_{jet}/\rho_{disk}$=5.5$\times 10^{-4}$)
than in the previous model (SYH-SNI-SB-JET).
Considering the momentum flux conservation at the shock that forms at the 
head of the jet, one can easily demonstrate that the jet propagation velocity 
into the disk, $v_h$, depends both on the jet speed and the density contrast, 
i.e. $v_h \simeq v_j/(1 + \eta^{-1/2})$ \citep[e.g.][]{bete93}. 
This indicates that the larger the density contrast and the jet speed the 
faster the jet propagates. For the model above with a light jet 
(SyH-SNI-SB-JET $light$), the much smaller density contrast gives a 
propagation velocity $v_h$ smaller than that of model SyH-SNI-SB-JET, in 
agreement with the results of the simulations 
(see Fig. \ref{fig:SNI-SB-JETlight}).  
Thus the corresponding time scale to cross the disk, of about 0.4 Myr, is 
longer than in the previous model, during which a slow shock wave
propagates into the disk transporting mass, momentum and heating to
the ISM arising its temperature to about $10^8$ K (with peaks of 10$^9$ K).
 
\begin{figure}    
\begin{center}   
\psfig{figure=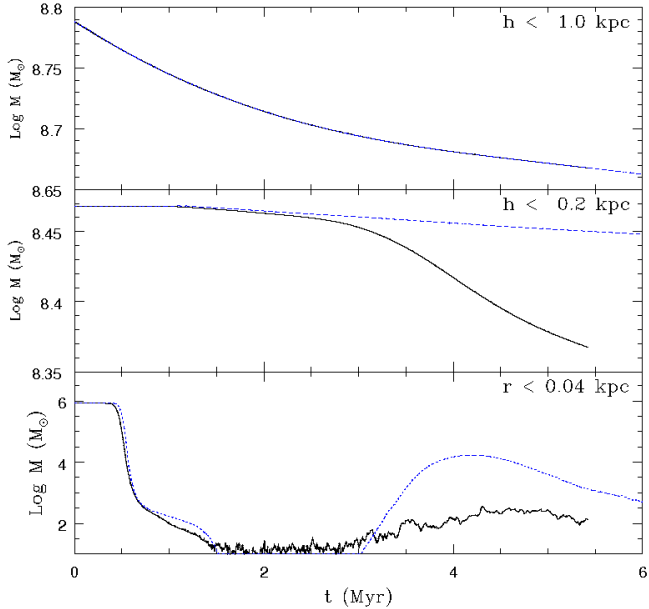,width=0.49\textwidth}    
\end{center}   
\caption{Time evolution of the mass of the gas within the whole system 
(z $\le$ 500 pc, upper panel), the thick disk (z $\le$ 200 pc, middle panel) 
and the central core of the galaxy ($r \le$ 40 pc, bottom panel) for the models 
SyH-SNI-SB-JET $light$ (solid-black lines) and SyH-SNI-JET $light$ 
(dotted-blue lines). Time is in Myr and mass is in units of M$_{\odot}$, 
logarithmic scale.}
\label{fig:M_jetlight} 
\end{figure}

\begin{figure}    
\begin{center}   
\psfig{figure=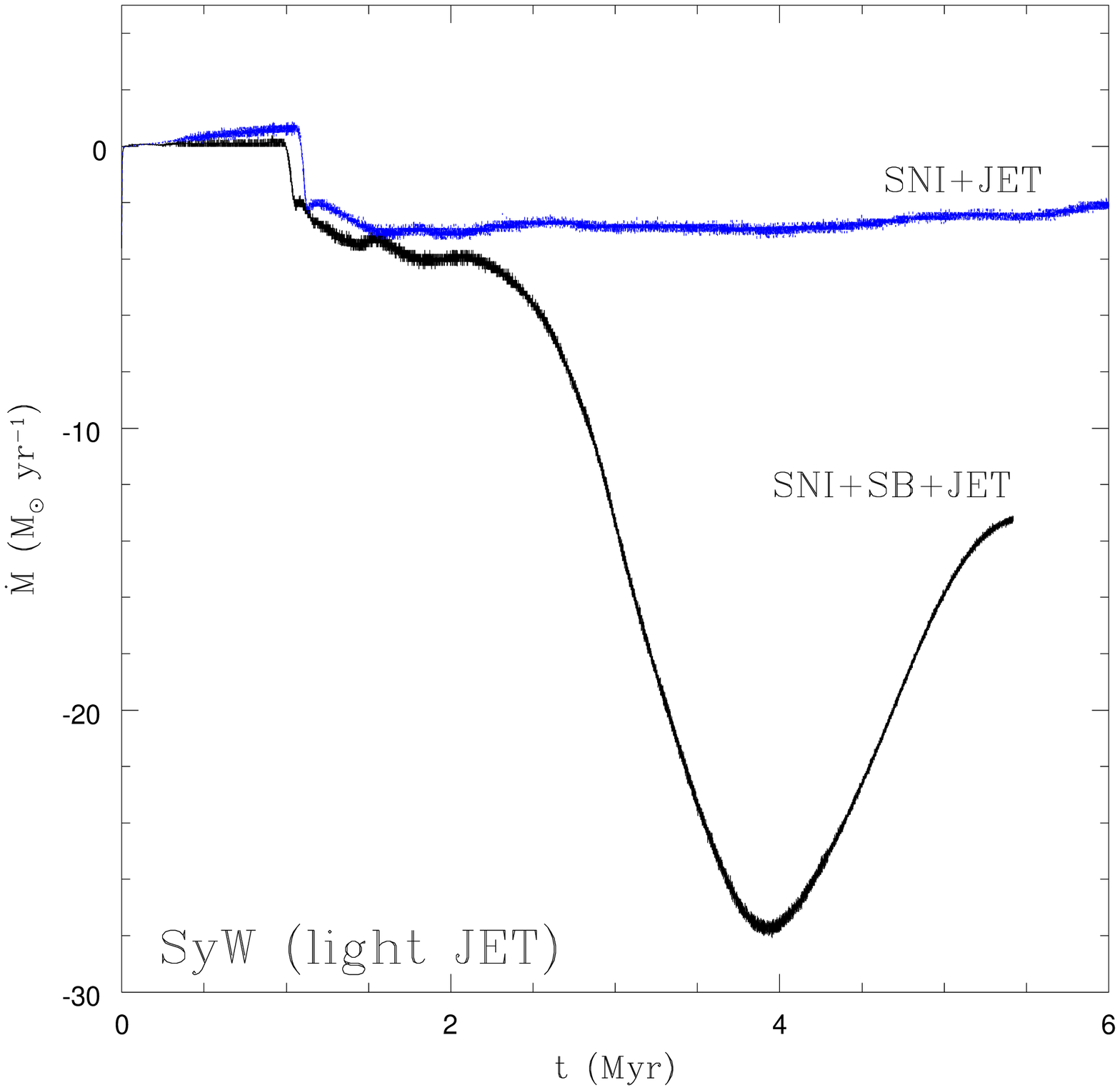,width=0.49\textwidth}    
\end{center}
\caption{Time evolution of the gas mass transfer and loss rate of the 
thick disk (z$\le$ 200 pc) for the models SyH-SNI-SB-JET $light$ 
(solid-black lines) and SyH-SNI-JET $light$ 
(dotted-blue lines). Time is in Myr and mass loss rate is in units of 
M$_{\odot}$ yr$^{-1}$.}
\label{fig:M_lost_jetlight} 
\end{figure}

However, after the jet breaks out into the halo the 
shock wave that propagates along the plane of the disk reaches a steady
state, separating the low-density, high-temperature region where the
jet propagates from the rest of the disk, within a distance of 80 pc from the
center of the galaxy. This region is larger than that obtained in the 
previous simulation  with a denser and slower jet 
(see Fig. \ref{fig:SNI-SB-JETlight}), but this difference does not affect 
much the evolution of the disk and confirms the results obtained in the 
previous models, as we can see in Figures \ref{fig:M_jetlight} and 
\ref{fig:M_lost_jetlight}, where we compared the mass evolution and the mass 
loss rate of the models with (SyH-SNI-SB-JET $light$) and without 
(SyH-SNI-JET $light$) a SB region.

Furthermore, since the faster the jet crosses the disk less energy is 
deposited into the ISM of the host galaxy, the jet with the higher density 
contrast (SyH-SNI-SB-JET in Figure \ref{fig:SNI-SB-JETlight}, right panels) 
produces a more collimated flow. 
On the other hand, the longer time of energy deposition into the disk 
by the lighter jet in SyH-SNI-SB-JET $light$ model due to the slower 
evolution, results a similar mass evolution to that of Model SyH-SNI-SB-JET, 
with differences of only a few  M$_{\odot}$ per year for the mass transfer 
rate of the thick disk and no significant differences in the mass loss rate 
of the whole system.

\subsection{Seyfert galaxy models with higher SFR (10SB)}

Finally, we have run  models  (not shown here) considering a SFR of the SB 
region 10 times higher than in the previous cases. We considered a $SyH$ setup 
with (SyH-SNI-10SB-JET model) and without (SyH-SNI-10SB model) the jet. 
These models were run either with higher or lower gas metallicity than the 
solar values\footnote{In the other models investigated here we employed  
solar metallicity.} and with different SN distribution, considering both 
single and clustered SN explosions. 
We have found similar results between these tests, nearly independent of the 
metallicity \citep[see, hoever,][for a more detail study 
on the effects of metalicity in galactic and SB winds]{melioli09, melioli13, 
melioli15}, which are also consistent with those presented above.
Of course, the mass transfer rate from the disk to the halo is higher and, 
almost insensitive to the presence of the jet, reaches a maximum value of 
$\sim$ 170 M$_{\odot}$  yr$^{-1}$ and a mean value of $\sim$ 90 M$_{\odot}$ 
yr$^{-1}$, resulting in a total mass exchange between the disk and the halo 
involving about 40\% of the mass of the system. 
Moreover, the higher energy injected by the SB region increases the 
thermal pressure of the gas in the nuclear region and consequently, the channel
dug by the jet is more confined and tighter. This result suggests that the 
higher the SFR and the column density of the disk, the greater is the 
collimation of the SMBH jet.

\section{Discussion and conclusions}

In this study our main goal was to understand the role of the collimated jet 
emerging from the center of a Seyfert-like galaxy in the development of a 
broader massive gas outflow in the nuclear region.
In contrast to previous studies 
\citep[e.g.][]{ciotti07, sijacki07, GaspMel11, fabian12}, we 
focused our attention on a small region of the host galaxy, characterized by a 
volume of 1 kpc$^3$, and performed high resolution (1.9 pc) hydrodynamical 
simulations which helped us to distinguish  the main driving mechanisms of the
gas evolution.

It has long been recognized the importance of a SMBH jet in balancing cooling
flows, inflating the X-ray cavities in the hot gas of galaxy 
cluster cores, or energizing the intergalactic medium 
\citep[e.g.,][]{Boehringer93, McNamara07, DiegoGall, Falceta_Cap, 
GaspMel11, GaspMel11b}. 
However, all these processes act at large scales, from 10 kpc to 1 Mpc. 
Less explored is the importance of the jet feedback on the small scales of 
the host galaxy.

Considering a rotating disk gas distribution initially in equilibrium with 
the galaxy gravitational potential with a given column density, we explored 
the evolution of this distribution in the surrounds of the nuclear region 
over $\sim$ 10 Myr, taking into account, either separately or all together, 
the effects of star formation, SN explosions, and the jet. In general lines, 
we have found that the formation of turbulent and clumpy outflows in the 
nuclear region of Seyfert-like systems is mainly driven by intense star 
formation. The jet feedback at these scales is less relevant.

In the next paragraphs we discuss our main results and their implications for 
the evolution of the nuclear regions of Seyfert galaxies, the formation 
of a multiphase ISM, the gas outflow, and the jet feedback on the host galaxy.

\subsection{Gas mass evolution and outflow collimation}

In all models studied, the gas of the disk is carried out in an 
outflow only in presence of star formation activity, regardless of the column 
density of the host galaxy. 
Nor the regular activity in the bulge (mainly due to SNI explosions), nor the 
jet, or both together are able to drive gas outflow at the small scales, 
particularly  after the first Myr, when the jet breaks through the disk and 
evolves along a low-density and high-temperature channel with a thickness of 
few tens of pc. 
We should stress, however, that we considered a radiative cooling optically 
thin gas and neglected the presence of dust and radiation pressure. 
Though this could be another driving mechanism, as remarked in \S 1 the 
typical gas column densities (between 10$^{21}$ and $\sim 10^{23}$ cm$^{−2}$) 
and the rapid destruction rate by SN shock waves  make it unlikely the 
survival of substantial amounts of dust in Seyfert nuclear regions 
\citep[see, e.g.][]{McKeeDust, JonesDust}.
\footnote{It is interesting to note that in a recent study, 
\citet[][]{gan14} addressed the problem of the AGN feedback in the nuclear 
regions of elliptical galaxies, and argued that Compton scattering can heat 
the gas in the surroundings of the AGN to temperatures as high as $10^9$ K and 
for this reason have taken into account the effects of the AGN radiative 
feedback on the system. In their simulations (which are performed for 
two-dimensional flows that arise from the center with a large opening angle), 
they find that this heating process will dominate only in the central 
regions over radial extensions of about 10 to 30 pc. Similarly, in our 
simulations with AGN feedback, although we neglected explicitly the AGN 
Compton heating effect, we also find that the temperature grows to  
$\sim 10^9$ K within a radial extension up to 20-35 pc due to the AGN jet 
shock heating. Therefore, the extension of the influence of the AGN heating 
feedback found in both studies is essentially the same.}

Though the analytical steady state approach indicates that the energy lost 
by radiative cooling in the denser regions of the disk would be high enough 
to balance the energy injected by the stellar feedback, thus preventing 
the formation of a galactic wind (see \S 3), the numerical hydrodynamical 
simulations have demonstrated that this steady state lasts only about 1$-$2 
Myr, after which the gas begins to flow out of the disk. 
In fact, at the beginning of the simulations the gas in the disk remains 
unperturbed and homogeneous, and its behaviour follows the one described by 
the analytical solutions in \S 3. However, with the energy injection by the 
stellar feedback, shock waves propagate into the ISM and the system becomes 
highly non-steady.
Shock heated, low density gas emerges from the disk and propagates into the 
stratified gas, transferring mass to heights greater than 200$-$300 pc.
The average mass transfer rate we obtained is between 3 and 25 M$_{\odot}$ 
yr$^{-1}$ for a SFR of 1 M$_{\odot}$ yr$^{-1}$ and a disk column density 
between $10^{21}$ and $10^{23}$ cm$^{-2}$, and between 60 and 90 M$_{\odot}$ 
yr$^{-1}$ for a SFR of 10 M$_{\odot}$ yr$^{-1}$ and a disk column density 
between $10^{22}$ and $10^{23}$ cm$^{-2}$. The gas outflow 
ends when about $\sim$50$-$70\% of the mass in the nuclear region 
(within a radius of $\sim$ 300 $-$ 400 pc, is removed.
At larger heights, the impact of the SN energy on the gas mass evolution is 
not so important. For instance, at a distance of 500 pc above the disk 
we find that the mass loss rate is only a few M$_{\odot}$ yr$^{-1}$ (for a SFR 
of 1 M$_{\odot}$ yr$^{-1}$) and this value is almost independent of the disk 
column density.
 
On the other hand, when considering only the propagation of the SMBH jet 
through the disk there is only a small fraction of gas removal from the core 
of the galaxy (within $r \sim$ 40 pc), but no relevant amount is carried 
away by shock waves and/or by gas heating to generate a wind. 

We have evaluated also the energy associated to the evolved outflow in 
the models and found that, in the cases without the AGN feedback,
the kinetic power of the outflow increases to a value between 
(2 to 4)$\times 10^{40}$ erg s$^{-1}$, that is, about 6.5\% to 13\% of the 
stellar feedback luminosity. At the same time, the thermal (or internal) 
energy rate of the gas outflow increases to about $10^{40}$ erg s$^{-1}$, 
that is, $\sim$ 3\% of the stellar feedback luminosity. 
In other words, about 15$-$20\% of the stellar 
luminosity goes to the outflow, and the remaining 80\% is radiated away. 
In the models where the jet feedback is included, we have obtained essentially 
the same luminosity values and for this reason we can conclude that 
we are observing a SF-driven wind.

The scenario would change completely in the case of an AGN outflow 
emerging with very large opening angles, as assumed, e.g., 
in \citet[][]{wagWind}. 
These authors (interested in exploring quasar-mode AGN feedback) have 
considered an AGN accretion disk outflow impinging into a very cloudy 
environment with a large opening angle of 30 degrees. 
In this case they find that the outflow can substantially affect the galactic 
ISM. Nevertheless, as in the present simulations, their AGN wind is injected 
within a scale of 2 pc. The only difference is that in our models the jet 
emerges very collimated, as expected, while in \citet[][]{wagWind} the AGN 
disk wind is assumed to emerge largely uncollimated from the nuclear region. 
This lateral expansion may be due to the wind pressure (which in a quasar-mode 
AGN may be dominated by radiation) overcoming the external pressure 
\citep[e.g.,][]{king03, Crenshaw03, Arav05}.
However, as stressed, we expect that in the nuclear regions of the Seyferts 
this quasar-mode feedback and therefore, the development of strong AGN disk 
winds with large opening angles should be inoperative.

In summary, the nuclear galactic gas evolution of a Seyfert is almost 
insensitive to the passage of the AGN outflow and, as demonstrated 
above, it needs an intense and more widespread source of energy injection in 
the disk (as a high SFR) to increase the gas pressure and drive shock waves 
and turbulence that generate an expanding, multiphase ISM over $\sim$ 500 pc 
above the disk.

As discussed in \S 2, the soft X-ray emission in Seyferts is likely dominated 
by photoionized gas \citep[e.g.,][]{guainazzi09}. Its location, mass and 
momentum structure resembles a biconical outflow and it has been suggested 
that this might be an evidence of an AGN-driven wind 
\citep[e.g.,][]{longinotti08, evans10}. Nevertheless, a SB-driven 
outflow may also have a biconical shape above the disk 
\citep[e.g.][]{melioli13}.
Given the small scales here simulated it is difficult to reproduce the angle 
of gas escape, but the distribution of the kinetic energy that will be 
discussed in \S 6.3 provides useful information about the morphology of the 
gas outflow. We should also note that a new set of numerical simulations at 
intermediate scales ($\sim$ 10 kpc) for these flows is in progress which will 
allow to identify more precisely these morphological features (Melioli et al. 
2015, in prep.).

\subsection{Multiphase ISM}

The SN explosions and the development of shock waves into the ISM of the host
galaxy determine the formation of a multiphase ambient where gas with 
different densities and temperatures coexists at a given pressure.
Indeed, almost independently of the presence of a SMBH at the center of the 
galaxy, the stellar feedback generates naturally regions at high densities 
which cool fast, and regions where the passage of shock fronts leaves low 
density and very high temperature gas tracks.
Thus a region characterized by a high SFR may inject enough energy to 
blow-out the gas from the
disk and at the same time lead to the formation of clouds and clumps 
which can then be continuously generated and steadily carried away roughly
preserving the total cloud number.

At the beginning of its evolution, right after launching, also the jet 
generates high density regions through a transverse shock wave that 
propagates radially until reaching an equilibrium position at about 
20$-$30 pc away from the galaxy center (see Figure \ref{fig:SNI-SB-JETlight}, 
right panels). This shock wave initially 
induces some turbulence \citep[as suggested by, e.g.,][]{Guillard2015, 
Alatalo2015} that may trigger either star formation or quenching 
depending on the shock strength and the ISM conditions 
\citep[see, e.g.,][]{melioli06, leao09}, but compared to the dynamical time 
scales of the system this strong interaction lasts too little (about 0.3 Myr)
in our simulations and it is hard to predict whether or not it may have a 
major impact  over the star formation history of the inner region of the 
galaxy.
\footnote{We should note that, in the case (not considered here) of an 
intermittent jet, this would naturally allow for a continuous repetition of 
events such as those described above, i.e., every time that the jet would 
resume, the digging of a new tunnel into the ISM would induce a new shock 
wave, turbulence and the boosting or the quenching of star formation, 
over the whole evolution of the galaxy or as long as the intermittent jet 
would survive.}

Clouds and clumps are therefore continuously and mostly formed
in the nuclear region of the galaxy by the fragmentation of the shocked gas
compressed by the supernova shock fronts, and we verify that in our 
simulations about 30\% of the gas mass goes to clouds 
characterized by high density, between 100 and 1000 cm$^{-3}$. 
These clouds, that are photoionized by the central source (and by
the stellar feedback too) and radially dragged out of the nuclear region 
mainly via ram-pressure, could be identified with the NLR observed in the 
Seyfert galaxy. 
On the other hand, we have found that the jet feedback alone is unable to 
produce a multiphase ambient and the dense clouds in the 
nuclear region of the host galaxy.

\begin{figure*}
     \begin{center}
        \subfigure{%
            \label{fig:first}
            \includegraphics[width=0.43\textwidth]{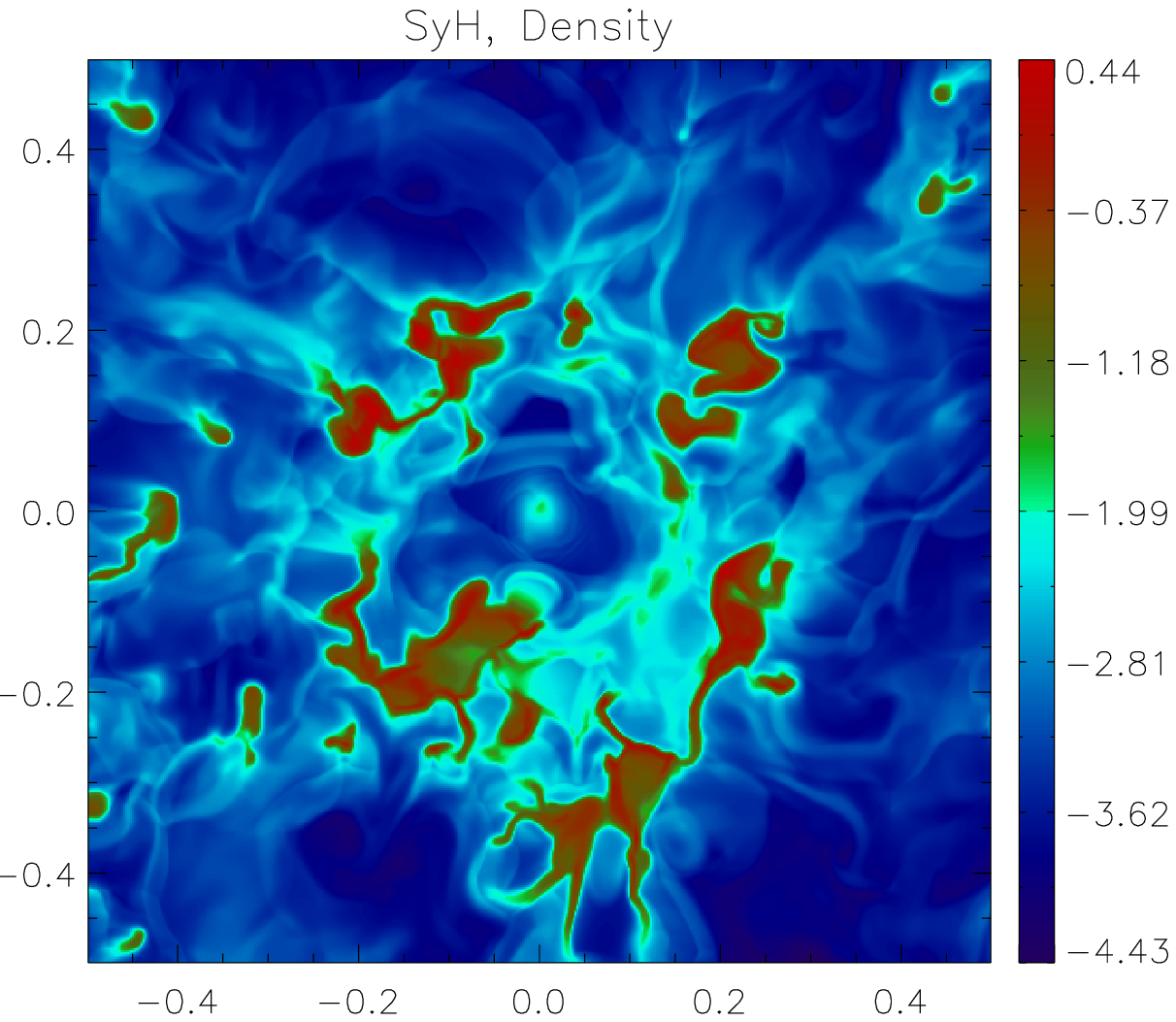}
        }%
        \hspace{0.7cm}
        \subfigure{%
           \label{fig:second}
           \includegraphics[width=0.43\textwidth]{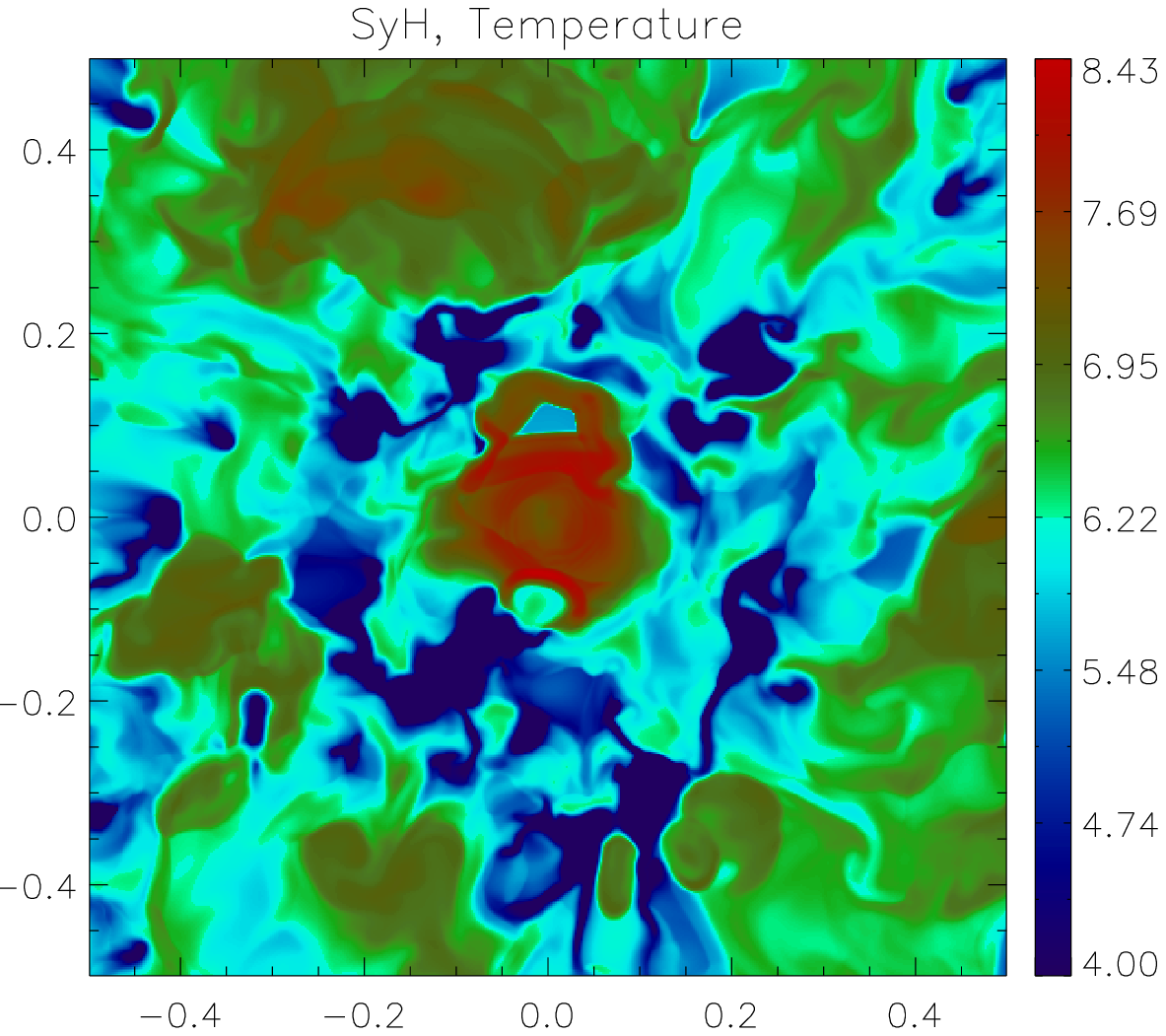}
        }\\ 
        \subfigure{%
            \label{fig:third}
            \includegraphics[width=0.43\textwidth]{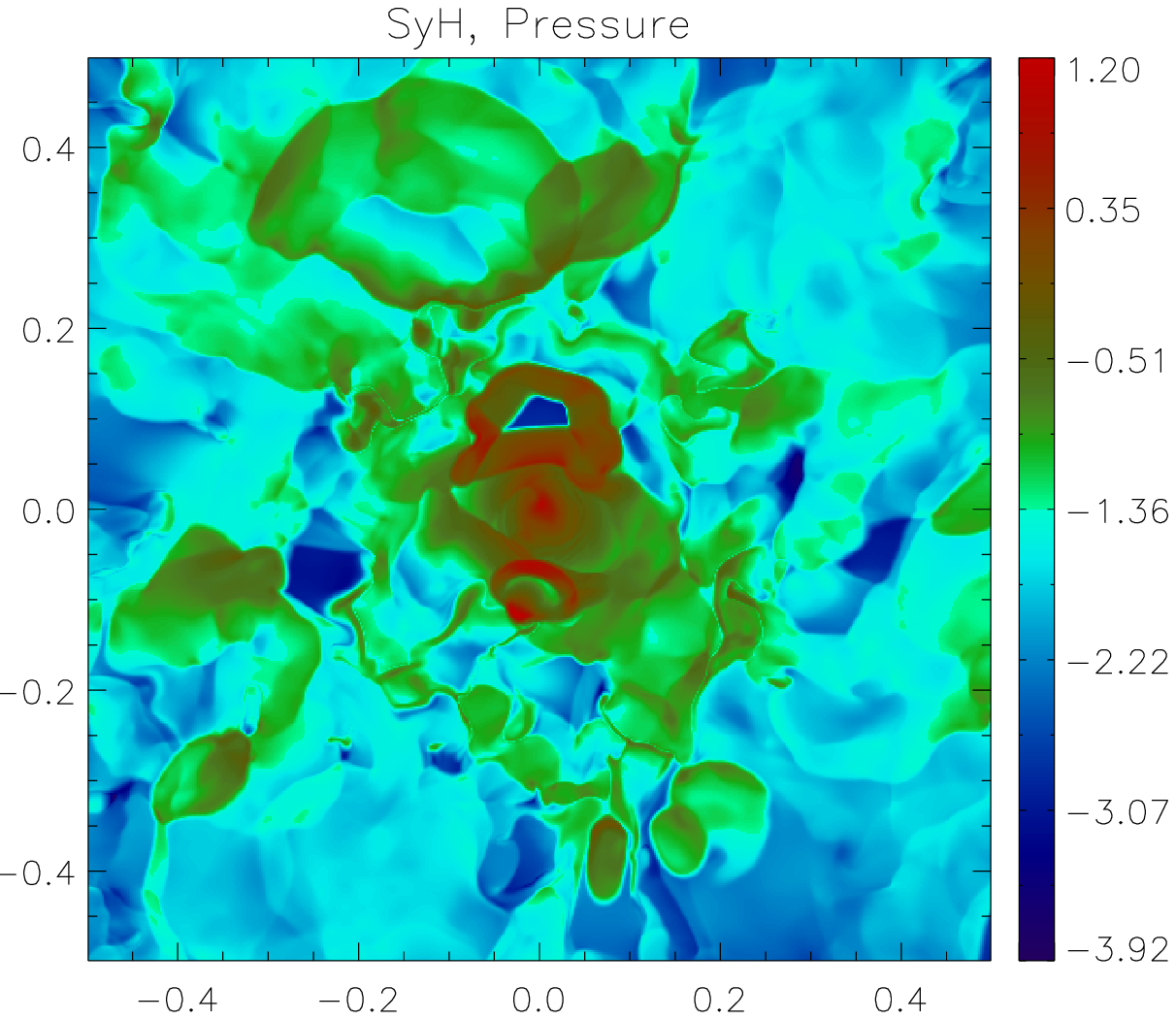}
        }%
        \hspace{0.7cm}
        \subfigure{%
            \label{fig:fourth}
            \includegraphics[width=0.43\textwidth]{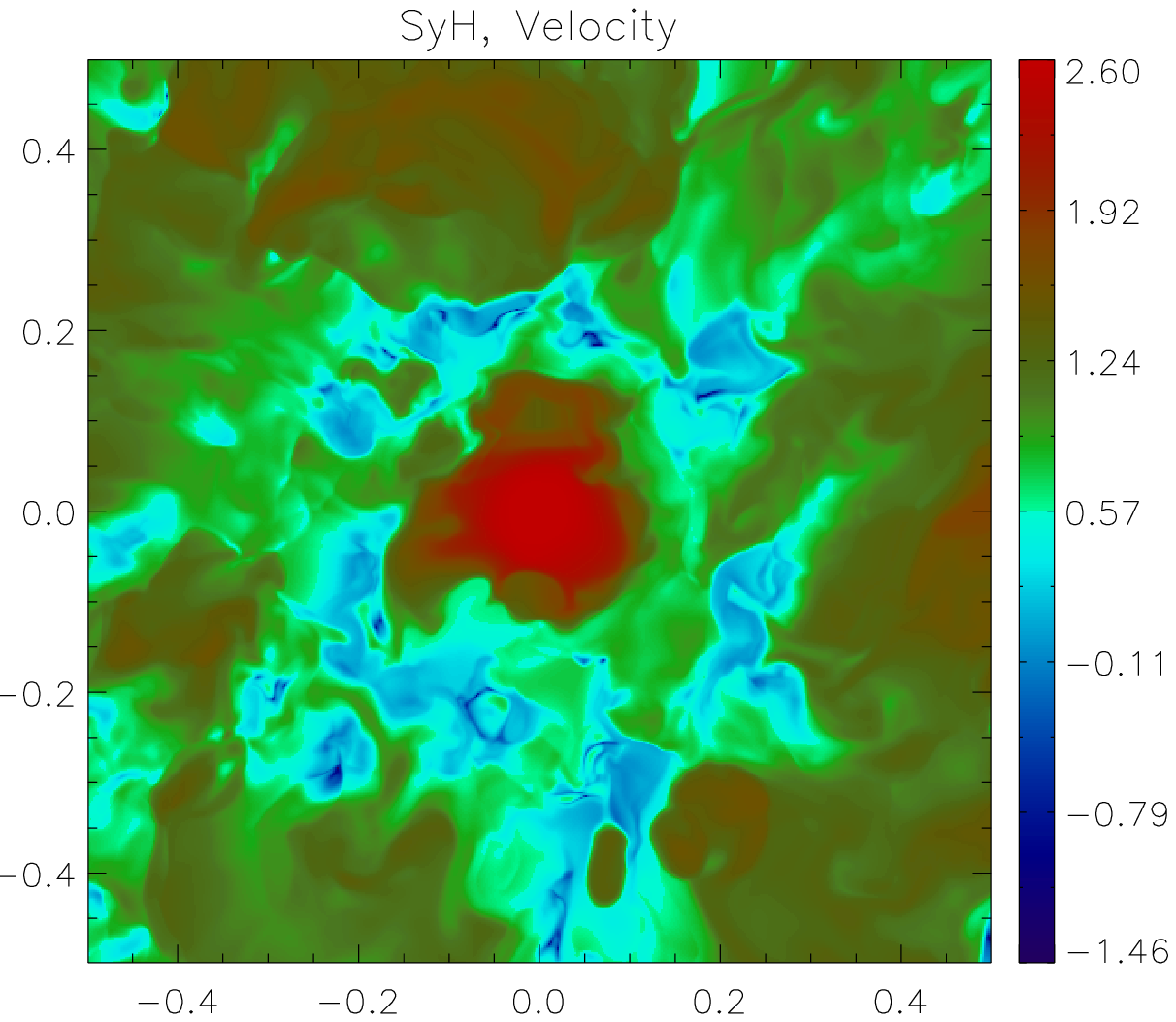}
        }%
    \end{center}
    \caption{Face-on logarithmic gas density (top-left panel), temperature 
(top-right panel), pressure (bottom-left panel) and velocity 
(bottom-right panel) distribution at $t$ = 6.5 Myr, for model SyH-SNI-SB-JET 
$large$ at a height $h$ = 430 pc. 
Distances are given in kpc, densities, temperatures and pressures are in cgs
and velocities are in units of the reference sound speed, computed at 
T=5$\times 10^4$ K, $c_{s,5\times 10^4}$ = 33 km s$^{-1}$.}
   \label{cloudmol}
\end{figure*}

At a given distance from the source of high energy photons, these clouds
may cool very fast to a temperature of few tens of K, since the cooling time
is $t_{cool}\sim 50/(\Lambda(T,z)_{10^{-24}} n_{10^3})$ yr, where 
$\Lambda(T,z)_{10^{-24}}$ is the value of the cooling function in units of 
$10^{-24}$ erg cm$^{3}$ s$^{-1}$ and $n_{10^3}$ is the gas density of the clouds 
in units of 10$^3$ cm$^{-3}$. 
From our simulations we note that these clouds are moving 
at velocities between -30 and 300 km s$^{-1}$, relative to the galaxy disk, 
depending on the height above the disk, and are characterized by a density 
contrast between 100 and 10$^4$, as shown in Figure \ref{cloudmol}.

The fact that some clouds have negative velocities implies that they may 
eventually fall back into the disk in a process that may resemble that of the 
galactic fountains \citep[e.g.][]{melioli08, melioli09}, while the clouds 
with positive velocities larger than the escape velocity can continue their 
journey through the halo. If dust forms in the way, it may 
eventually also originate molecular clouds, as recently observed, 
for example, in the western radio lobe of the Seyfert galaxy IC 5063 
\citep[][]{Morganti5063}. We note however, that in this system they detected 
clouds with higher velocities than those found in our simulations 
(which are around 1200 km s$^{-1}$). This may be an indication that the 
observed clouds in this case could be produced in the SB-driven wind, as 
described above, and then dragged by the much faster jet flow. 
Another possibility is that they result from direct interactions 
of the jet head front with the dense environment. We actually see this 
interaction in the very beggining of the simulations (in the first 10$^5$ yr) 
when the jet breaks into the ISM for the first time and sweeps the 
interstelar matter (see also, e.g., Falceta-Gon{\c c}alves et al., in prep.).

\subsection{Velocity distribution}

The velocity distributions depicted in Figures \ref{vel_V2} clearly show that 
one of the main differences between models with and without a SMBH jet is the 
gas with velocity signatures larger than 1500 km s$^{-1}$.
Depending on the column density of the disk of the host galaxy, the amount 
of very high velocity gas when a jet is present is between 5 and 20 times 
larger than in models without a jet. However, the fastest moving gas is
concentrated around the jet and the gas mass with
these very high velocities corresponds to about 500 $-$ 1000 M$_{\odot}$ only, 
an amount too low to be able to modify the overall evolution of the system, 
though it may influence in the drag of high velocity clouds 
\citep[][]{Morganti5063,tombesi15}, as remarked in \S 6.2. 

Therefore, we may conclude that the jet plays an important role in producing 
the very high gas velocities observed in the expanding nuclear regions of the 
Seyfert galaxies, but at the same time these velocities involve such a 
negligible amount of gas  mass compared to the total amount in the outflow 
that it is unable to drive the gas exchange between the disk and the halo 
within the kpc scale.

According to recent observations (see \S. 2 and references therein), we 
note that our results basically reproduce the structures of the gas outflow, 
that is, an extended gas outflow with systemic velocity around the 
nucleus having a broader biconical component perpendicular to the disk 
and an inner component due to the interaction between the jet and the 
galactic disk material.

Our simulations suggest that in order to achieve this structure, three 
different contributing factors are necessary: $i$) a diffuse region 
characterized by high SFR (or by a SB) which 
is responsible for the increase of thermal pressure, the driving of shock 
waves and the formation and acceleration of the denser features; $ii$) a SMBH 
jet, responsible for a little fraction of very high velocity clouds and for 
the gas at very high temperatures (see Fig. \ref{tem_V2}); and $iii$) the 
presence of high density gas within the thin disk (of the order of 10$^2$ 
cm$^{-3}$) in order to constrain most of the SB driven gas outflow to the 
inner radii (since the SB activity is smaller in the outer radii) and allow the 
formation of the biconical shape as observed.

\begin{figure}
     \begin{center}
        \subfigure{%
            \label{fig:first}
            \includegraphics[width=0.43\textwidth]{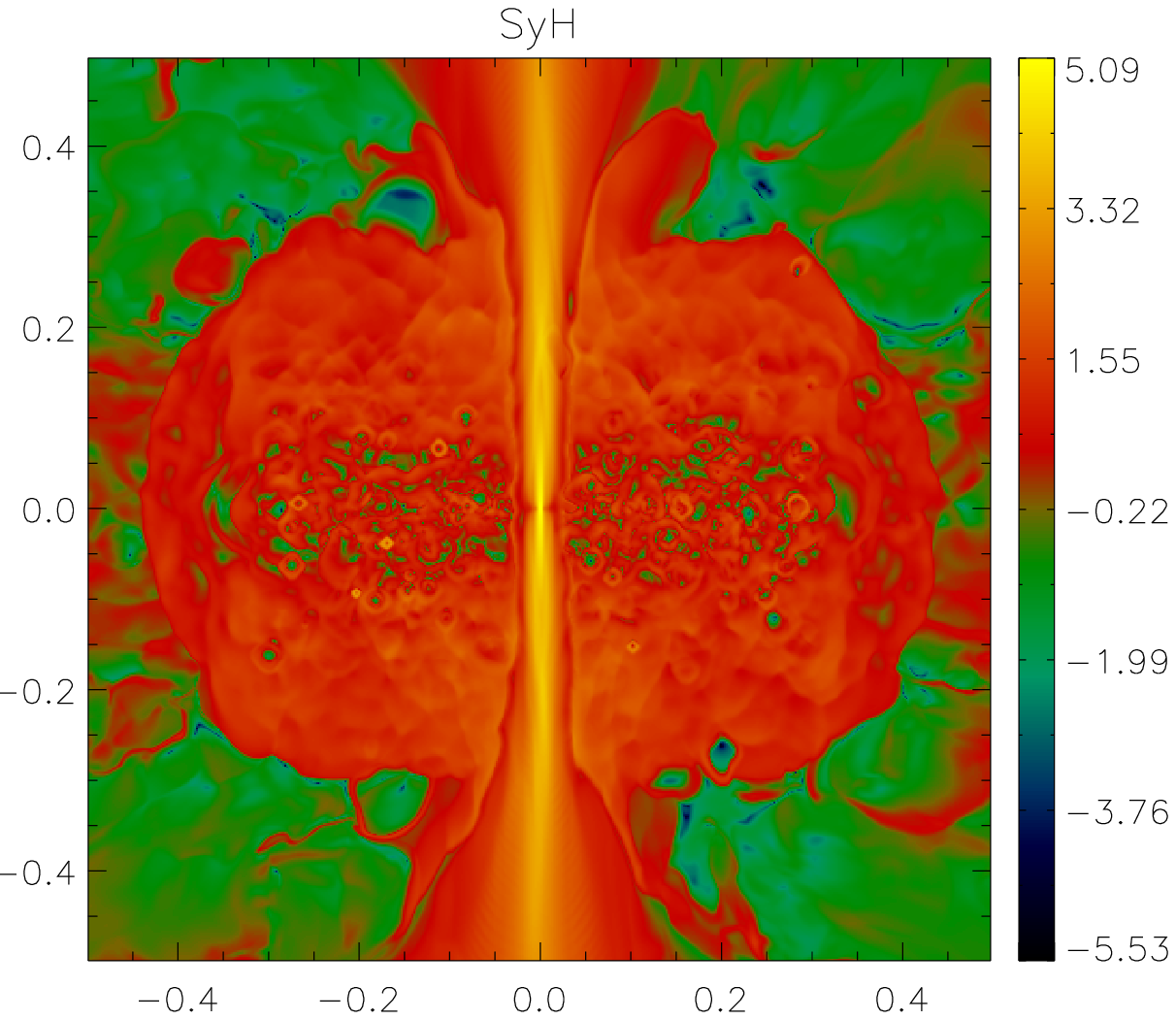}
        }\\ 
        \subfigure{%
            \label{fig:third}

            \includegraphics[width=0.43\textwidth]{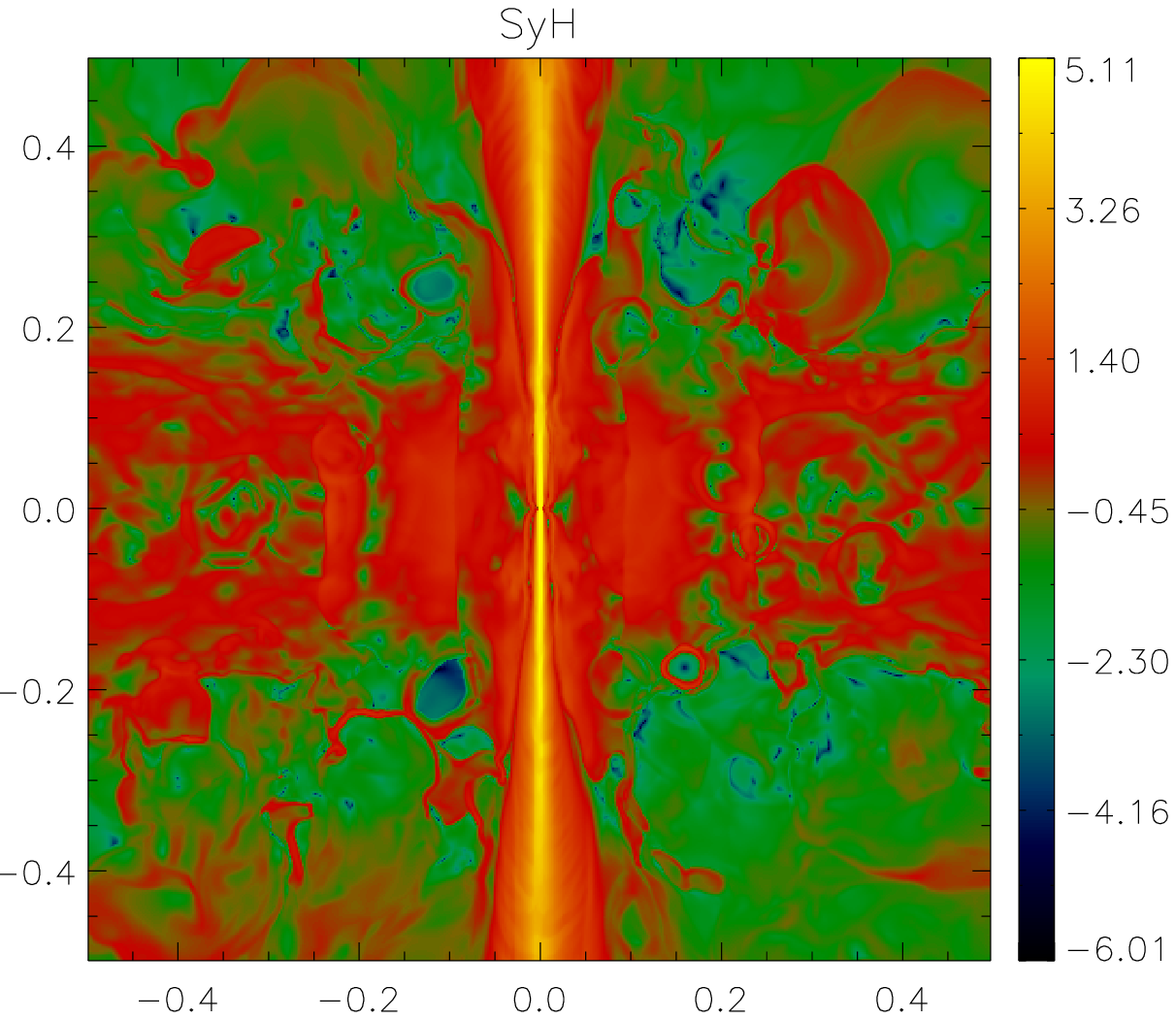}
        }%
    \end{center}
    \caption{Edge-on logarithmic kinetic energy gas distribution for  models 
SyH-SNI-SB-JET $large$ (left panel) and SyH-SNI-JET (right panel) at
$t$ = 4.8 Myr. Distances are given in kpc and energy is in units of 
5$\times 10^{45}$ erg (code units)}
   \label{logvel}
\end{figure}

Figure \ref{logvel} that depicts the edge-on logarithmic distribution of the 
kinetic energy of the gas for models SyH-SNI-JET and SyH-SNI-SB-JET $large$ 
(which differ only by the extension of the SB region), illustrates 
the scenario above.
The gas along the jet is clearly at very high velocity, but its high
collimation makes negligible the amount of kinetic energy that is transported
from the disk to the halo at $h_z \sim$ 500 pc. 
At the same time, the presence of a larger SB region causes an almost 
uniform expansion of the gas within 300$ - $500 pc from the center of the 
galaxy. In some regions the flow carries an amount of kinetic energy about 
1000 times larger than that associated to the jet propagation only. 
This energy transfer causes the development of a nearly biconical gas 
outflow characterized by a width of $\sim$ 200 pc and an opening angle of 
$\sim$ 60$^{\circ}$ at $h_z$ = 400 pc.

\begin{figure}
     \begin{center}
        \subfigure{%
            \label{fig:first}
            \includegraphics[width=0.35\textwidth,angle=270]{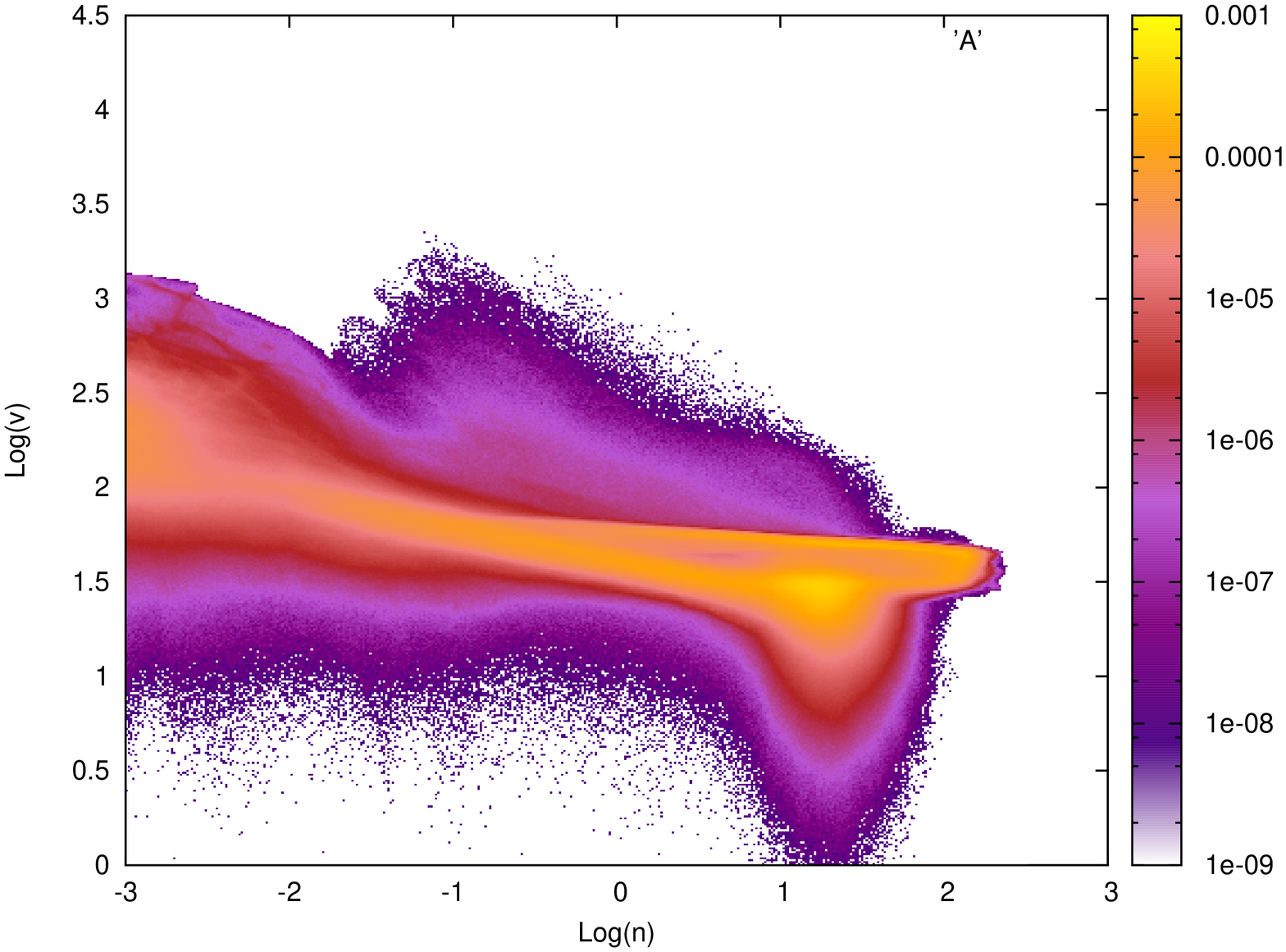}
        }\\ 
        \subfigure{%
            \label{fig:third}

            \includegraphics[width=0.35\textwidth,angle=270]{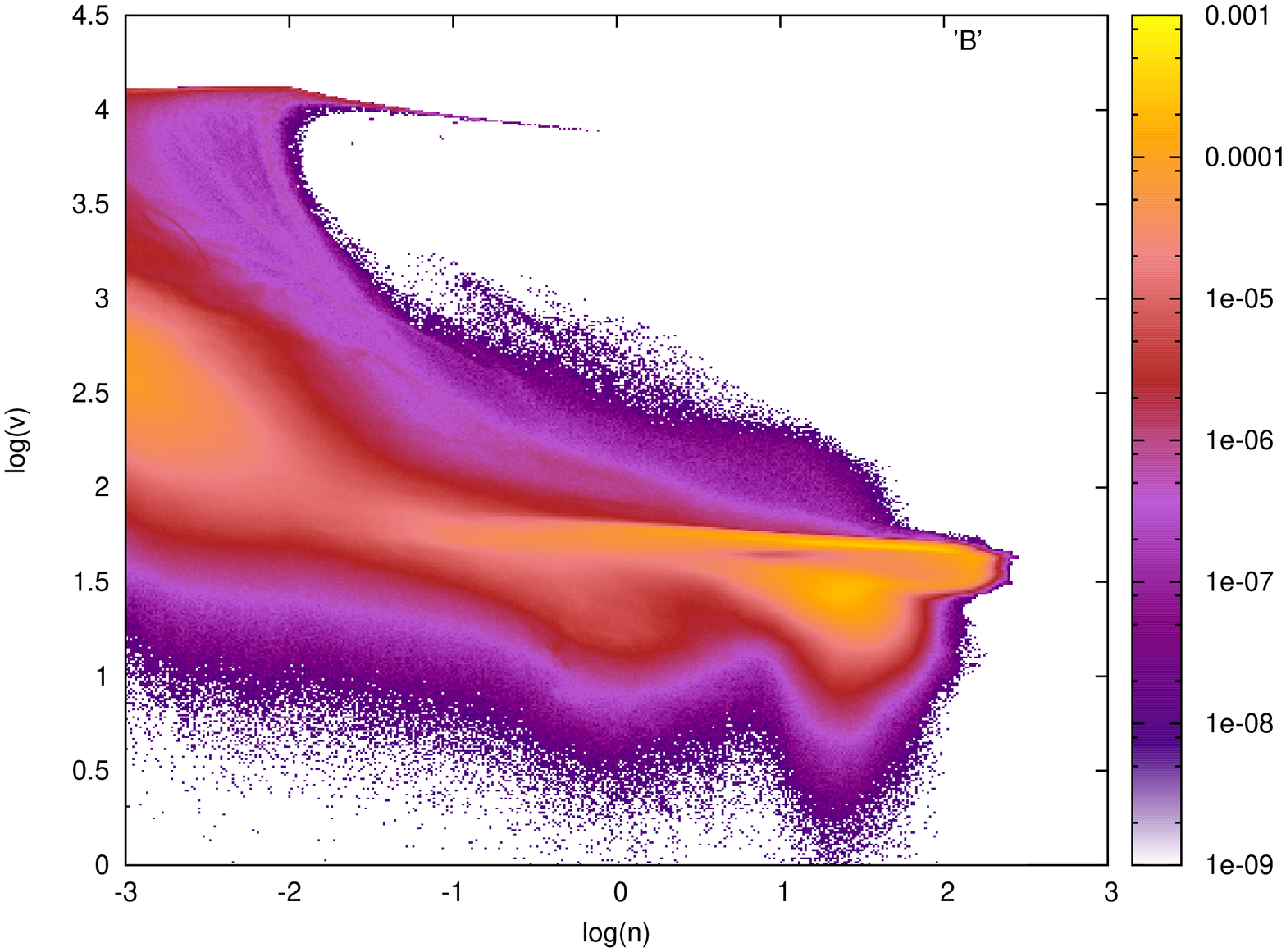}
        }%
    \end{center}
\caption{Two dimensional histogram of the vertical velocity versus 
density} calculated considering every cell within 
the whole simulated system for the models SyH-SNI-SB $large$ {\bf (top panel) 
and SyH-SNI-SB-JET $large$ (bottom panel)} at a time 
t=5 Myr. Velocities are in units of km s$^{-1}$ (logarithmic scale) and 
densities are in units of cm$^{-3}$ (logarithmic scale). {\bf The colour bar
indicates the cell number normalized to their total number.}
\label{histVEL} 
\end{figure}

We should also notice that several earlier studies have ruled out 
completely SF, or more specifically, SNe as the main source of wind feedback 
based on the high observed outflow speeds, larger than or around  600 km/s 
\citep[e.g][]{maiolino12, fabian12}. 
However, our simulations indicate speeds between 700 and 1500 km/s even for 
models with no AGN feedback (see SyH-SNI-SB $large$ case in Fig. \ref{vel_V2}). 
As a matter of fact an individual SN-driven gas outflow is expected to have 
velocities of a few hundred km/s. However, these 
velocities are related to the colder, denser part the gas that is pushed by 
the external shock shell of the expanding bubble inflated by the SN itself. 
In a more complex system, like the one we are investigating here, the stellar 
feedback generates several SN bubbles that interact which each other giving 
rise to a complex multi-phase ambient formed by low velocity, high density 
gas mixed with low density, very high temperature gas moving at much higher 
speeds. 
The histogram of the gas density distribution versus velocity, in Figure 
\ref{histVEL}, illustrates this point for the jetless model SyH-SNI-SB $large$
(upper diagram). 
We clearly see that the very small fraction of gas that is accelerated to 
velocities of about 1000 km/s has densities between 10$^{-2}$ and 10$^{-3}$ 
cm$^{-3}$. In other words, the highest outflow speeds are related to a little 
fraction of gas at very low density and very high temperature. In the same 
Figure (bottom diagram), we see that the introduction of the jet increases a 
little the gas fraction in the high velocity tail (to values larger than 
$10^4$ km s$^{-1}$), but the remainder of the gas behaves similarly as the 
jetless model (see also Figures \ref{vel_V2} and \ref{fig:MlostV0}, 
bottom-right panel).

\subsection{Conclusions}

In summary our results indicate that:

\begin{itemize}

\item The nuclear galactic gas evolution of a Seyfert is almost 
insensitive to the passage of the jet and it needs an intense and more 
widespread source of energy injection in the disk to increase the gas 
pressure and drive shock waves and turbulence;

\item The SN explosions and the development of shock waves into the ISM of 
the host galaxy determine the formation of a multiphase ambient, and therefore 
a SB and/or a region with high SFR may inject enough energy to blow-out the 
gas from the disk and at the same time lead to the formation of clouds and 
clumps which can be continuously generated and steadily carried away. 
Therefore, we may roughly expect a preservation of the total cloud 
number;

\item The fact that some clouds have negative velocities implies that they may 
eventually fall back into the disk, while the clouds 
with positive velocities larger than the escape velocity can continue their 
journey through the halo. If dust forms in the way, it may 
eventually also originate molecular clouds;

\item The velocity distribution reproduced by our simulations clearly show that 
one of the main differences between  models with and without a SMBH jet is the presence of
gas with velocity signatures larger than 1500 km s$^{-1}$.
Depending on the column density of the disk of the host galaxy, the amount 
of very high velocity gas when a jet is present is between 5 and 20 times 
larger than in models without a jet;

\item The jet plays an important role in producing 
the very high gas velocities observed in the expanding nuclear regions of the 
Seyfert galaxies, but at the same time these velocities involve such a 
negligible amount of gas mass (compared to the total amount in the outflow) 
that it is unable to drive the gas evolution within the kpc scale. 
Nevertheless, they may be responsible for recent observed high velocity 
features in Seyfert and ULIRGs \citep[e.g.][]{Morganti5063,tombesi15};

\item Our results basically reproduce the structures of the gas outflow, 
that is, an extended gas outflow with systemic velocity around the 
nucleus having a broader biconical component perpendicular to the disk 
and an inner component due to the interaction between the jet and the 
galactic disk material.

Finally, we should remark that we have neglected the effects of magnetic 
fields in the present study. Although they are known to be particularly 
important in the formation and propagation of the jet outflow, we have 
neglected their effects here for simplicity. Nevertheless, we may expect 
that its presence would not affect much the overall conclusions of this work. 
In fact, the inclusion of magnetic fields in the shaping of the jet would 
help on one side to confine the beam even more in the central narrow channel, 
therefore further constraining its interaction with the surroundings. 
On the other side, the enhanced collimation and magneto-centrifugal 
acceleration implied by the presence of magnetic fields in the jet would help 
to push captured clumps from the underlying star-formation driven wind to 
much higher speeds as those suggested by the recent observations mentioned 
above.

\end{itemize}  

\section*{Acknowledgements}
This work has been partially supported by the Brazilian Funding Agencies 
FAPESP (CM grant 2011/22078-6, and EMGDP grant 2013/10559-5) and CNPq 
(EMGDP grant 300083/94 − 7). The authors also acknowledge useful comments and 
advices of G. Kowal on the numerical code. We are also indebted to an 
(anonymous) referee and to R. Morganti and F. Tombesi for
their profitable comments and insightful suggestions that
have greatly helped to improve this manuscript. The numerical simulations were 
performed in the cluster of the Laboratory of Astro-informatics of the 
Astronomy Department of IAG-USP (FAPESP (grant 2009/54006-4) and the cluster 
GAAE of the High Energy and Plasma Astrophysics group of IAG-USP.

\appendix
\section{Seyfert galaxy models with low column density (SyL)}

\begin{figure*}
     \begin{center}
        \subfigure{%
            \label{fig:first}
            \includegraphics[width=0.43\textwidth]{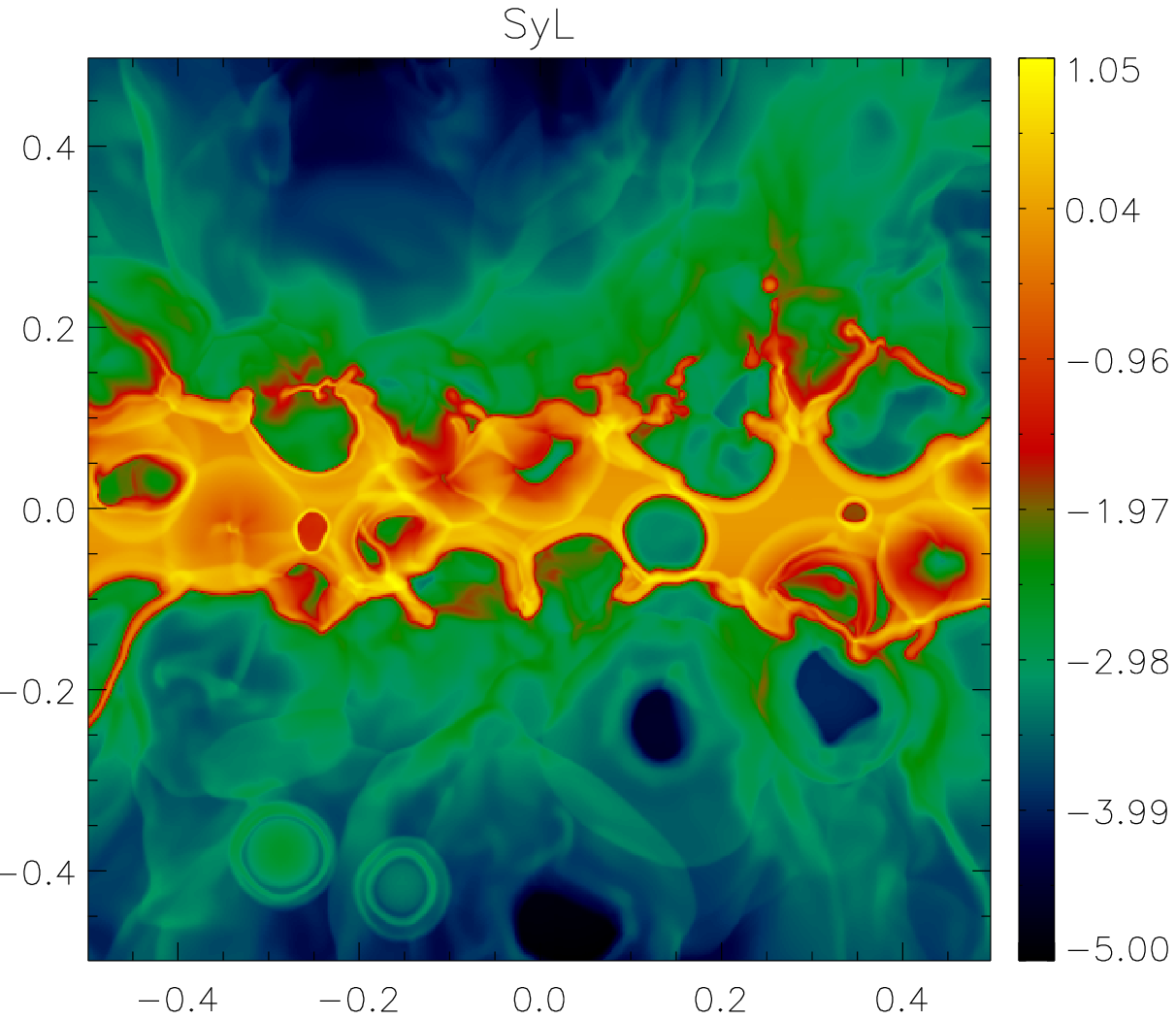}
        }%
        \hspace{0.7cm}
        \subfigure{%
           \label{fig:second}
           \includegraphics[width=0.43\textwidth]{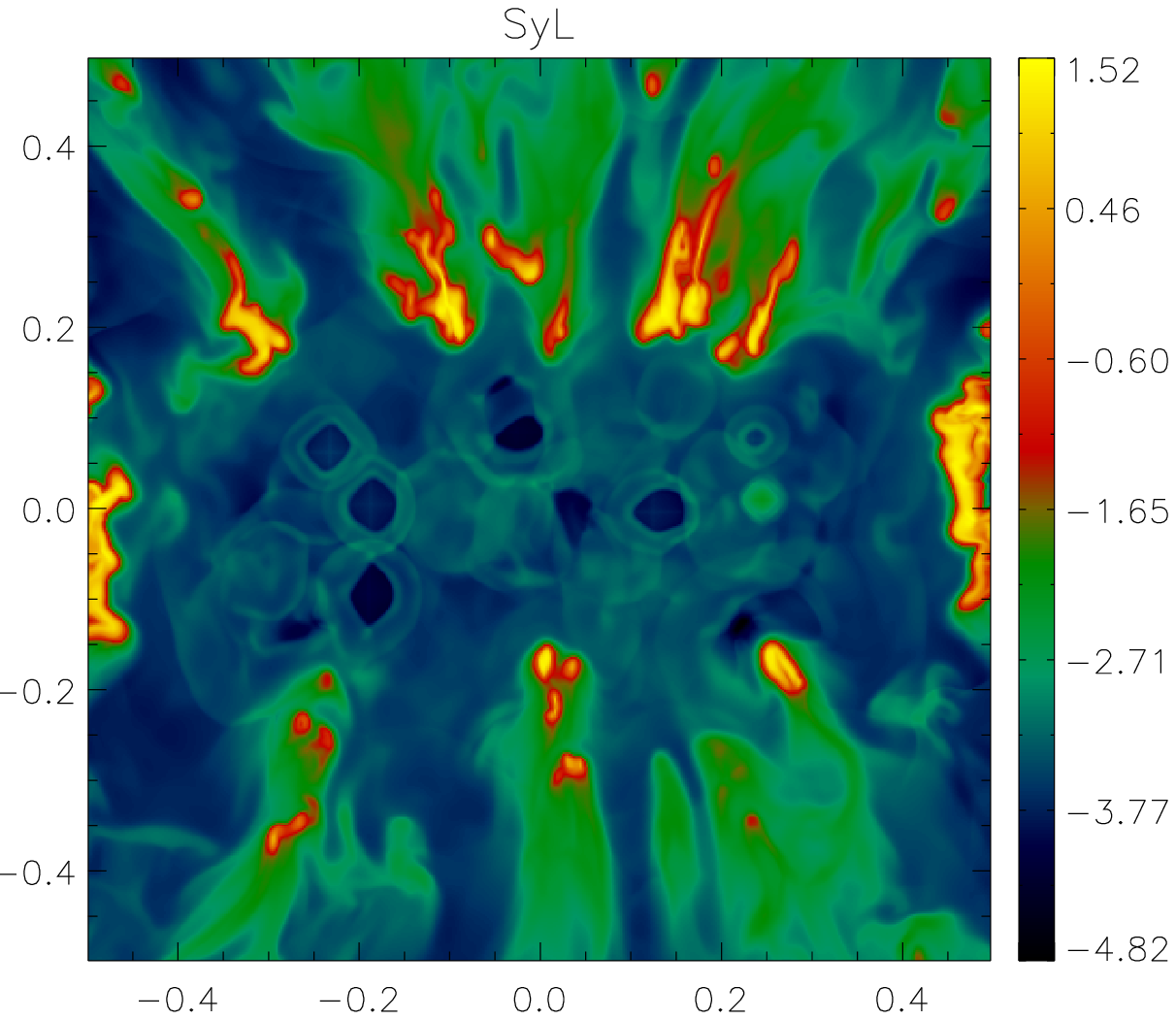}
        }\\ 
        \subfigure{%
            \label{fig:third}
            \includegraphics[width=0.43\textwidth]{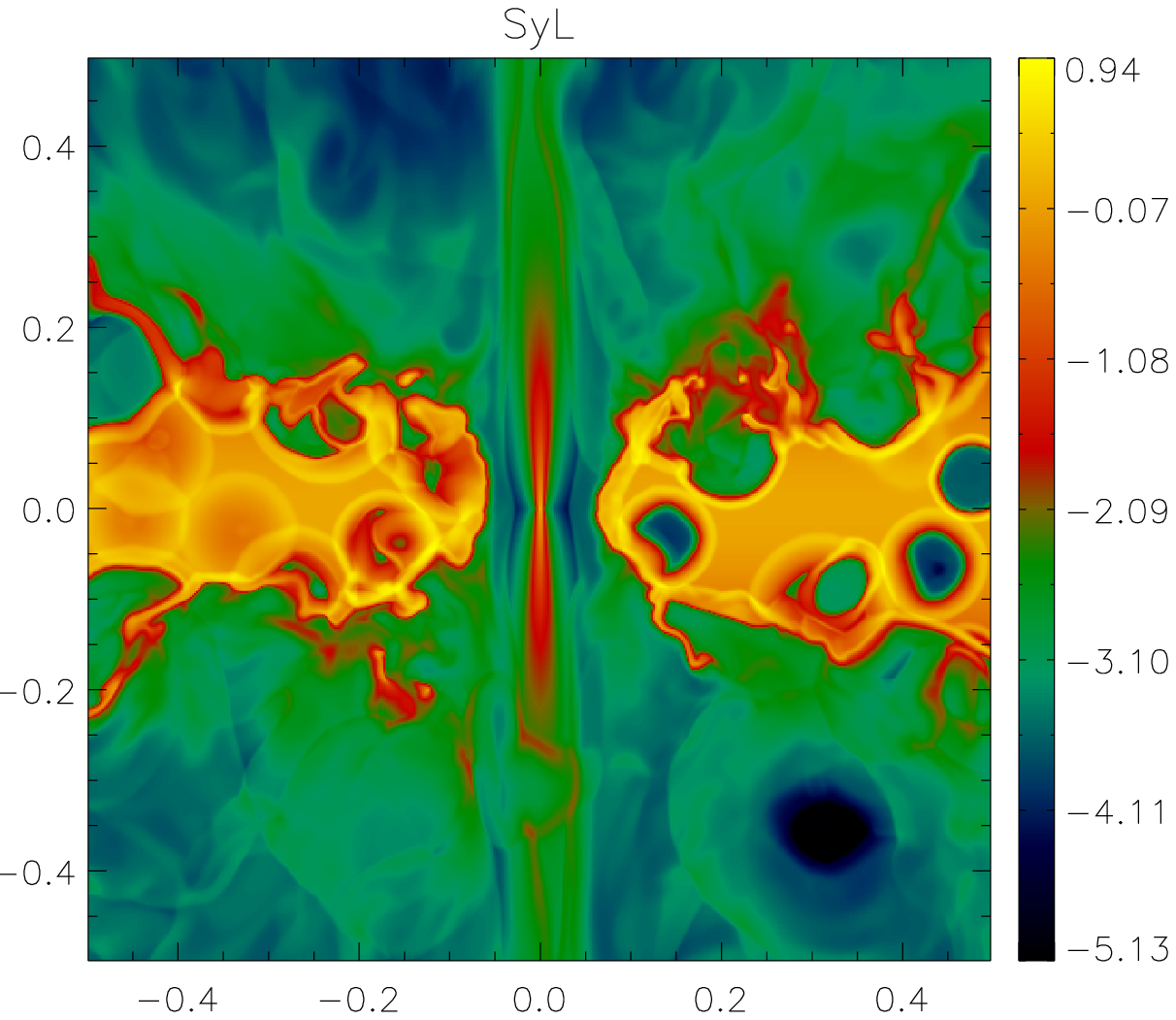}
        }%
        \hspace{0.7cm}
        \subfigure{%
            \label{fig:fourth}
            \includegraphics[width=0.43\textwidth]{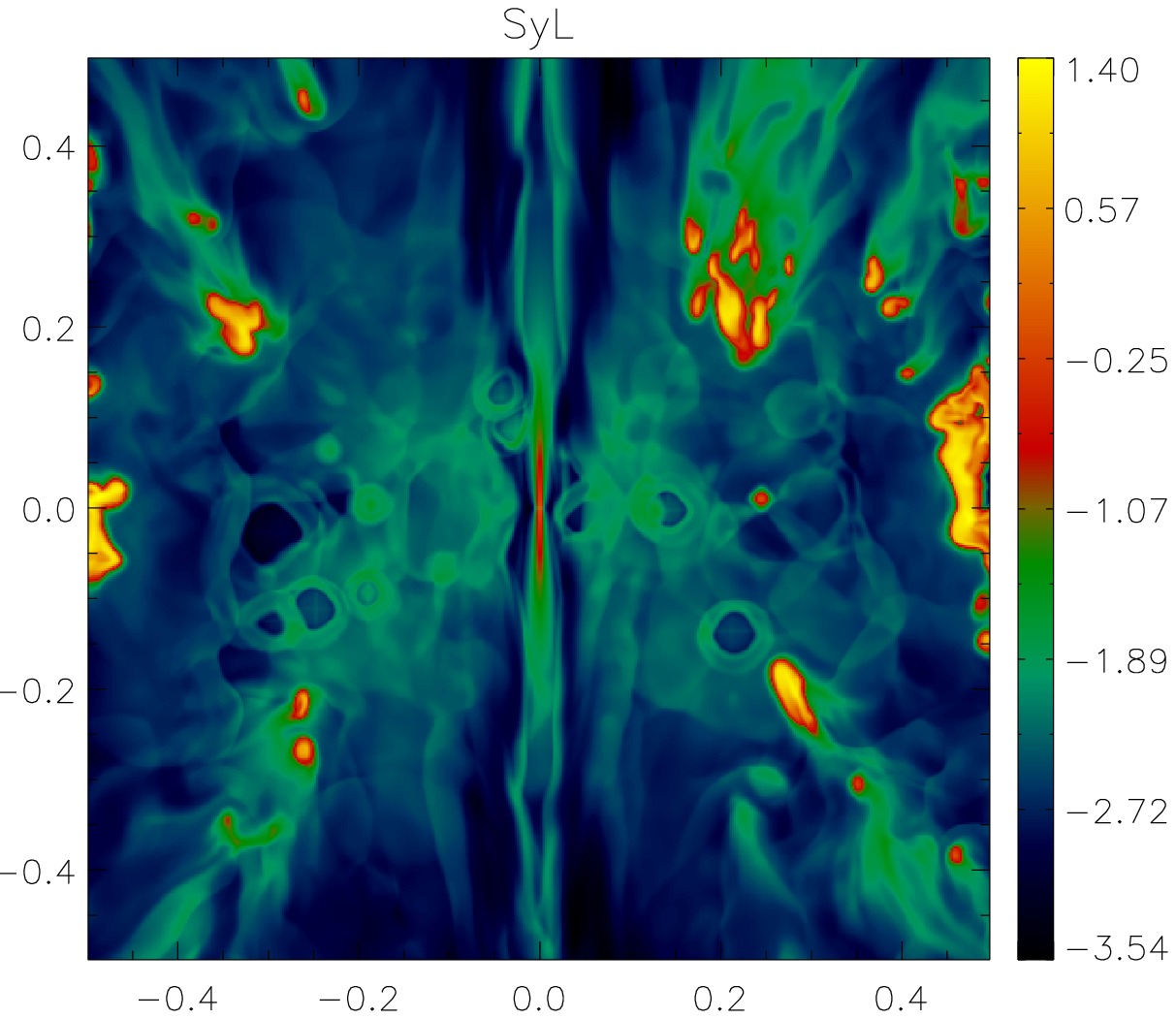}
        }%
    \end{center}
    \caption{Edge-on logarithmic gas density distribution at $t$ = 1.5 Myr, 
for the models SyL-SNI (top-left panel), SyL-SNI-SB (top-right panel),
SyL-SNI-SB-JET (bottom-left panel) and SyL-SNI-JET (bottom-right panel). 
Distances are given in kpc and densities are in cm$^{-3}$}
   \label{FigV0}
\end{figure*}

We have also run numerical models considering the presence (or 
not) of SNI, SB and JET in a disk galaxy characterized by lower values 
of the column density (SyL). 
The results are consistent with those presented above, and the gas
evolution in the disk is completely driven by the SB, as can be seen in Fig.
\ref{FigV0}, where the gas density evolution for each model (SyL-SNI, 
SyL-SNI-SB, SyL-SNI-SB-JET and SyL-SNI-JET) at $t$=1.5 Myr is shown.
 
\begin{figure*}
     \begin{center}
        \subfigure{%
            \label{fig:first}
            \includegraphics[width=0.43\textwidth]{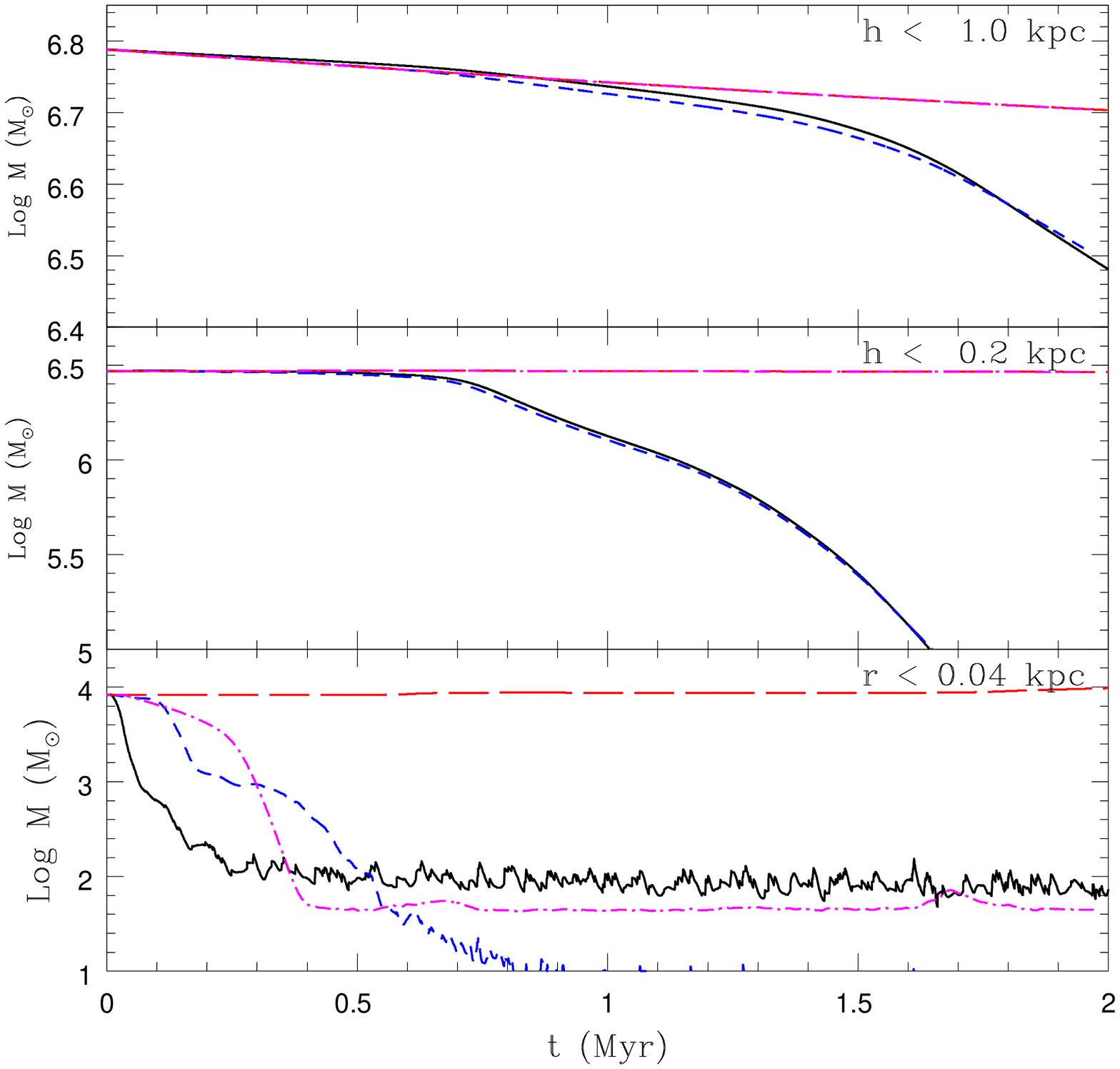}
        }%
        \subfigure{%
           \label{fig:second}
           \includegraphics[width=0.43\textwidth]{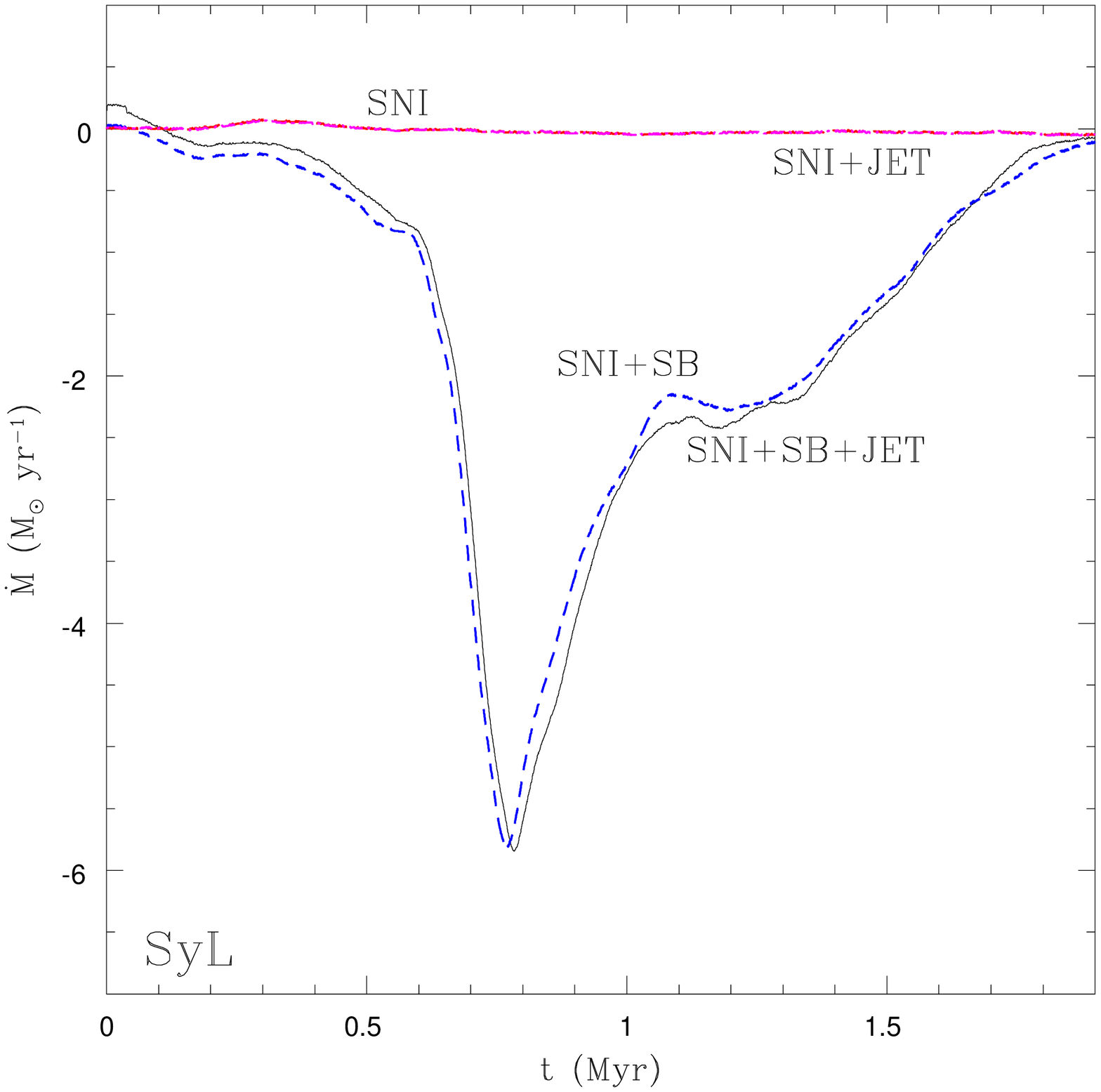}
        }\\ 
        \subfigure{%
            \label{fig:third}
            \includegraphics[width=0.43\textwidth]{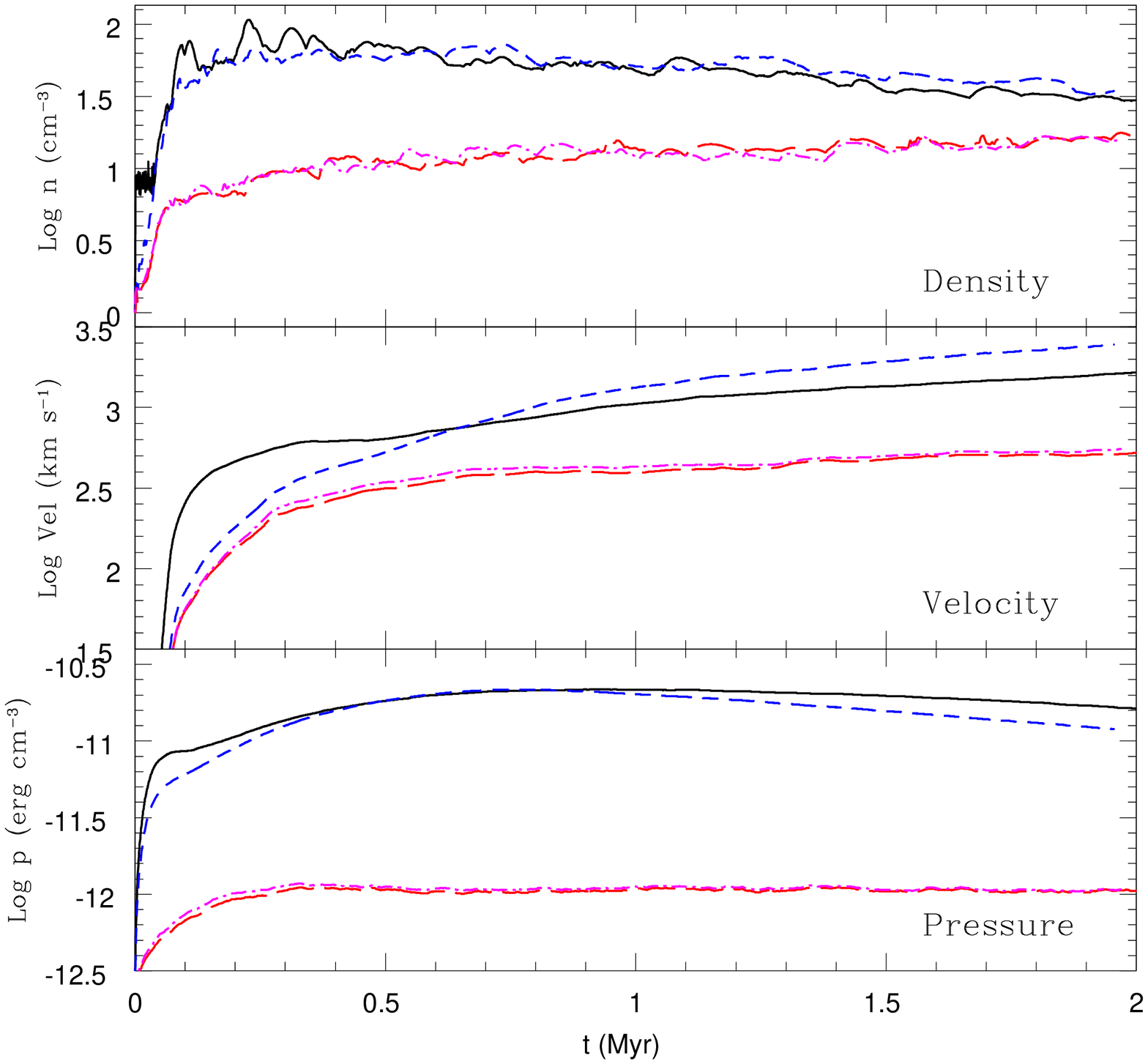}
        }%
        \subfigure{%
            \label{fig:fourth}
            \includegraphics[width=0.43\textwidth]{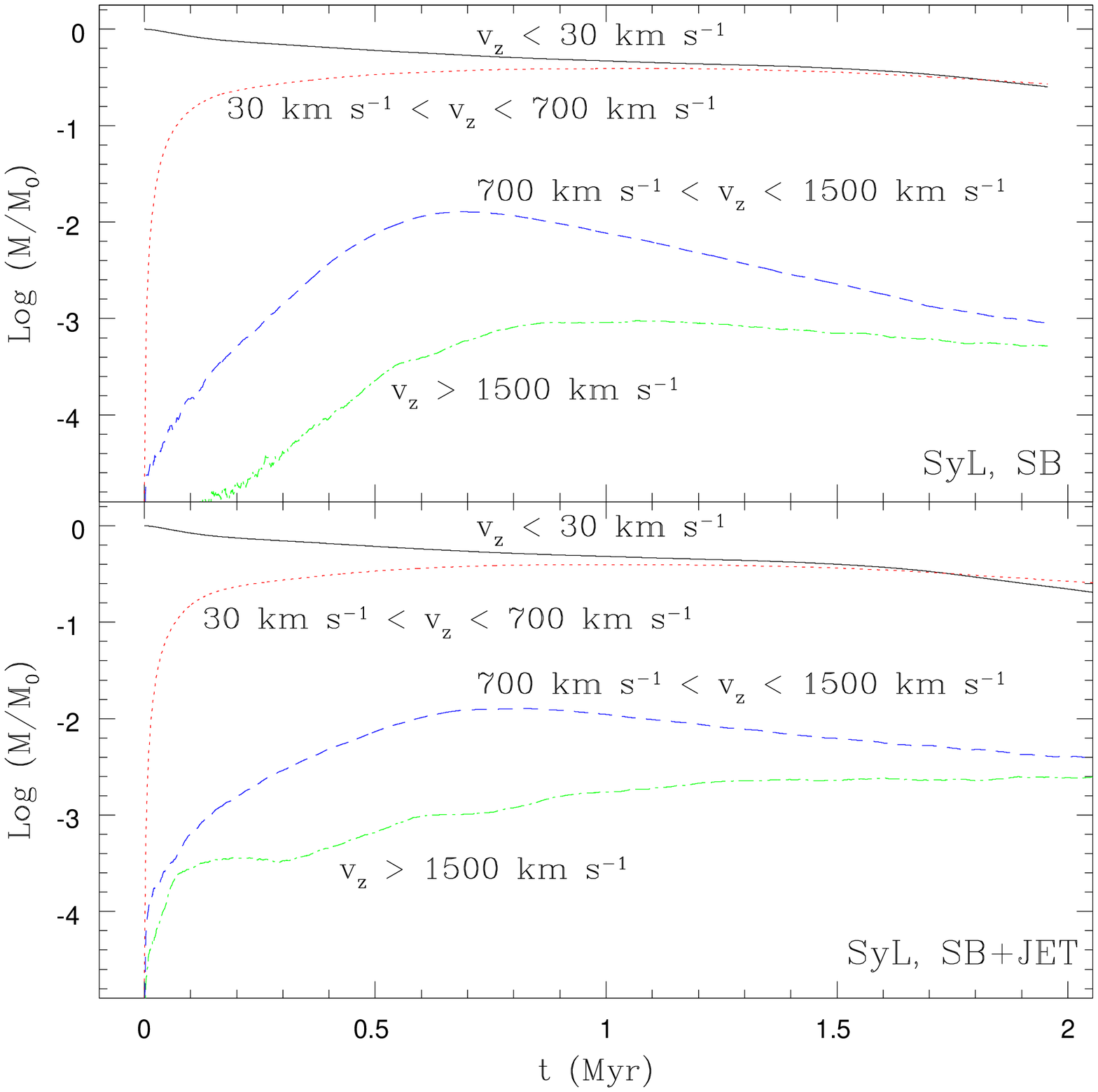}
        }%
    \end{center}
    \caption{Time evolution of the mass of the gas (top-left panel), of 
the gas mass transfer and loss rate of the thick disk (z$\le$ 200 pc) 
(top-right panel), of the main physical variables within
the system (bottom-left panel) and of the mass of gas within the whole
system for different vertical velocities (bottom-right panel) for the models
with the SyL setup. Time is in Myr, mass loss rate is in units of 
M$_{\odot}$ yr$^{-1}$, densities and pressures are in cgs and velocities are in 
units of the reference sound speed, computed at T=5$\times 10^4$ K, 
$c_{s,5\times 10^4}$ = 33 km s$^{-1}$.}
   \label{fig:MlostV0}
\end{figure*}

Examining Figure \ref{fig:MlostV0}
we note that when a SB region is active, regardless of the presence of a SMBH 
jet, the mean mass transfer rate from the the disk is about 3 M$_{\odot}$ 
yr$^{-1}$ over a time of $\sim$2 Myr, resulting in a net total gas mass lost 
by the system of about 2 $\times 10^6$ M$_{\odot}$. 
Also the velocity distribution, shown in bottom-right panel of Fig. 
\ref{fig:MlostV0}, confirms the great similarity between the two models with 
and without the SMBH jet. 
Both models lost about 40\% of their mass, and have about 95\% of the
remaining mass characterized by the same velocity profile. 

\bibliographystyle{apj} \bibliography{agn_ref}

\end{document}